\documentclass[]{aa}
\usepackage{graphicx}
\usepackage{lscape}
\usepackage[maxfloats=40]{morefloats}
\usepackage{hyperref}
\usepackage{longtable}
\usepackage{natbib}
\bibpunct{(}{)}{;}{a}{}{,}

\begin{document}

\title{The first GRB survey of the IBIS/PICsIT archive}

\author{V.~Bianchin\inst{1} \and S.~Mereghetti\inst{2} \and C.~Guidorzi\inst{3} \and L.~Foschini\inst{4} \and G.~Vianello\inst{5} \and G.~Malaguti\inst{1} \and G.~Di~Cocco\inst{1} \and F.~Gianotti\inst{1} \and F.~Schiavone\inst{1}}

\institute{INAF/IASF-Bologna, Via Gobetti 101, I-40129 Bologna, Italy\\\email{bianchin@iasfbo.inaf.it}
\and INAF/IASF-Milano, Via Bassini 15, I-20133 Milano, Italy
\and Dipartimento di Fisica, Universit\`a di Ferrara, Via Saragat 1, I-44100, Ferrara, Italy
\and INAF - Osservatorio Astronomico di Brera, Via Bianchi 46, I-23807 Merate, Italy
\and W. W. Hansen Experimental Physics Laboratory, Kavli Institute for Particle Astrophysics and Cosmology, Department of Physics and SLAC National Accelerator Laboratory, Stanford University, Stanford, CA 94305, USA}
\date{Received --; accepted --}
\abstract
{The multi-purpose \emph{INTEGRAL} mission is continuously contributing to Gamma Ray Burst (GRB) science, thanks to the performances of its two main instruments, IBIS and SPI, operating in the hard X-ray /  soft $\gamma$-ray domain.}
{We investigate the possibilities offered to the study of GRBs by PICsIT, the high-energy detector of the IBIS instrument. }
{We searched for transient episodes in the PICsIT light curves archive from May 2006 to August 2009, using stringent criteria optimized for the detection of long events. In the time interval under examination PICsIT provides an energy coverage from 208 to 2600~keV, resolved in eight energy channels, combined with a fine time resolution of 16~ms. }
{PICsIT successfully observes GRBs in the $260-2600$~keV energy range with an incoming direction spread over half the sky for the brightest events. 
We compiled a list of 39 bursts, most of which are confirmed GRBs or simultaneous to triggers from other satellites/instruments. 
We produced light curves with a time sampling down to 0.25~s in three energy intervals for all events. 
Because an adequate response matrix is not yet available for the PICsIT burst sample, we obtained a calibration coefficient in three selected energy bands by comparing instrumental counts with physical fluences inferred from observations with different satellites. 
The good time resolution provided by the PICsIT data allows a spectral variability study of our sample through the hardness ratio. 
}
{}


\titlerunning{The first GRB survey of the IBIS/PICsIT archive}
\authorrunning{V.~Bianchin et al.}
\maketitle
\section{Introduction}
The \emph{INTEGRAL} satellite \citep{winkler03}, launched in October 2002 and still in operation, is an international observatory exploiting a set of coded mask instruments to explore the sky in the hard X-ray /  soft $\gamma$-ray domain with fine imaging and high spectral resolution.
Although \emph{INTEGRAL} was conceived as a multi-purpose high-energy mission not specifically optimized for the observation of GRBs, it has been making a relevant contribution to the study of these fascinating and in many respects still mysterious events thanks to the high sensitivity and good imaging performances of its two main instruments IBIS  \citep{ubertini} and SPI \citep{vedrenne03}.  
Gamma ray bursts are detected in the IBIS field of view at a rate of $\sim$10 per year.
Their positions, with an accuracy of few arcminutes, are automatically distributed in near real time thanks to the \emph{INTEGRAL} burst alert system (IBAS, \citealt{mereghetti03}), enabling rapid follow-up observations. 
A systematic spectral analysis of all GRBs detected with IBIS until 2008 has been presented  by \citet{vianello09}, while \citet{foley08} carried out a comprehensive study of a sample  of 46  bursts, focusing mainly on their timing properties using IBIS and SPI. 
\emph{INTEGRAL} is also providing light curves for about 15 triggers per month \citep{rau}, detected with the anticoincidence shield (ACS) of the SPI instrument. 
These light curves, with 50 ms resolution, are used for GRB triangulations with other satellites of the inter planetary network (IPN, \citealt{hurley04}). 
Although SPI-ACS provides no spectral information, \citet{vigano09} derived a calibration factor to convert the instrumental count rate to physical flux on the basis of the observation of a large sample of GRBs. 

In this paper we explore the possibilities for GRB studies offered  by PICsIT \citep{dicocco03}, the high-energy detector of the IBIS instrument, by presenting the results of a search for GRBs detected from May 2006 to August 2009.

\section{Data analysis}
\begin{figure*}[htb!]
  \centering
  \includegraphics[width=1.2\textwidth]{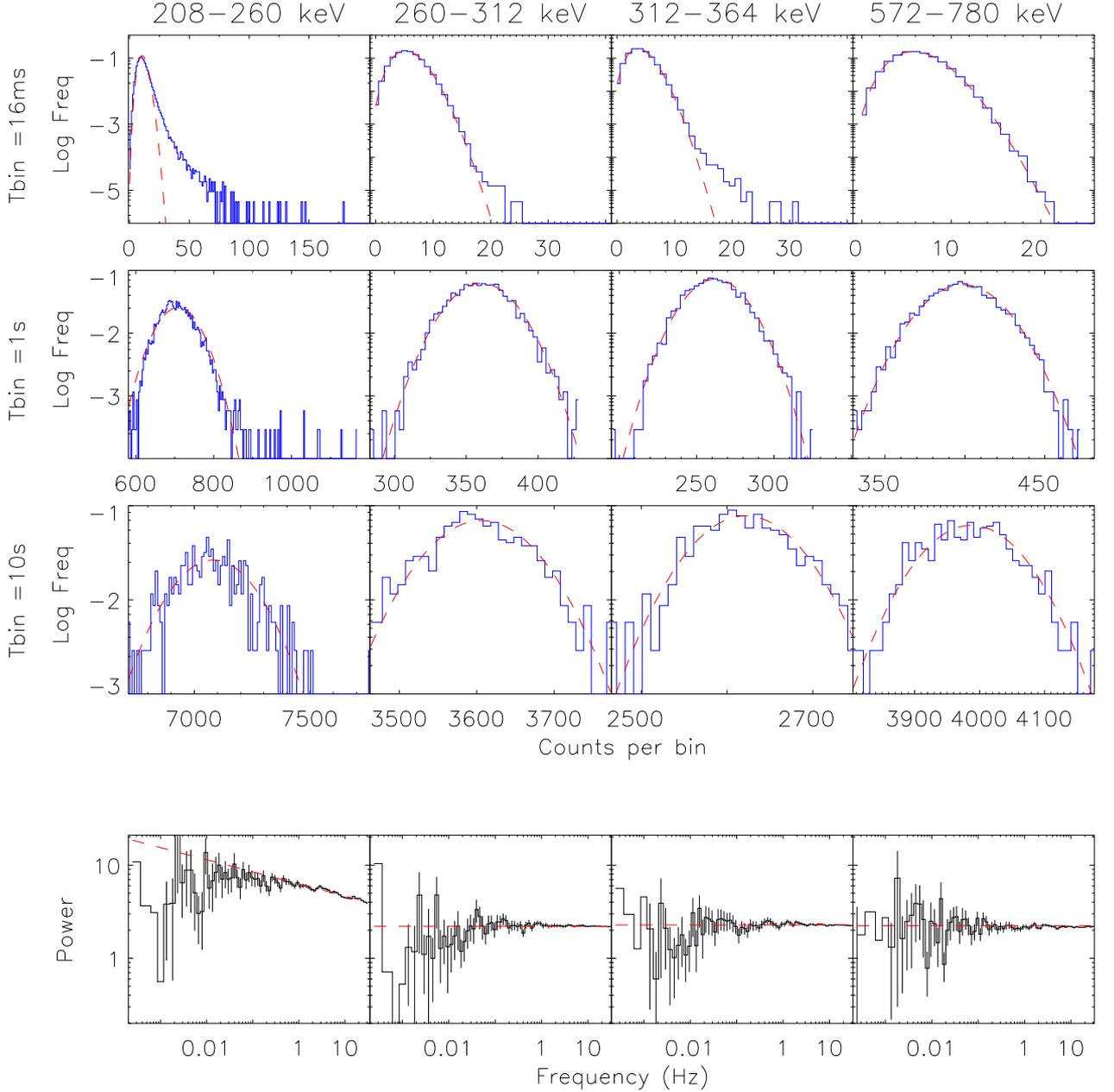}
  \caption{\label{count_distrib} Count distribution and power spectrum in four energy bands for ScW 04820018. Count distributions are obtained with a time resolution of $\sim 16$~ms (\textit{first row}), $1$~s (\textit{second row}) and $10$~s (\textit{third row}). Poissonian (\textit{upper panels}) and Gaussian (\textit{second and third rows}) models are superimposed as (red) dashed lines. The number of bins is 222256, 3462, 345, for $16$~ms, $1$~s and $10$~s, respectively. The power spectra (\textit{bottom panels}) show a white noise background in all energy bands with the exception of the 208-260~keV channel.}
\end{figure*}

\subsection{PICsIT data}
\begin{table}[!tb]
\centering{
\begin{tabular}{cc}
\hline
\hline
Channel &   Energy Band \\
       &          (keV)      \\
\hline
1              & 208-260            \\
2                & 260-312            \\
3             & 312-364            \\
4        & 364-468            \\
5            & 468-572            \\
6          & 572-780            \\
7       & 780-1196           \\
8       & 1196-2600          \\
\hline
\hline
\end{tabular}}
\caption{\label{spt_conf} On-board configuration of the energy channels of spectral timing data since May 2006. The time binning is 16 ms for all energy channels.} 
\end{table}
\emph{INTEGRAL} moves along a highly eccentric orbit with a revolution period around the Earth of 3 days,  perigee of $\sim 10,000$~km and apogee of $\sim 150,000$~km. Observations are carried out when the satellite is above the radiation belts, at an altitude higher than $\sim 40,000$~km. 
To optimize the image reconstruction a dithering strategy is adopted during most scientific acquisitions: each observation is split into a set of pointings (science windows - ScWs) covering a grid of directions spaced $2^\circ$ apart according to a rectangular or an hexagonal pattern. 
The typical duration of each ScW is $\sim 2,000$~s. 

The PICsIT detector consists of  a $64\times64$ pixels array of  CsI(Tl) scintillators, operating in the nominal energy band 175~keV-10~MeV. 
A combination of active and passive shieldings partially surrounds the detector to reduce the background and  to limit its field of view to the directions corresponding to the coded mask ($29^\circ \times 29^\circ$). 
With increasing energy the passive shielding becomes progressively transparent to the incoming radiation and PICsIT is expected to detect GRBs even at far off-axis directions, outside its nominal field of view. 

The first GRB detected by PICsIT (GRB~021125-\citealt{malaguti03}) occurred during the in-flight calibration phase, when data acquisition (in photon-by-photon mode) maintained the full spatial, temporal and energy information. 
When PICsIT is operated in standard scientific mode, its data are pre-processed on-board and organized in different kinds of histograms before they are sent to the ground. 
This is required to comply with the tight telemetry rate available, which does not permit to transmit the whole information (time, energy and position) of each detected count. 
The standard histograms consist of spectral imaging (SI) and spectral timing (ST) data, optimized for spatial and temporal resolution, respectively. 
Spectral imaging histograms preserve the full spatial information and can be used to derive sky images in different energy bands. 
However, they are integrated over the whole duration of each ScW (typically a few thousand seconds), and are in general useless for GRBs, which are bright on shorter time scales and therefore diluted and not detectable in the SI data (the only exception to date is the very bright and long GRB~041219A, which has been detected in both SI and ST data sets - \citealt{dicocco07}). 

In this work we used the ST data, which provide the total rate of events on the detector plane with a configurable time binning down to 1~ms, and an energy resolution in up to eight energy channels. 
The ST data do not permit to produce images because the spatial information is not preserved. 
The on-board configuration of the ST data has been changed a few times during the mission lifetime to refine the energy sampling and enhance the signal-to-noise level of the detected sources. 
We restricted the analysis to the time period from May 2006 to August 2009 (revolutions 440--834), during which the ST time binning was set to 16~ms and the energy channels, covering the range 208-2600~keV, remained fixed at the nominal values given in Table~\ref{spt_conf}. 

\begin{table*}[!ht]
\caption{\label{tab_sample} Observables of the PICsIT burst sample}
\begin{tabular}{lclcccclcc}
\hline
\hline
Burst & ScW & UT       & T$_{90}$   & Fluence         & Peak Flux        & Energy range & Instrument & Ref & Notes\\
      &     &(hh:mm:ss)& (s)        & (ct)            &  (ct/s)          & (keV)        &            &     & \\
\hline
\hline 
060805B& 4650034 & 14:27:15 &   4.7$\pm$  0.4&  6315$\pm$186 & 3600$\pm$208 & 260-2600 & IA, K, R       & (1)  & \\
060819 & 4700017 & 18:28:13 &  14  $\pm$  2  & 12228$\pm$305 & 2394$\pm$195 & 260-2600 & IA, K          &      & (a) \\
060901 & 4740058 & 18:43:56 &   8  $\pm$  1  &  2199$\pm$182 & 1671$\pm$186 & 260-780  & II, K          & (2)  & (b) \\
060905 & 4750078 & 14:48:51 & 239  $\pm$ 47  & 25055$\pm$849 & 1918$\pm$188 & 260-572  &                &      &   \\
060928 & 4830025 & 01:20:06 &  31  $\pm$  4  & 38936$\pm$462 & 4966$\pm$220 & 260-2600 & B, IA, K, R, S & (3)  & (c) \\
061031 & 4940040 & 12:19:47 &  34  $\pm$  3  &  4707$\pm$329 & 1367$\pm$186 & 260-572  & K              &      & (d) \\
061122 & 5010073 & 07:56:52 &   $>$1.50      &  $>$2196      & $>$ 3212     & 260-780  & II, K          & (4)  & (e) \\
061222A& 5110078 & 03:30:15 &  10.5$\pm$  1.5&  3509$\pm$236 & 1871$\pm$191 & 260-1196 & B, K           & (5)  & (f) \\
070207 & 5270055 & 21:00:48 &  10.0$\pm$  0.5&  6178$\pm$255 & 3700$\pm$210 & 260-2600 & B, IA, K, S    & (6)  & \\
070227B& 5340043 & 21:39:23 &  32$\pm$   13  &  3341$\pm$316 &  672$\pm$179 & 260-572  & IA, R          &      & (g) \\
070326 & 5430018 & 00:45:26 &  26.2$\pm$ 0.7 &  3080$\pm$302 & 1466$\pm$189 & 260-468  & IA             &      & (f) \\
070329 & 5440037 & 14:59:48 &  40$\pm$    3  &  8720$\pm$495 & 1048$\pm$185 & 260-2600 & IA, K          &      & (g) \\
070403 & 5460003 & 12:40:28 &  47$\pm$    3  &  5788$\pm$347 &  701$\pm$181 & 260-468  &                &      & \\
070418 & 5510013 & 17:16:19 &  31$\pm$    2  &  5471$\pm$369 & 1461$\pm$192 & 260-780  & K              &      & (f)\\
070429 & 5540054 & 02:40:28 &  20.50$\pm$0.4 &  3081$\pm$326 & 1634$\pm$192 & 260-2600 &                &      & (f) \\
070829 & 5950047 & 20:08:36 &  73$\pm$    1  &  9766$\pm$618 & 1762$\pm$194 & 260-1196 & IA, K          &      & \\
070917 & 6010058 & 04:41:24 &   7$\pm$    2  &  2946$\pm$202 & 1119$\pm$95  & 260-780  & IA, K          &      & (f)\\
071003 & 6070014 & 07:40:54 &12.0$\pm$    0.3& 16175$\pm$302 & 4013$\pm$218 & 260-2600 & B, K, II       & (7)  & \\
071006 & 6080009 & 06:42:08 &  47$\pm$    8  &  3332$\pm$401 &  940$\pm$188 & 260-572  & B, K, IA       & (8)  & \\
\hline
\hline
\end{tabular}
{\bf{Column description}}: \textit{Burst}: name of the event (YYMMDD); \textit{ScW}: pointing ID in which the event occurred; \textit{UT}: trigger time; \textit{$T_{90}$}: duration of the event; \textit{Fluence}: fluence in units of counts, referring to the $T_{90}$ interval and to the energy interval given in column 7; \textit{Peak Flux}: in units of count/s in the same time and energy intervals as for fluence; \textit{Instrument}: telescopes/instruments that observed the event (A = AGILE, B= Swift/BAT, F = Fermi/GBM, II = \emph{INTEGRAL}/ISGRI, IA = \emph{INTEGRAL}/SPI-ACS, K= Wind/Konus, R = RHESSI, S = Suzaku/WAM); 
\textit{Ref}: GRBs have been documented by: 
(1) = \cite{hurley06a}, 
(2) = \cite{mereghetti06a}, 
(3) =  \cite{hurley06b}, 
(4) = \cite{mereghetti06b}, 
(5) = \cite{grupe06}, 
(6) = \cite{golenetskii07a}, 
(7) = \cite{schady07}, 
(8) = \cite{cummings07}, 
{\bf{\textit{Notes}}}: 
(a) = T$_{90}$ refers to the 260-312~keV range, that gives a better signal to noise ratio. 
(b) = The bright GRB~060901, detected by \emph{INTEGRAL} 
over $\sim$20~s, saturated ISGRI telemetry but did not affect PICsIT data. In PICsIT light curve the burst shows a multipeak structure extending over 8~s, in accordance with Konus main event (followed, in Konus data, by a faint tail up to 15~s). 
(c) = GRB~060928 is structured in two main episodes at 01:17:05 and 01:17:20, with duration $\sim 14$~s and $\sim 25$~s, respectively. PICsIT observed both events (Fig.~\ref{060928_1}) but our analysis has been restricted to the main second event (Fig.~\ref{060928_2}). 
(d) = A linear fit of the background was required for T$_{90}$ computation. 
(e) = PICsIT light curves are severely affected by data gaps at all energies (Fig.~\ref{lc_061122}). A $T_{90}$ of 11~s was obtained by \citet{mcglynn09} with SPI. 
(f) = For T$_{90}$ computation the background was fitted to a polynomial function of the third degree. 
(g) = For T$_{90}$ computation the background was fitted to a polynomial function of the second degree. 
\end{table*}
\setcounter{table}{1} 
\begin{table*}[!ht]
\caption{Observables of the PICsIT burst sample - Continued.}
\small{
\begin{tabular}{lclcccclcc}
\hline
\hline
Burst & ScW & UT       & T$_{90}$   & Fluence         & Peak Flux        & Energy range & Instrument & Ref & Notes\\
      &     &(hh:mm:ss)& (s)        & (ct)            &  (ct/s)          & (keV)        &            &     & \\
\hline
\hline 
071108 & 6190025 & 21:40:29 & 20.0$\pm$   0.5&  9976$\pm$372 & 5072$\pm$229 & 260-2600 & IA, K, A       &      & (h) \\
080122 & 6440034 & 18:32:43 & 105$\pm$    4  & 10796$\pm$743 & 2112$\pm$199 & 260-1196 & IA, K, S, A, B & (9)  & \\
080204 & 6480067 & 13:56:35 &  14$\pm$    3  &  9901$\pm$312 & 4254$\pm$220 & 260-2600 & K, IA, S, A, B & (10) & \\
080303 & 6580014 & 21:34:45 &29.2$\pm$    1.5& 31507$\pm$461 & 8283$\pm$254 & 260-2600 & IA             &      & \\
080319B& 6630025 & 06:12:50 &  42$\pm$    1  & 29392$\pm$536 & 1905$\pm$198 & 260-2600 & B, K           & (11) & \\
080328 & 6660038 & 08:03:12 &  33$\pm$    2  &  6593$\pm$419 & 2243$\pm$200 & 260-1196 & B, K, S, IA    & (12) & \\
080408 & 6700004 & 10:23:25 &  40$\pm$    1  &  6355$\pm$424 & 1464$\pm$194 & 260-780  &                &      & (i, f) \\
080514B& 6820005 & 09:55:58 &   6$\pm$    1  &  5611$\pm$195 & 3495$\pm$214 & 260-2600 & A, K, S,IA     & (13) & \\
080607 & 6900002 & 06:07:27 &  10$\pm$    1  &  4748$\pm$195 & 2225$\pm$201 & 260-2600 & B, A, K, IA    & (14) & \\
080613B& 6920007 & 11:12:40 &  31.7$\pm$  0.5&  5181$\pm$417 & 1712$\pm$197 & 260-1196 & B, K, IA       & (15) & \\
080615 & 6920047 & 04:07:33 &   9.5$\pm$  0.3&  6769$\pm$239 & 2768$\pm$206 & 260-1196 & K, R           &      & \\
080721 & 7040083 & 10:25:14 &  16.0$\pm$  0.3& 14627$\pm$340 & 4888$\pm$227 & 260-2600 & B, K, IA, R    & (16) & \\
080723B& 7050036 & 13:22:34 &  45$\pm$    2  & $>$9936       & $>$3030      & 260-780  & II, A, K       & (17) & (j) \\
080817A & 7130052 & 03:52:16 &  42$\pm$    2  & 10461$\pm$528 & 1509$\pm$196 & 260-1196 & F, K, IA       & (18) & \\
080918 & 7240065 & 09:44:37 &  2.5$\pm$  0.3 &  3986$\pm$147 & 3916$\pm$220 & 260-2600 & K, IA          &      & \\
090528B& 8080058 & 12:22:57 &  81 $\pm$   7  &  6665$\pm$622 & 1141$\pm$199 & 260-780  & F, S, K, IA    & (19) & (f)\\
090618A& 8150054 & 08:29:26 &  15$\pm$    7  &  3032$\pm$301 &  865$\pm$198 & 260-1196 & B,A,F,K,S,IA   & (20) & (k, f) \\
090623 & 8170026 & 02:34:30 &  40$\pm$    2  &  6517$\pm$489 & 1540$\pm$204 & 260-1196 & F, S, K, IA    & (21) & \\
090626A& 8180029 & 04:32:11 &  42$\pm$    1  & 13167$\pm$545 & 3577$\pm$221 & 260-2600 & F, S, K, IA, B & (22) & \\
090809B& 8330036 & 23:28:17 &   9$\pm$    1  &  3399$\pm$257 & 1501$\pm$203 & 260-1196 & F, K IA        & (23) & (f) \\
\hline
\hline
\end{tabular}}
\textit{Ref}: GRBs have been documented by: 
(9) = \cite{hurley08}, 
(10) = \cite{golenetskii08a}, 
(11) = \cite{racusin08}
(12) = \cite{perri08},  
(13) = \cite{rapisarda08}, 
(14) = \cite{mangano08}, 
(15) = \cite{markwardt08}, 
(16) = \cite{marshall08}, 
(17) = \cite{gotz08}, 
(18) = \cite{bissaldi08}, 
(19) = \cite{vkienlin09a}, 
(20) = \cite{schady09}, 
(21) = \cite{rau09}, 
(22) = \cite{vkienlin09b}, 
(23) =  \cite{horst09}.
{\bf{\textit{Notes}}}: 
(h) = We analyzed the main pulse of the event, which is composed of three peaks over $\sim$70~s. 
(i) = The burst was observed by PICsIT in two main episodes separated by 50~s and extending over 120~s. The first smaller peak seems to be affected by count drops and therefore the analysis was conducted on the second event only. 
(j) = The bright GRB~080723B fell in the ISGRI FoV, with an off-axis angle $\Theta \sim 8^\circ$, and saturated the available telemetry. PICsIT data show gaps at all energies that are limited to few time bins, however, and did not affect the T$_{90}$ computation. We report the lower limits of the peak flux and fluence. 
(k) = in PICsIT data the burst occurs $\sim$60~s after the Swift/BAT trigger and coincides with the brightest portion of this long event ($\sim 130$~s). 
\end{table*}

\subsection{GRB search}
In order to search for possible GRBs detected in the PICsIT ST data, we examined the light curves looking for  statistically significant excesses above the background counts. 
The detector background level is not constant: intra-revolution variations are visible when the satellite is close to the radiation belt passage, the relative position of strong sources can cause different background level as a function of the pointing direction, and solar activity can also produce background variations. 
In addition, the background count rate shows a long-term evolution, which can be ascribed to the detector activation and to the variable cosmic ray environment, modulated with the solar cycle. 

We performed our GRB search analyzing each ScW individually, because these are the longest homogeneous periods with a fixed satellite attitude, and are sufficiently short, compared to the timescales of orbital background variations, to approximate the background with a constant value. 
Therefore, we evaluated the expected background level as follows for any given ScW, energy channel and time rebinning. 
We first removed all time intervals with large spikes or drops verifying: $|C_i - <C>| > 6 \cdot \sigma_C$, where $C_i$ is the number of counts in bin $i$ and $<C>$ and $\sigma_C$ are the mean and standard deviation of counts, respectively, computed over the whole ScW. 
We then defined $<B>$ and $\sigma_B$ as the mean and standard deviation over the cleaned ScW. 
Once the energy band and the time binning were selected, we defined a burst candidate as an excess in the light  curve where the  number of counts $C_i$ is higher than a threshold given by  $<B> +  N_{th} \cdot  \sigma_B$. 
The value of $N_{th}$ and the sample size (in units of bins) establish the probability $P$ of detecting statistical fluctuations above the threshold. 

In order to select the most convenient energy range and time binning for the GRB search, we inspected the distribution of counts within ScWs where no GRBs occur for different time binnings and in all available energy channels. 
Some examples of these count distributions are shown in Fig.~\ref{count_distrib}. 
We found that the count distribution strongly deviates from the Poissonian statistics in the lowest energy channel. 
This is confirmed by the Fourier analysis of the light curves. 
As shown in Fig.~\ref{count_distrib} (bottom panels), the power spectra above 0.2~Hz are consistent with a constant model ($c = 2.23\pm 0.08$, $2.31\pm 0.09$, $2.17\pm 0.07 $, in $260-312$, $312-364$ and $572-780$~keV, respectively), with the exception of the first energy band, which follows a power law trend above 0.1~Hz. 
Above $260$~keV the count distribution is still not consistent with the expected Poissonian statistics, because power spectra describe a white noise process with observed variance exceeding the Poissonian of at least 10\%. 
For this reason the threshold level for the GRB search is expressed in terms of standard deviation. 
The different behavior of the first energy band can be ascribed to a significant contribution of charged particles interacting with the detector CsI crystals and inducing long-lasting phosphorescence emission, which is not eliminated by the anti-coincidence system \citep{segreto}. 
On the basis of this analysis we decided to perform the GRB search in the $260-312$~keV energy band, which seems to be the least affected by the systematic background component and, for the typical spectra of GRBs, maximizes the signal-to-noise ratio. 

A preliminary analysis was conducted on the data sample to explore several rebinning and several threshold levels. 
We selected a threshold level $N_{th} = 8$ applied to light curves rebinned on $10$~s time bins. 
This value guarantees a probability  $P < 10^{-8}$ of statistical fluctuation above the threshold, over the whole data set, of $\sim 8 \cdot 10^6$ time bins. 
The search with $10$~s bins was carried out twice, with a half-bin offset. 

We searched  for the  GRB candidates using the pre-processed data of all ScWs, without an \textsl{a priori} selection to eliminate the ScWs with potential problems caused by, e.g., high solar activity, bright variable sources, high particle background close to the radiation belt entrance/exit, particular instrumental settings or deviations from the nominal operating parameters. 
The resulting ScWs with GRB candidates were then manually screened for the above conditions and re-analyzed with the   specific standard analysis software \texttt{osa v.7.0}, described in \citet{goldwurm03} to derive the relevant physical quantities of the events. 
The time of occurrence of each burst was refined using light curves with 1~s resolution and then cross-checked with solar flares or high activity revealed by the \emph{INTEGRAL} radiation environment monitor (IREM, \citealt{hajdas03}), the GEOS net\footnote{http://www.sec.noaa.gov} and \emph{RHESSI} monitoring \citep{lin02} to exclude particle events. 

\begin{figure}[tb]
  \includegraphics[width=0.38\textwidth,angle=90]{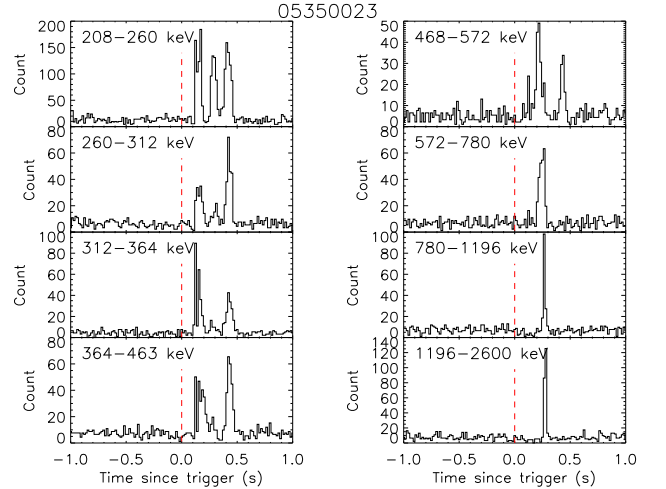}
  \caption{\label{short} Typical light curve of short fake events, probably caused by instrumental effects. Light curves in eight energy bands and 16~ms time resolution refer to the event in ScW~5350023. }
\end{figure}

We also performed a search with binning of 1~s and $N_{th} = 12$   on a subset of our data (revs.~$440-680$). This resulted in 21  events, 10 of which were also detected with the 10 s bin search.
However, a deeper analysis of the remaining 11 events suggests that they are not GRBs. 
Indeed, the light curves of all these events with the on-board resolution ($\sim 16$~ms) show a similar pattern, consisting of three peaks separated by $\sim 0.1$~s, the central one dominating the high-energy bands and progressively dimming as energy decreases (Fig.\ref{short}). 
Although the origin of these short events is uncertain, they are probably an instrumental effect caused by charged particles interactions. 
Since the search for short episodes did not add genuine GRBs to the sample of bursts, we restricted the search to the 10 s binning. 
\section{Results}
The PICsIT burst sample selected with the 10~s binning as described above in the period May 2006 - August 2009 consists of the $39$ events listed in Table \ref{tab_sample}. 
Most of these events (23) have also been detected by other satellites, which provided their localizations, confirming their GRB nature. 
We are confident that the other events, for which no localizations have been obtained, are also most likely GRBs. 
Indeed, 11 of them have also been detected by other instruments such as the \emph{INTEGRAL}/SPI-ACS (sensitive to photons above $\sim$80 keV), Wind/Konus (energy range 10~keV-10~MeV), and RHESSI (50~keV-17~MeV). 
Only four events were detected exclusively by PICsIT. 
Sixteen bursts are hard events detected up to the highest energy channel (1196-2600~keV), 26 reach the 780-1196~keV band, 5 are not visible above 572~keV, and 2 are observed only up to 468~keV. 
\begin{figure}
\includegraphics[width=0.5\textwidth]{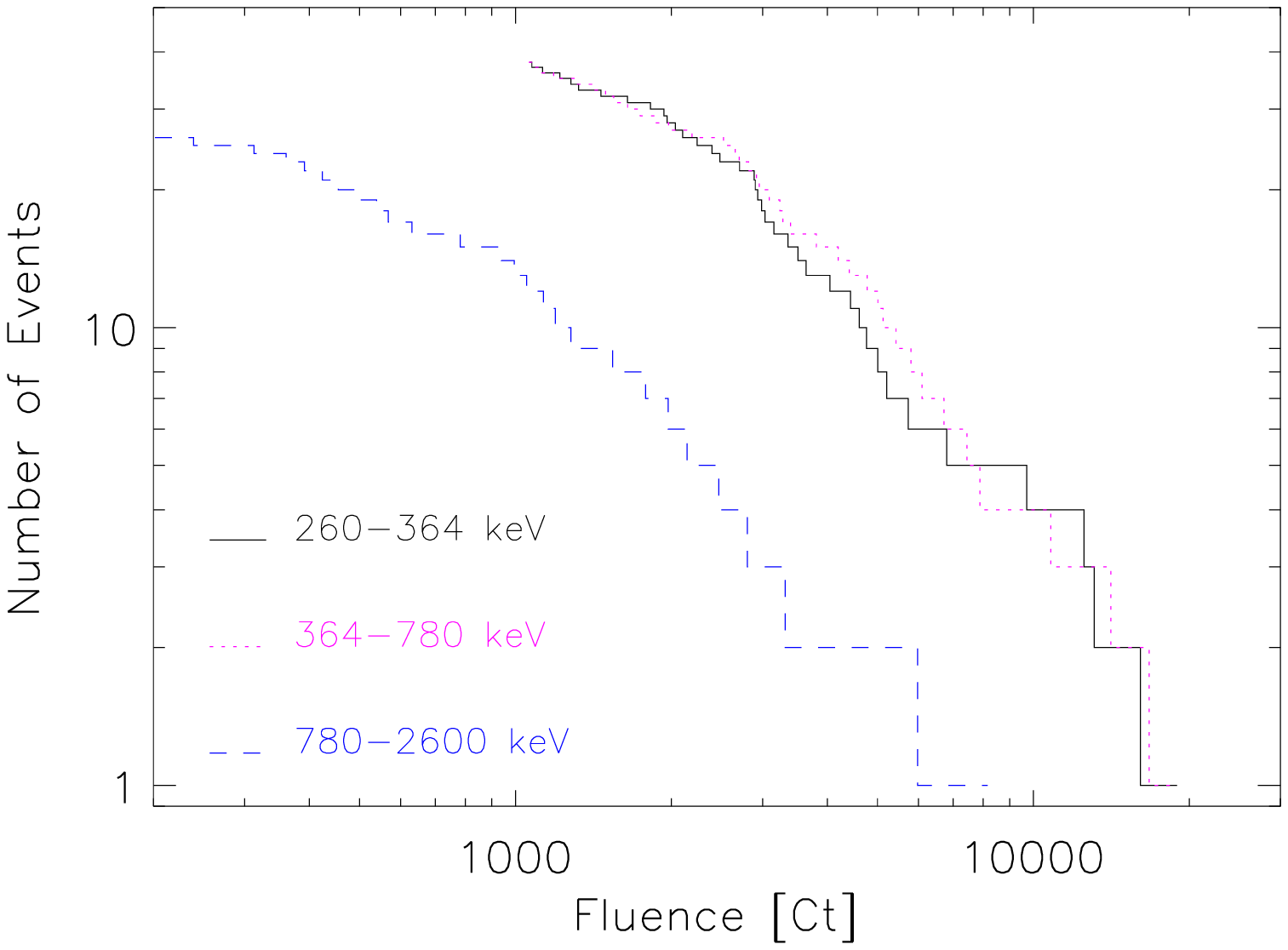}
\includegraphics[width=0.5\textwidth]{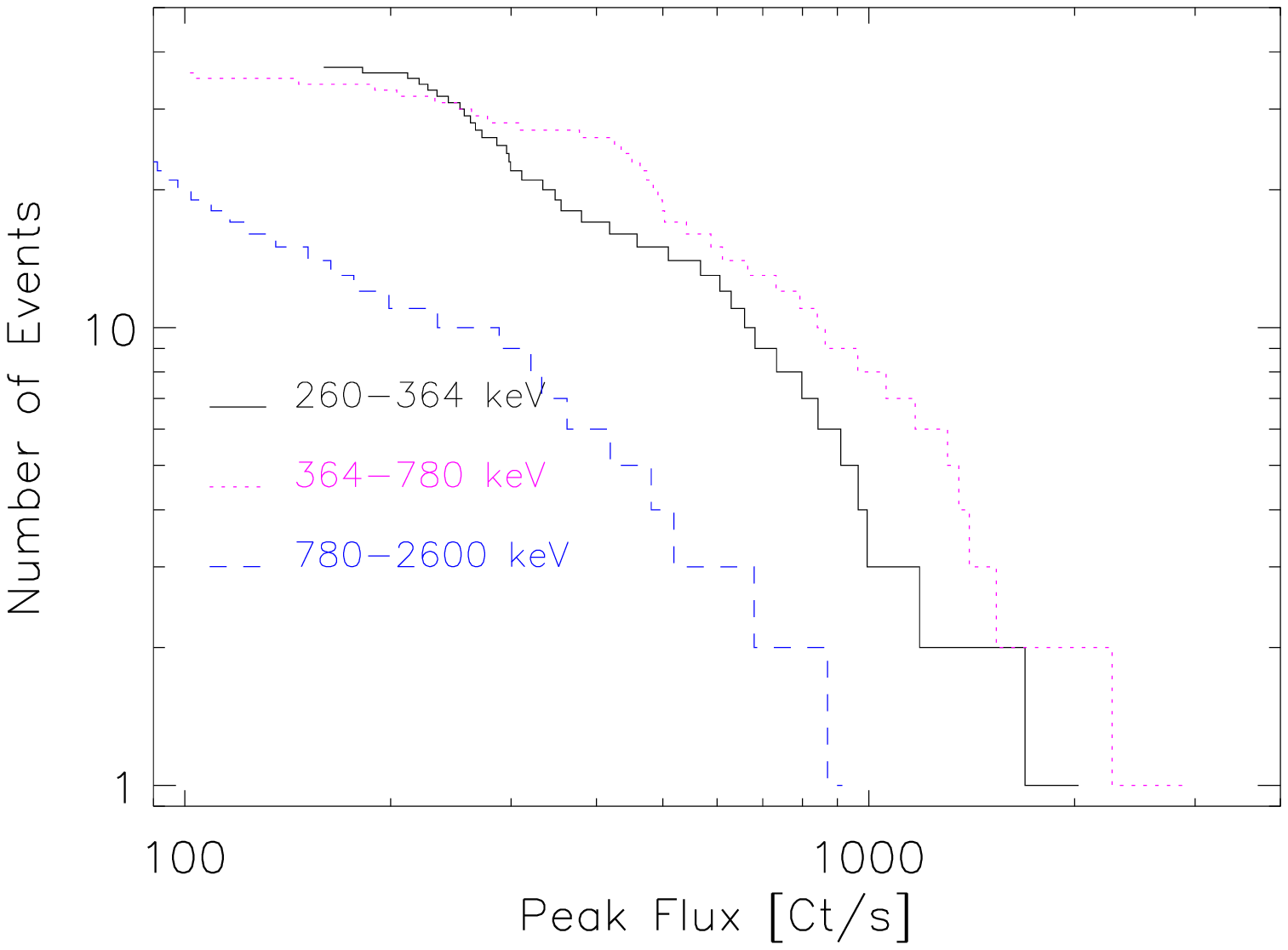}
\caption{\label{fpt_dist} Integral distribution of fluence (upper panel) and peak flux (lower panel) in instrumental units of the PICsIT burst sample.}
\end{figure}
\begin{figure}
\includegraphics[width=0.5\textwidth]{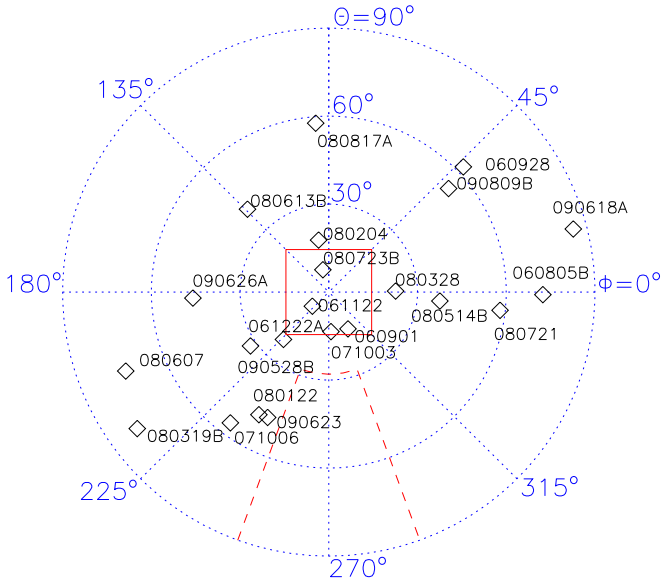}
\caption{\label{theta-phi} Instrumental coordinates of 22 GRBs in the PICsIT sample. The (red) square gives the approximate PICsIT FoV and dashed line borders the area where PICsIT is partially shielded by the spectrometer SPI. }
\end{figure}
\begin{figure}
\includegraphics[width=0.5\textwidth]{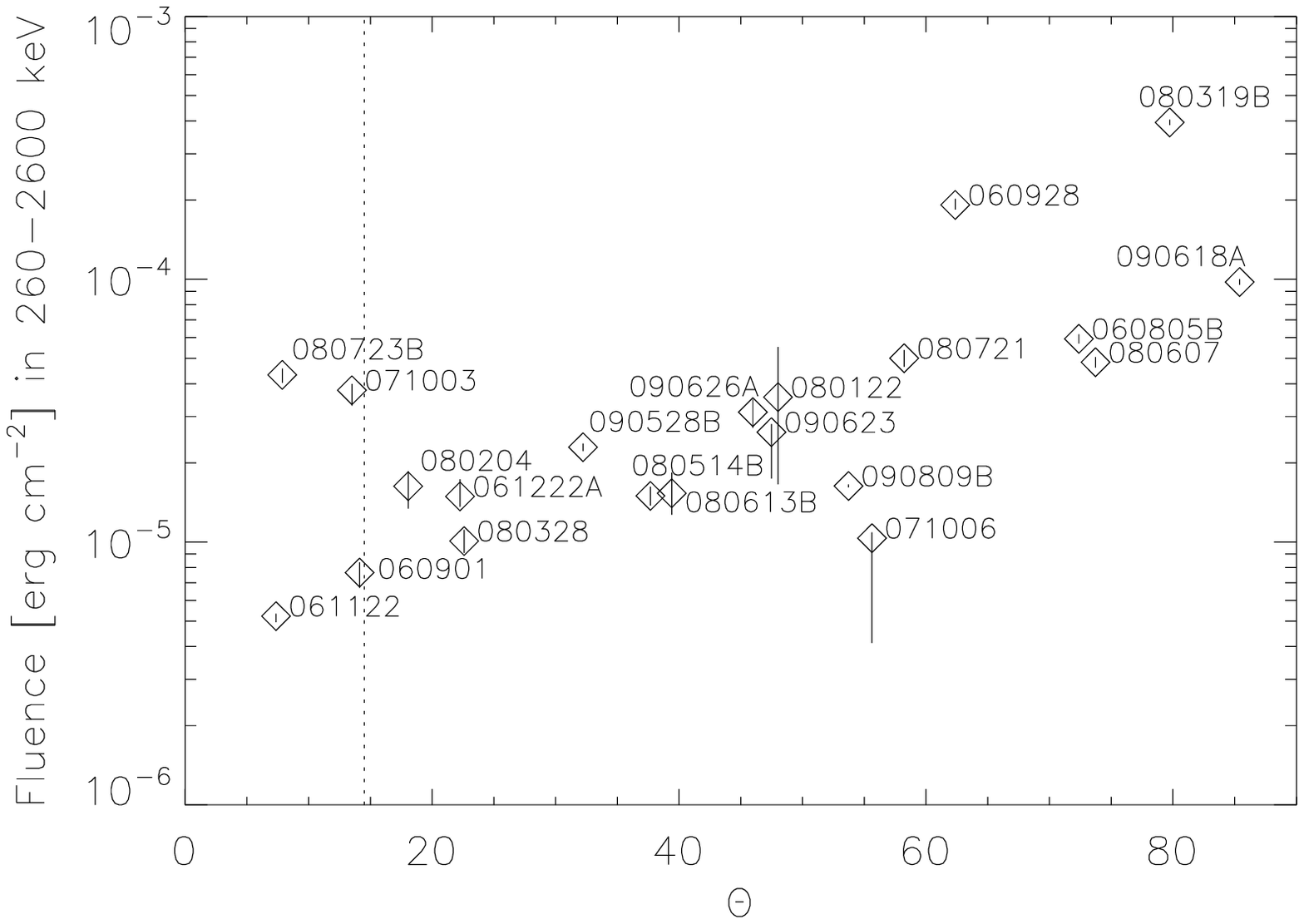}
\includegraphics[width=0.5\textwidth]{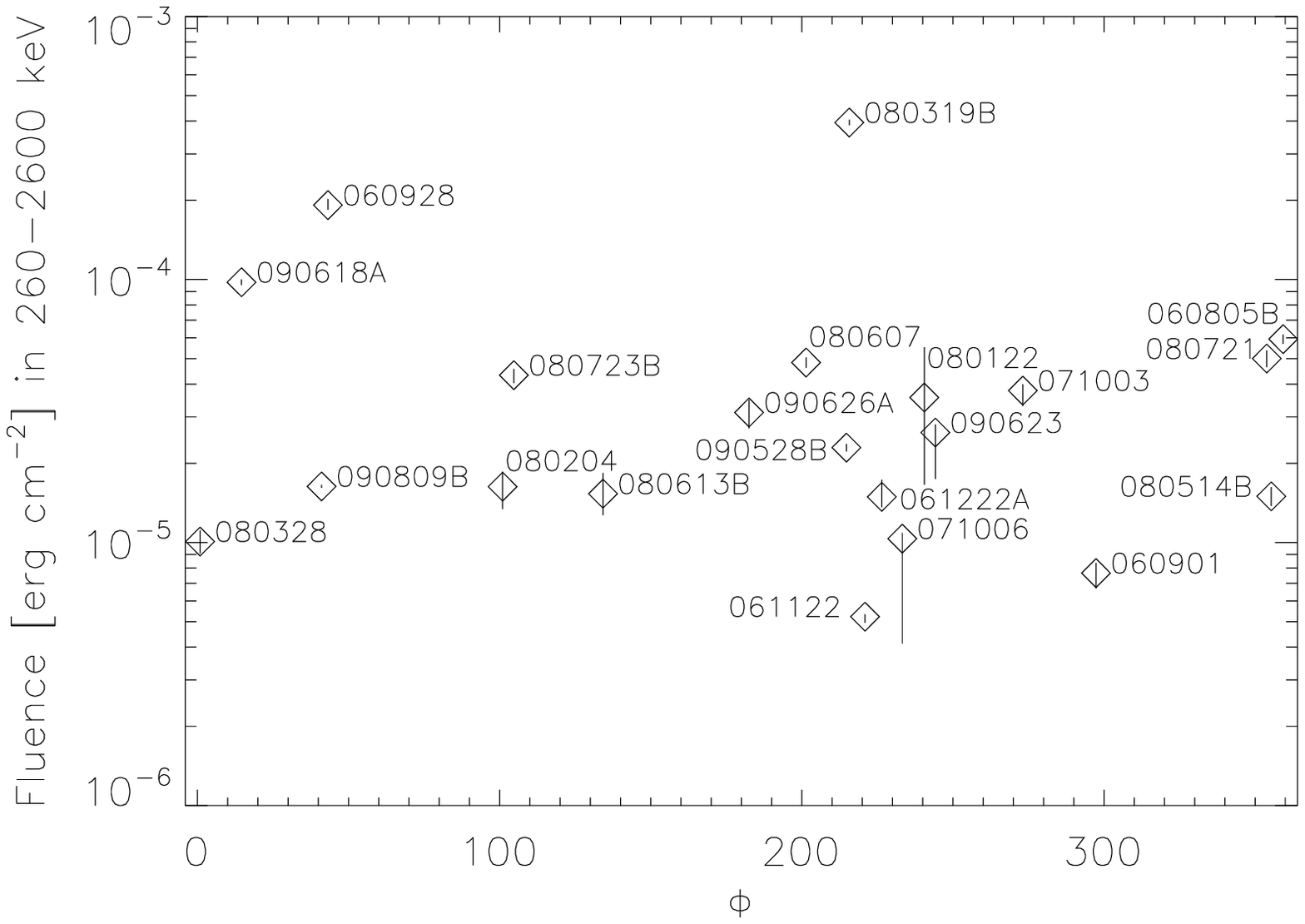}
\caption{\label{flang1} Fluence 260-2600 keV as a function of instrumental coordinates $\Theta$ (upper panel) and $\phi$ (lower panel) for the 21 GRBs with known spectral parameters . The telescope aperture is shown as a dotted line in the upper figure.}
\end{figure}
\begin{figure}
\includegraphics[width=0.5\textwidth]{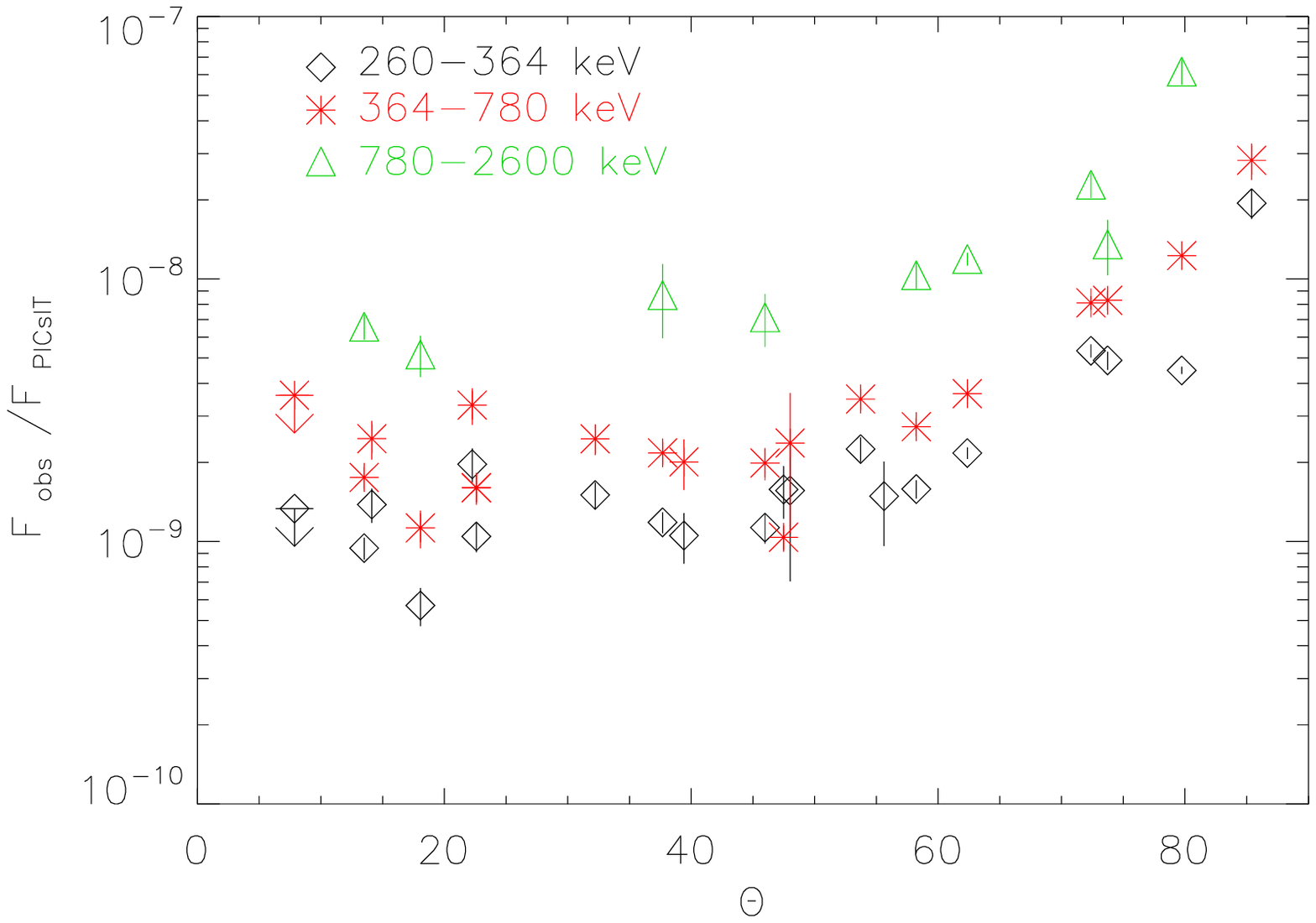}
\includegraphics[width=0.5\textwidth]{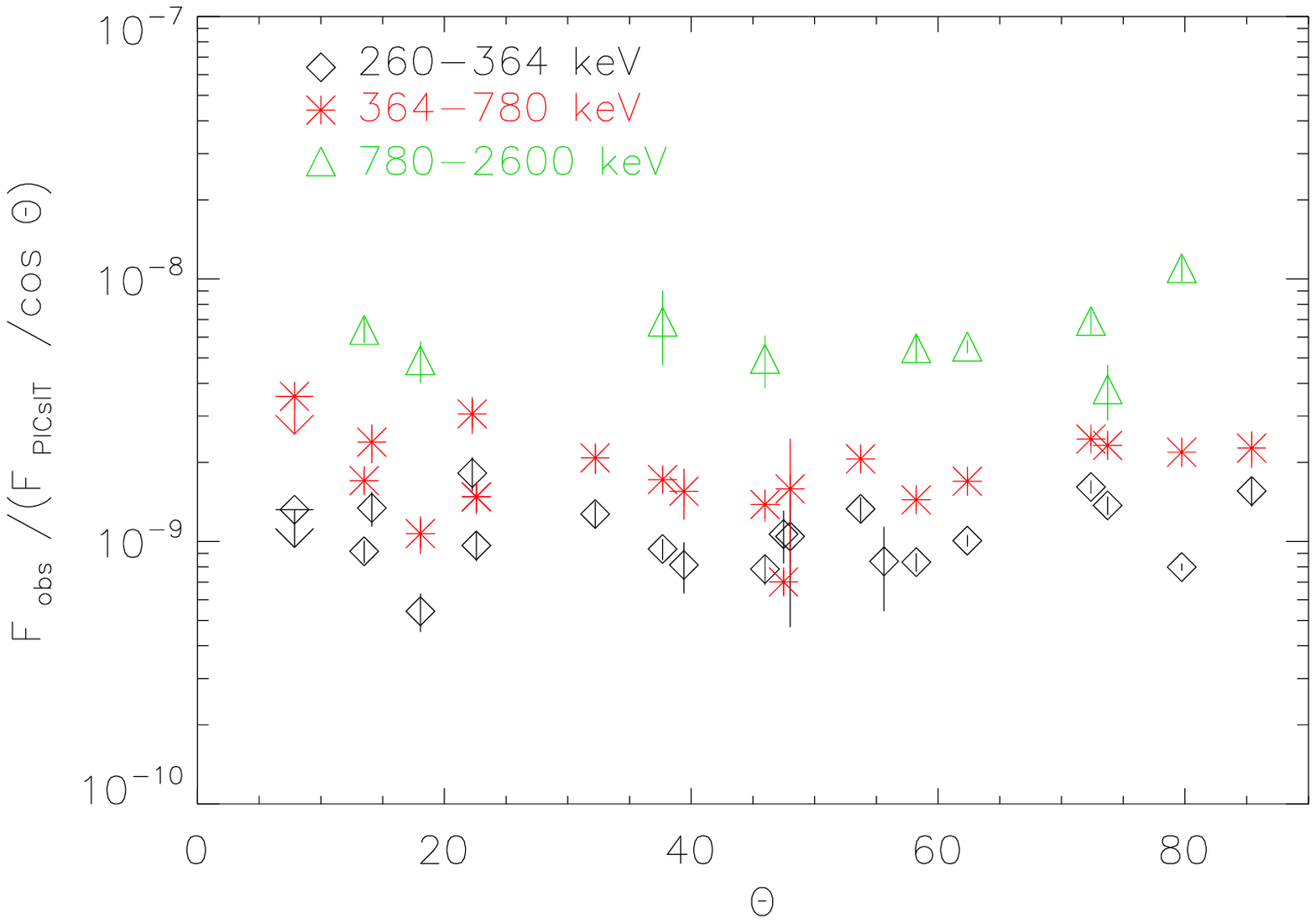}
\caption{\label{ratio} Ratio of fluences in physical and instrumental units in three energy bands for a subset of 20 GRBs. In the lower panel PICsIT counts were corrected for the geometrical factor $\cos(\Theta)$.}
\end{figure}
\begin{figure}
\includegraphics[width=0.5\textwidth]{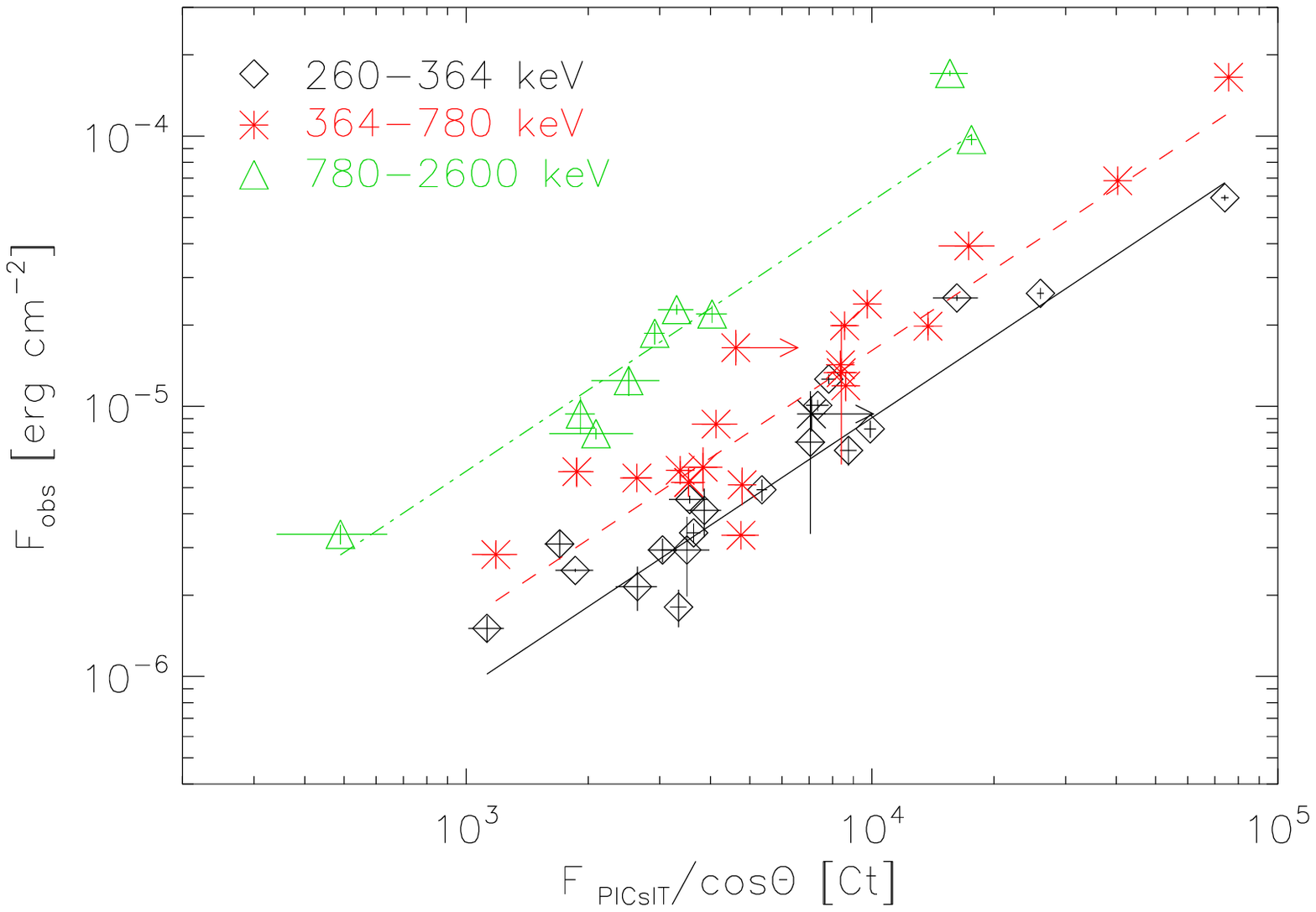}
\caption{\label{calib} Correlation between PICsIT measured fluence and observed fluence from other instruments for 20 GRBs in our sample in three energy bands. Lines give the expected fluences derived from the conversion parameters in Table~\ref{rfit}. }
\end{figure}

We evaluated the background level in each energy channel by fitting a polynomial to counts measured in two time intervals of a few hundred seconds manually selected before and after the GRB. 
In all cases a constant was adequate, while a polynomial function up to the third degree was required for a better determination of the T$_{90}$ interval of 12 events (see notes in Table~\ref{tab_sample}). 
The T$_{90}$ duration was computed as described in \citet{koshut} from the background-subtracted light curves in the energy range 260-2600~keV, with the exception of 060819, whose duration was evaluated in the 260-312~keV light curve. 
The background-subtracted light curves in three energy bands are reported in Figs.~\ref{lcfirst}-\ref{lclast}. 

The values of fluence and peak flux in Table~\ref{tab_sample} are reported in instrumental units. 
Indeed, in the next section we argue that a large part of bursts is expected to occur far off-axis, where a response matrix is not available yet. 
The fluences refer to the T$_{90}$ intervals and to the energy range reported in column 7: the energy coverage for each event is given by the energy channels where the fluence $F$ verifies the condition $F > 2 \cdot \sigma_F$, where $\sigma_F$ is the error on the fluence. 
The first energy channel (208-260~keV) was excluded because it is dominated by the systematic background component described in the previous section. 
The peak fluxes in the same energy range are for an integration time of 1~s.

\subsection{Spectral characterization and PICsIT effective area}
\begin{table}[t]
\caption{\label{tab_ang} Position in instrumental coordinates and fluence in 260-2600~keV for the subsample of documented GRBs.}
\begin{tabular}{cccccc}
\hline
\hline
Burst & $\theta $    & $\phi $      & F (260-2600~keV) & Ref. \\
      & ($^{\circ}$) & ($^{\circ}$) & ($10^{-5}$~erg~cm$^{-2}$) &  \\
\hline
\hline
060805B& 72.4 &  359.3  & 5.9$\pm$0.2 & (1) \\
060901 & 14.1 &  297.4  & 0.76$^{+0.07}_{-0.10}$ & (2) \\
060928 & 62.4 &   43.2  &19.17$^{+1.03}_{-0.76}$ &  (3) \\
061122 & 7.4  & 220.9   & 0.52$^{+0.01}_{-0.03}$ & (4) \\
061222 & 22.3 & 226.5  & 1.5$^{+0.2}_{-0.1}$ &(5)\\
071003 & 13.5 &   273.1 & 3.8$^{+0.2}_{-0.5}$ & (6) \\
071006 & 55.6 &   233.3 & 1.03$^{+0.05}_{-0.6}$ & (7)\\
080122 & 48.0$\pm0.3$&   240$\pm1$& 4$\pm2$ & (8)\\
080204 & 18$\pm1$    &   101$\pm2$ & 1.6$^{+0.2}_{-0.3}$ & (9)\\
080319B& 79.7 & 215.7 & 39.5$^{+1.0}_{-0.9}$ & (10)\\
080328 & 22.6 &  0.9  & 1.0$\pm0.1$ & (11) \\
080514B& 37.7 & 355.4 & 1.5$\pm0.1$ & (12) \\
080607 & 73.7 & 201.4 & 4.8$\pm0.2$ & (13) \\
080613B& 39.4 & 134.2 & 1.5$\pm0.3$ & (14) \\
080721 & 58.2 & 353.9 & 5.0$\pm0.4$ & (15) \\
080723B& 7.9  & 104.7 & 4.3$\pm0.3$ & (16)\\
080817A& 57.7$\pm0.6$ & 94$\pm1$ & &\\
090528B& 32$\pm1$ & 215$\pm1$ & 2.29$\pm0.07$ & (17) \\
090618A& 85.4 & 14.6& 9.8$\pm0.2$ & (18) \\
090623 & 47$\pm1$ & 244$\pm1$ & 2.6$^{+0.2}_{-0.9}$ & (19) \\
090626A& 46$\pm1$ & 182$\pm1$ & 3.1$\pm0.4$ & (20) \\
090809B&  53.7 & 41.1 & 1.63$\pm0.02$ & (21) \\
\hline
\hline
\end{tabular}

The errors in polar coordinates were neglected when $\le 0.1^\circ$. Fluences were extrapolated in the 260-2600~keV energy range from spectral parameters reported in: (1) = \citet{bellm06}, (2) = \citet{golenetskii06a}, (3) = \citet{golenetskii06b}, (4) = \citet{golenetskii06c}, (5) = \citet{golenetskii06d}, (6) = \citet{golenetskii07b}, (7) = \citet{golenetskii07c}, (8) = \citet{golenetskii08b}, (9) = \citet{golenetskii08c}, (10) = \citet{golenetskii08d}, (11) = \citet{golenetskii08e}, (12) = \citet{golenetskii08f}, (13) = \citet{golenetskii08g}, (14) = \citet{golenetskii08h}, (15) = \citet{golenetskii08i}, (16) = \citet{golenetskii08j}, (17) = \citet{vkienlin09a}, (18) = \citet{golenetskii09a}, (19) = \citet{nakagawa09}, (20) = \citet{golenetskii09b}, (21) = \citet{horst09}
\end{table}
\begin{table}[t]
\caption{\label{rfit} Conversion coefficient from PICsIT counts to physical units in three energy bands.}
\centering{
\begin{tabular}{ccc}
\hline
\hline
Energy & K   & $\sigma$\\
(keV)  & \multicolumn{2}{c}($10^{-9}$~erg ~ cm$^{-2}$ ~ ct$^{-1}$) \\
\hline
\hline
260-364   & 0.9 
& 0.3 \\
364-780   & 1.6 
& 0.6 \\
780-2600  & 5.7 
& 2.0 \\
\end{tabular}}
\end{table}

In Fig.~\ref{fpt_dist} we show the cumulative distributions of the instrumental fluences and peak fluxes of the PICsIT burst sample in three energy intervals: 260-364~keV, 364-780~keV and 780-2600~keV. 
The irregular behavior is likely caused by a non linear relation between instrumental observables and physical fluences and peak fluxes. 

Although not provided by the ST data, the arrival direction of the photons can be expressed in PICsIT coordinates for GRBs with known sky positions from observations with other satellites. 
This can be made using the instrument attitude parameters stored in the housekeeping data on-board. 
In Fig.~\ref{theta-phi} we show the spatial distribution in PICsIT coordinates of 22 events in our sample. 
The polar angle $\Theta$ is the off-axis angle, while $\phi$ is the azimuth, defined so that the direction toward the Sun is at $\phi=90^\circ$. 
In the figure we indicated the IBIS FoV and the approximate region occupied by the spectrometer SPI ($\phi \in [250^\circ,290^\circ]$, $\Theta > 28^\circ$). 
Despite the low number of events, we note that the region that includes the directions shielded by SPI in the ($\Theta$, $\phi$) plane is scarcely populated.  

For 21 GRBs in our sample the fluences in physical units and the average spectral parameters are available from observations with other satellites. 
The spectrum of six events in this subsample is described by a cut-off powerlaw, while 15 events follow a Band function with mean parameters $\alpha = -0.9 \pm 0.2$, $\beta = -2.5 \pm 0.4$, $E_p = 318 \pm 174$~keV. 
On the basis of the spectral parameters reported for each single GRB, we converted the fluence to the PICsIT energy range (260-2600~keV), assuming no spectral evolution and neglecting differences in the reported burst duration. 
When the same event was measured by more than one instrument, we chose the one whose energy coverage overlapped the most with that of PICsIT.
The derived fluences are reported in Table~\ref{tab_ang} and plotted in Fig.~\ref{flang1} as a function of PICsIT angular coordinates.
We note that the four bursts within the PICsIT FoV ($\Theta<15^\circ$) span about one order of magnitude in fluence, while at wider off-axis angles only bursts with higher fluences were detected. 
No significant dependence on the azimuthal angle $\phi$ is evident. 

With the exception of GRB~061122, whose light curve is compromised by telemetry saturation, we compared the fluences observed by PICsIT to those derived in the same energy range for 20 GRBs with known spectral parameters. 
The ratio of the fluences in physical and instrumental units shows a strong dependence on the off-axis angle (see Fig.~\ref{ratio}, upper panel), which can be partly compensated by a simple $\cos\Theta$ factor, accounting for the geometrical projection effect of the  PICsIT detection plane (see Fig.~\ref{ratio}, lower panel). 
The spread (in the worst case of a factor $\sim3$) is caused by a more complicated energy- and direction-dependent instrumental response as well as to the spectral parameters assumed for each single GRB. 
Neglecting the residual dispersion, the strong correlation shown in Fig.~\ref{calib} between physical and instrumental fluences corrected for the geometrical projection can be assessed by a simple conversion coefficient: 
\begin{equation}
F(\Delta E) =  k(\Delta E)  \times \frac{F_{PICsIT}(\Delta E)} {\cos \Theta} \quad \rm{erg ~ cm}^{-2} \quad ,
\end{equation}
computed as the weighted average of the ratio of the two quantities. 
The derived values of $k$ in the three selected energy bands are given in Table~\ref{rfit}, with errors $\sigma$ estimated as the standard deviation. 

The rough calibration derived above allows us to compare the number of GRBs in our sample with the expected burst rate based on  results from previous experiments.
From the peak flux cumulative distribution in the 260-364~keV energy band (Fig.~\ref{fpt_dist}), we can estimate a completeness flux limit at $\sim$600~ct/s, where  the slope of the distribution flattens because of the decreasing efficiency to detect fainter GRBs at different off-axis angles. 
Assuming $\cos \Theta$=0.5, and a Band spectrum with average values of the parameters $\alpha$= -1, $\beta$=-2.3, and E$_{peak}$=300 keV \citep{kaneko06}, this count-rate limit corresponds to a flux of approximately 1~ph~cm$^{-2}$~s$^{-1}$ in the 50-300~ keV energy domain. 
From the BATSE LogN-LogP in this energy range (\citealt{kommers00}, \citealt{fenimore95}) the expected rate of GRBs with flux above the PICsIT limit is $\sim$10 events/yr over a solid angle of 4$\pi$. 
This agrees reasonably well with the PICsIT detection rate of $\sim$5.5 bursts per year (13 bursts in a  net exposure time of $\sim$850 days) in approximately half sky. 
\begin{figure}
\includegraphics[width=0.5\textwidth]{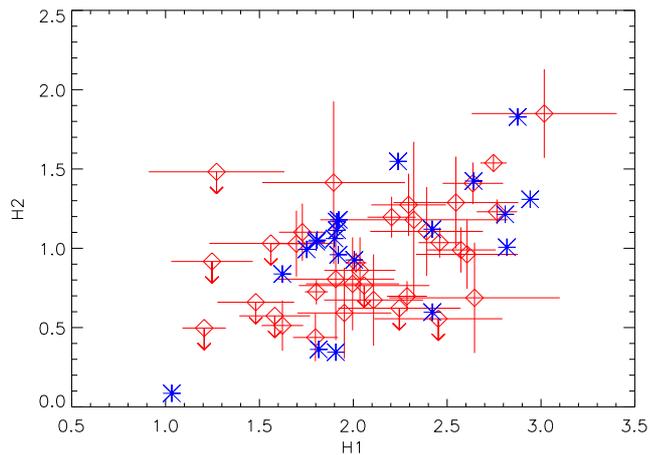}
\caption{\label{allhr} Hard and soft color for 35 bursts in the PICsIT sample. $H1=F2/F1$ and $H2=F3/F2$ are derived as the ratio of the physical fluences in 260-364~keV (F1), 364-780~keV (F2) and 780-2600~keV (F3). Diamonds refer to PICsIT instrumental quantities, and stars are obtained from documented spectral parameters. }
\end{figure}

Because most of the dependence on the burst incoming direction is accounted for by the detector area projection coefficient, the ratio between fluences in different energy bands is almost direction-independent. 
We defined the quantities $H1 = F2/F1$ and $H2=F3/F2$, where $F1$, $F2$, and $F3$ are the fluences observed by PICsIT in the three selected energy bands, multiplied for the conversion coefficients in Table~\ref{rfit}. 
Figure~\ref{allhr} shows (red diamonds) the hardness ratios H1 and H2 for the 35 bursts in our sample with non-null fluence in the 780-2600~keV band, including also those events without localization. 
Star (blue) symbols mark the same quantities derived from spectral parameters of GRBs in Table~\ref{tab_ang}. 
The region populated by PICsIT observables is well confined within the expected values and characterizes the spectral colors of a typical burst observed by PICsIT. 

For those spectra described by a Band function we inspected the dependence of instrumental hardness ratio on the spectral parameters, but we found no straightforward correlation of the soft color with the $\alpha$ parameter, nor for hard color with the $\beta$ and $E_p$ parameters. 

\subsection{Spectral variability}
\begin{table}[t]
\caption{\label{var} Burst variability: mean hardness ratio, $\chi^2$, number of bins and associated null-hypothesis probability.}
\centering{
\begin{tabular}{ccccc}
\hline
\hline
Burst & Const & $\chi^2$ & N & $P(\chi^2 > \chi^2_{obs})$\\
\hline
\hline
\multicolumn{5}{c}{H21}\\
\hline
060928 & 1.6 & 28.4 & 14 & $8\cdot10^{-3}$ \\
070829 & 1.4 & 15.0 &  9 & $6\cdot10^{-2}$ \\
071003 & 1.6 & 22.9 & 12 & $2\cdot10^{-2}$ \\
080303 & 1.0 &106.4  & 23 & $0$            \\
080319B& 1.0 & 37.0 & 21 & $1\cdot10^{-2}$ \\
080615 & 1.0 & 17.5 &  9 & $2\cdot10^{-2}$ \\
080721 & 1.3 & 19.0 & 13 & $9\cdot10^{-2}$ \\
080723B& 1.0 & 38.0 & 13 & $2\cdot10^{-4}$ \\
\hline
\multicolumn{5}{c}{H32}\\
\hline
060805B& 0.3 &  13.8 &  8 & $5\cdot10^{-2}$ \\
060928 & 0.4 & 120.1 & 13 & $0$             \\
071003 & 0.4 &  18.5 & 11 & $5\cdot10^{-2}$ \\
080303 & 0.3 &  46.4 & 17 & $8\cdot10^{-5}$ \\
080319B& 0.3 &  44.6 & 15 & $5\cdot10^{-5}$ \\
080721 & 0.4 &  53.2 & 10 & $6\cdot10^{-8}$ \\
\end{tabular}}
\end{table}
While a spectral characterization of a burst can be attempted only for events with known incoming direction, the hardness ratio evolution along the event is not dependent on the off-axis angle. 
From the background-subtracted count curves in the three selected energy bands 260-364 keV (soft), 364-780 keV (medium), and 780-2600 keV (hard), we computed the hardness ratios $H21$ (medium to soft) and $H32$ (hard to medium) along each burst. 
The hardness ratios were derived over the $T_{90}$ interval only for bins where counts are at least $2\cdot \sigma$ above the background. 
The spectral variability was estimated by a simple $\chi^2$ test:
\begin{equation}
\chi^2 = \Sigma_{i=0}^{N} \frac{(HR_i-\langle HR \rangle)^2}{\sigma_i^2} \quad ,
\end{equation}
where $\langle HR \rangle$ is the mean value over the $N$ bins verifying the condition above, $\sigma_i$ is the error on the hardness ratio in the $i-$th time bin. 
When allowed by the signal-to-noise ratio, the hardness evolution for each burst is plotted in Figs.~\ref{lcfirst}-\ref{lclast} together with the event light curves. 
Note that a different binning time is usually adopted for light curve and HR plots. 

Table~\ref{var} reports the mean hardness ratio, the number of bins $N$, 
the value of $\chi^2$ and the associated null hypothesis probability of no-variability. 
Only bursts with a $\chi^2$-probability higher than $90$\% and at least 8 bins are given. 

Despite the strong limitation owing to the low SNR in short time bins, we identified nine events with significant variability ($P>90$\%), among which six are known GRBs and three are bursts without a documented position. 
The spectral variability in both hard to medium and medium to soft bands does not show a common evolution, as observed in different energy regimes by other instruments (see e.g. \citealt{guidorzi11}). 

\section{Conclusions}
We have investigated the capabilities of the IBIS/PICsIT instrument applied to the study of GRBs. 
We analyzed spectral timing light curves over about three years, provided by PICsIT with energy resolution of eight channels and time sampling of $\sim$16~ms. 
Under stringent selection criteria, optimized for the search of long events, we identified 39 bursts, most of which are documented GRBs or events detected by other telescopes. 
We showed that PICsIT successfully detects GRBs extending up to $\sim3$~MeV at a rate of $\sim5.5$ events/year above the instrumental threshold of $\sim 600$~ct/s in the 260-364~keV energy band. 
Bright events are observed even extremely off-axis, well beyond the nominal FoV. 
For all detected events we produced light curves with a time resolution down to 0.25~s. 
We used the spectral properties available in the literature of a subsample of GRBs to attempt a rough calibration between fluences in physical and instrumental units by separating the off-axis and energy dependence and computing an average calibration coefficient in three selected energy bands. 
For sufficiently intense events the fine time and energy sampling of ST data allowed us to trace spectral evolution. 
We showed that it is possible to exploit the PICsIT archive for GRB spectral and temporal characterization in the hard X-ray /  soft $\gamma$-ray regime. 

\begin{acknowledgements}
  This paper is based on observations of INTEGRAL, an ESA project with instrument and science data center funded by ESA member states (principal investigator countries: Denmark, France, Germany, Italy, Switzerland and Spain), the Czech Republic and Poland, and with participation of Russia and US. 
V.~B. thanks E.~Maiorano, M.~Orlandini and L.~Amati for useful suggestions and discussion. 
This work has been partially supported  by ASI contract I/008/07/0.
\end{acknowledgements}

\clearpage
\twocolumn

\onlfig{9}{
\begin{figure}
\includegraphics[angle=90,width=0.45\textwidth]{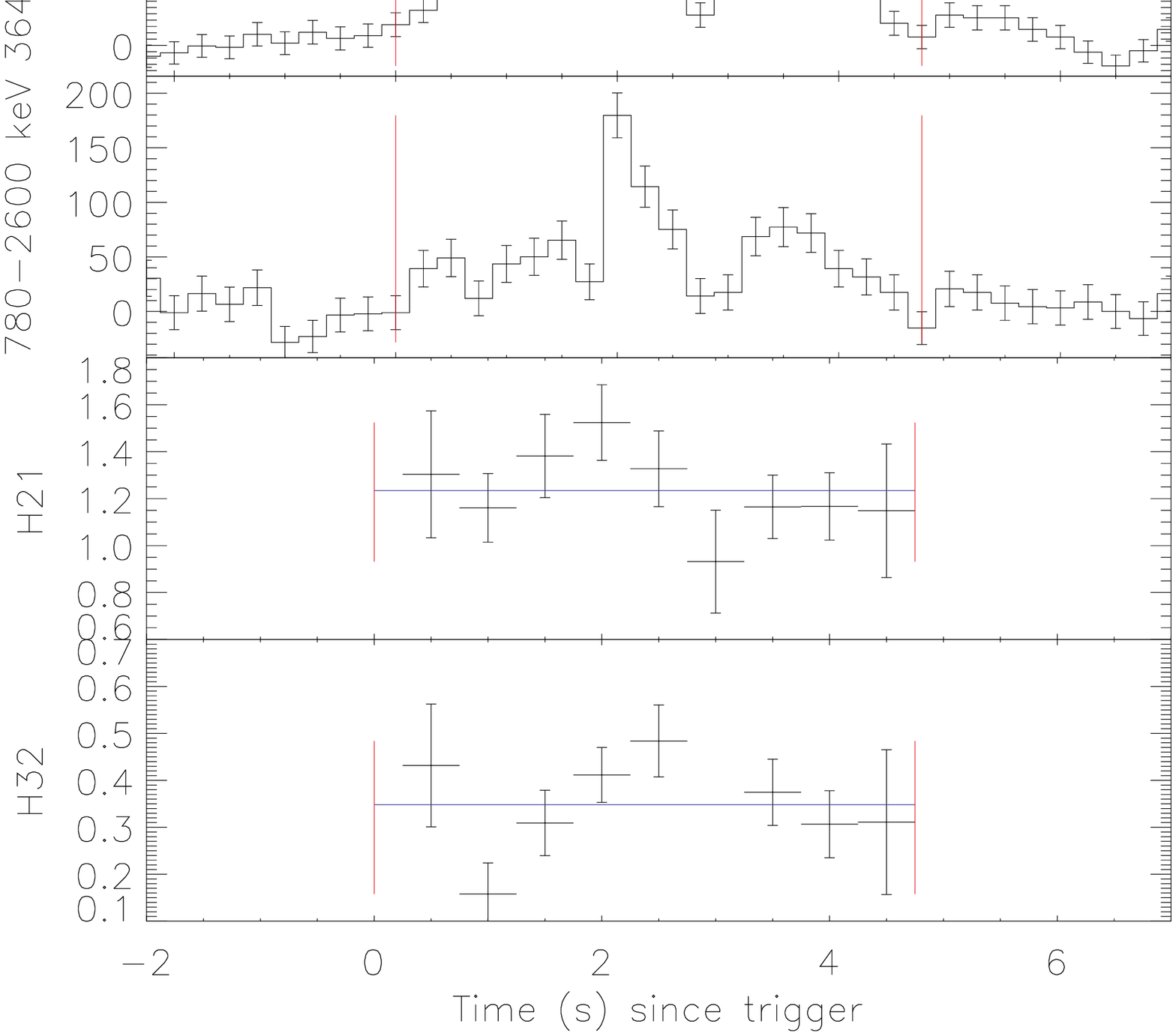}
\caption{\label{lcfirst} GRB~060805B background-subtracted light curve with 0.25~s binning time and hardness ratio with 0.5~s binning time. Vertical (red) lines define the T$_{90}$ interval and horizontal (blue) lines show the average hardness ratios.}
\end{figure}
}

\onlfig{10}{
\begin{figure}
\includegraphics[angle=90,width=0.45\textwidth]{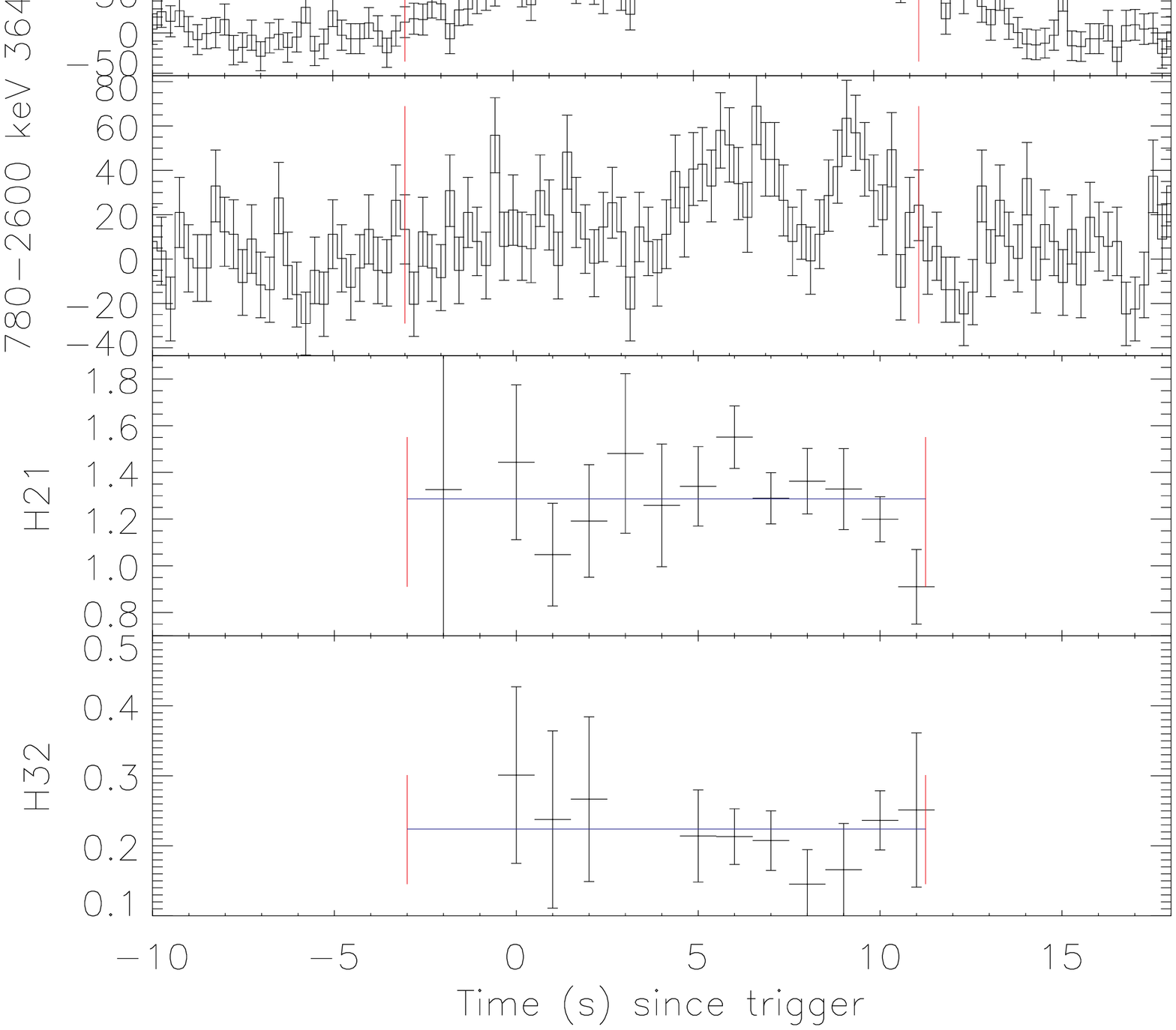}
\caption{Burst~060819 background-subtracted light curve with 0.25~s binning time and hardness ratio with 1~s binning time. See caption of Fig.~\ref{lcfirst} for a description of the lines.}
\end{figure}
}

\onlfig{11}{
\begin{figure}
\includegraphics[angle=90,width=0.45\textwidth]{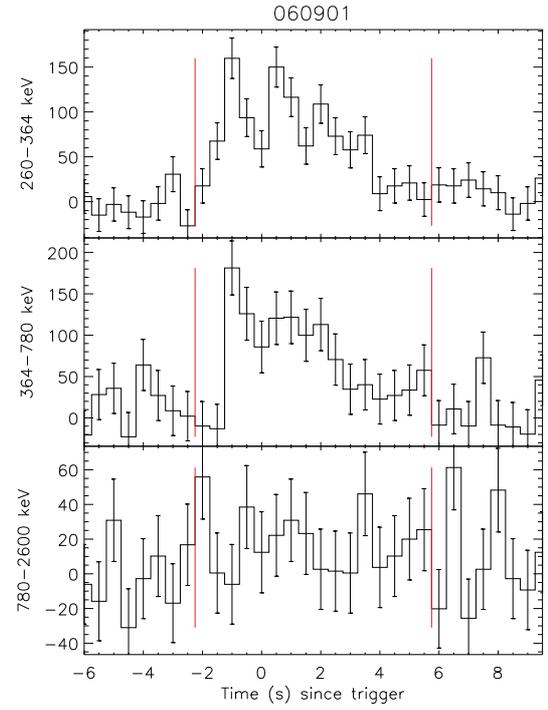}
\caption{GRB~060901 background-subtracted light curve with 0.5~s binning time. }
\end{figure}
}

\onlfig{12}{
\begin{figure}
\includegraphics[angle=90,width=0.45\textwidth]{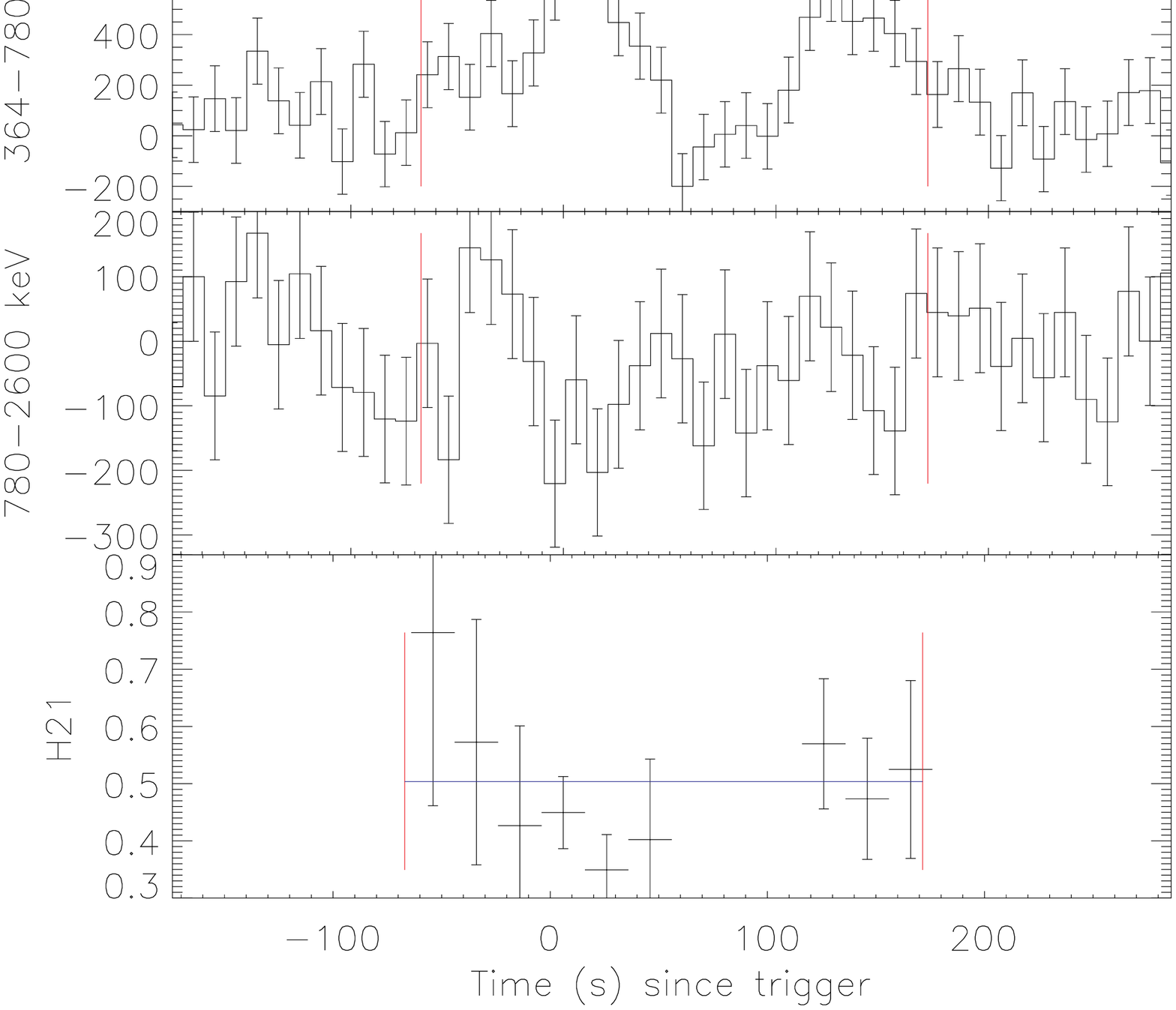}
\caption{Burst~060905 background-subtracted light curve with 10~s binning time and hardness ratio with 20~s binning time. See caption of Fig.~\ref{lcfirst} for a description of the lines.}
\end{figure}
}

\onlfig{13}{
\begin{figure}
\includegraphics[angle=90,width=0.45\textwidth]{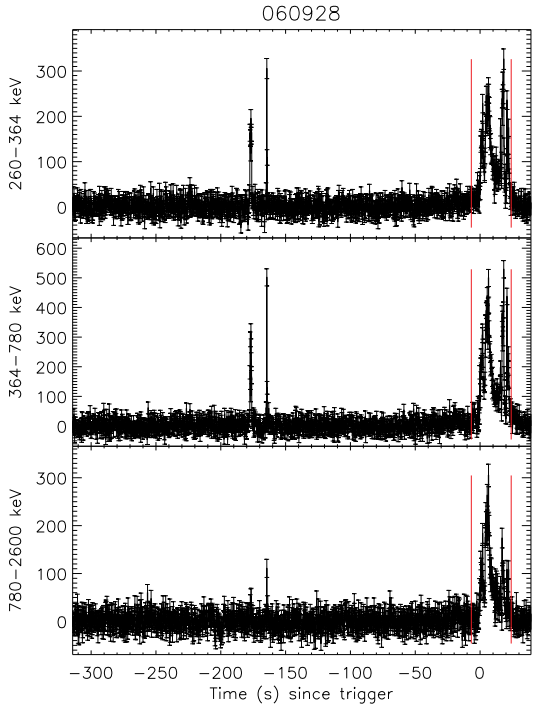}
\caption{\label{060928_1} GRB~060928 - both episodes.  Vertical (red) lines define the T$_{90}$ interval of the main episode.  }
\end{figure}
}

\onlfig{14}{
\begin{figure}
\includegraphics[angle=90,width=0.45\textwidth]{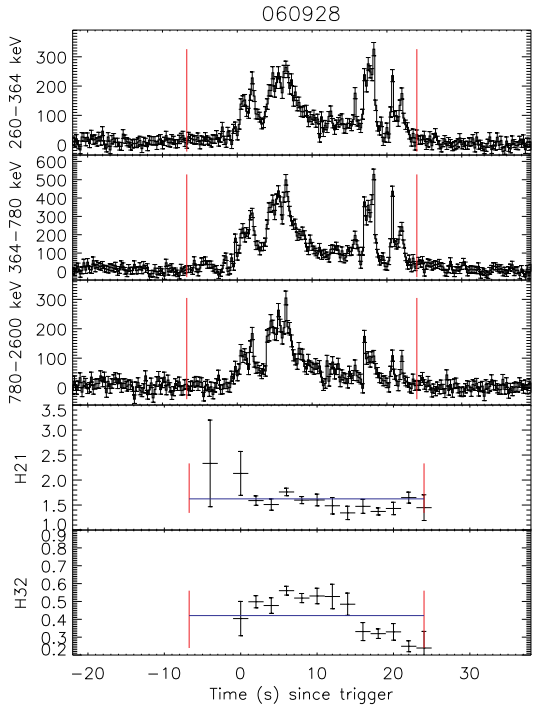}
\caption{\label{060928_2} Second episode of GRB~060928: background-subtracted light curve with 0.25~s binning time and hardness ratio with 2~s binning time. See caption of Fig.~\ref{lcfirst} for a description of the lines.}
\end{figure}
}

\onlfig{15}{
\begin{figure}
\includegraphics[angle=90,width=0.45\textwidth]{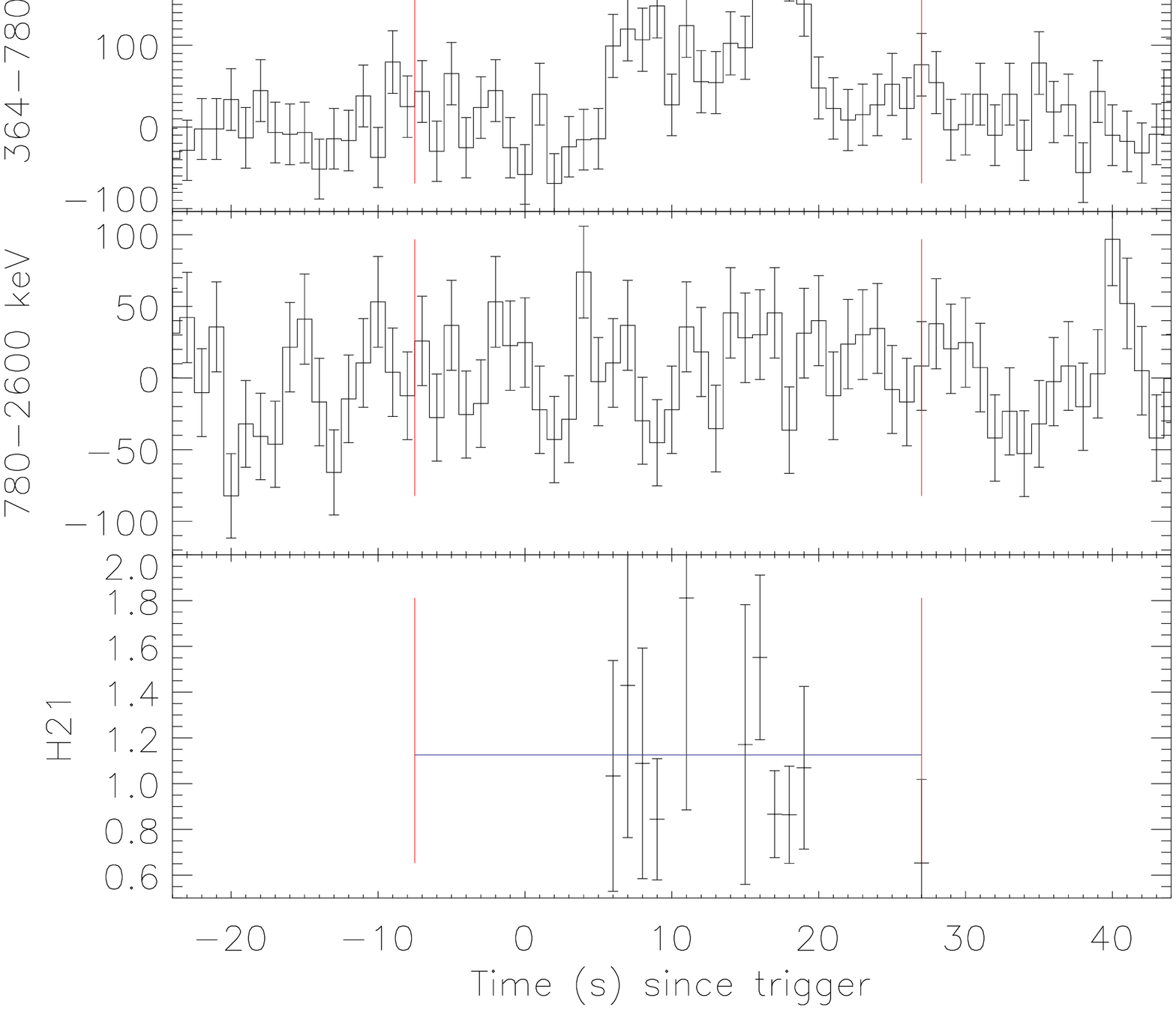}
\caption{Burst~061031 background-subtracted light curve with 1~s binning time and hardness ratio with 1~s binning time. See caption of Fig.~\ref{lcfirst} for a description of the lines. }
\end{figure}
}

\onlfig{16}{
\begin{figure}
\includegraphics[angle=90,width=0.45\textwidth]{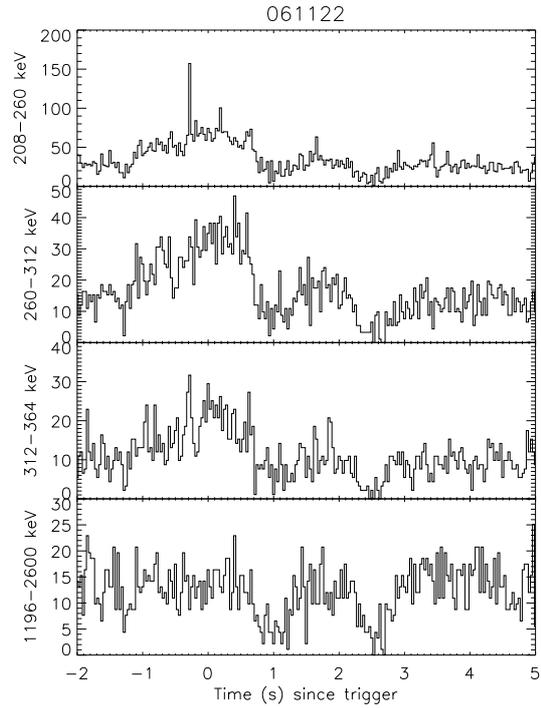}
\caption{\label{lc_061122} Light curve of GRB~061122 in four energy channels and with $\sim$31~ms time resolution. Data gaps are recognized in all energy bands.}
\end{figure}
}

\onlfig{17}{
\begin{figure}
\includegraphics[angle=90,width=0.45\textwidth]{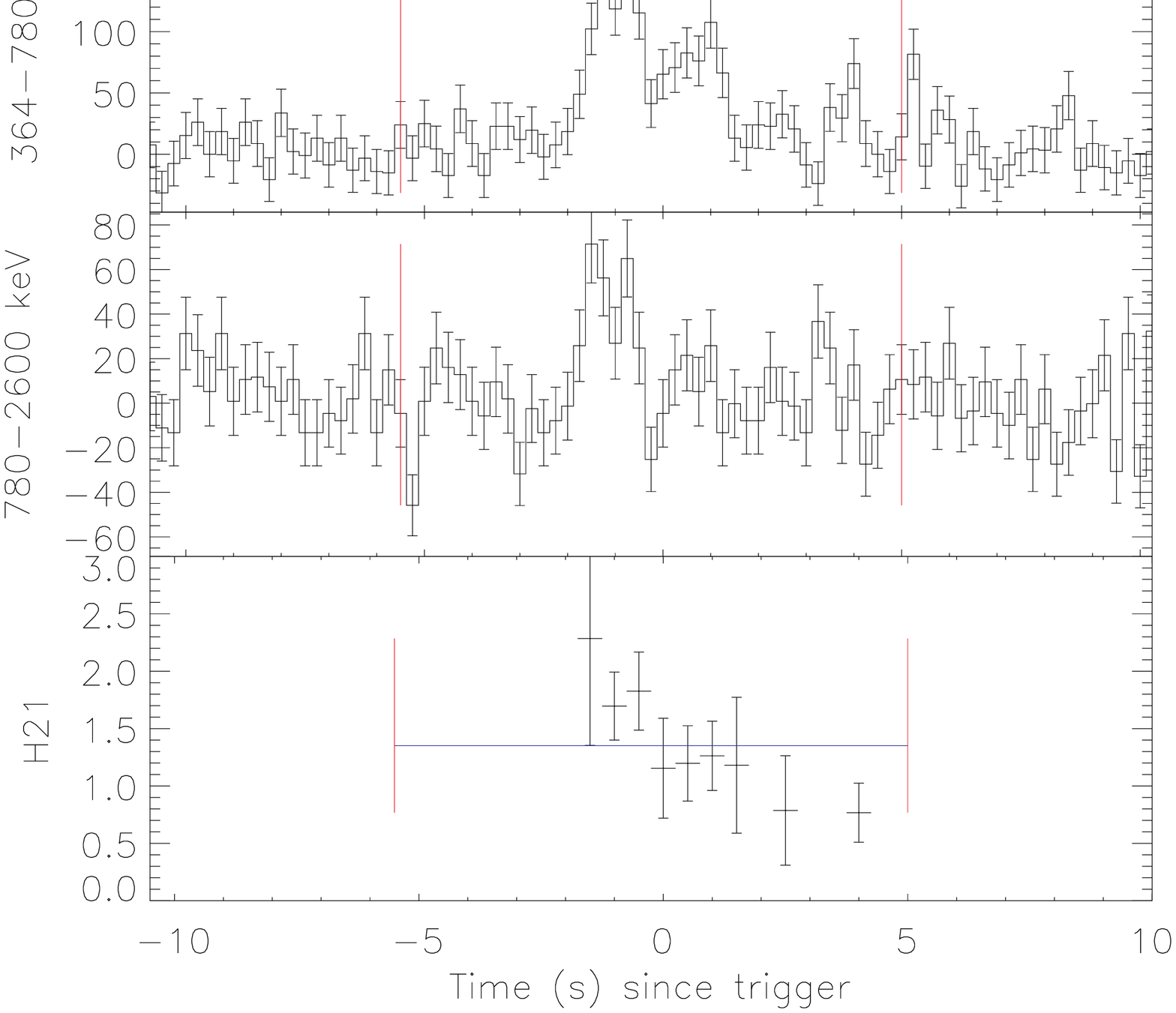}
\caption{GRB~061222A background-subtracted light curve with 0.25~s binning time and hardness ratio with 0.5~s binning time. See caption of Fig.~\ref{lcfirst} for a description of the lines. }
\end{figure}
}

\onlfig{18}{
\begin{figure}
\includegraphics[angle=90,width=0.45\textwidth]{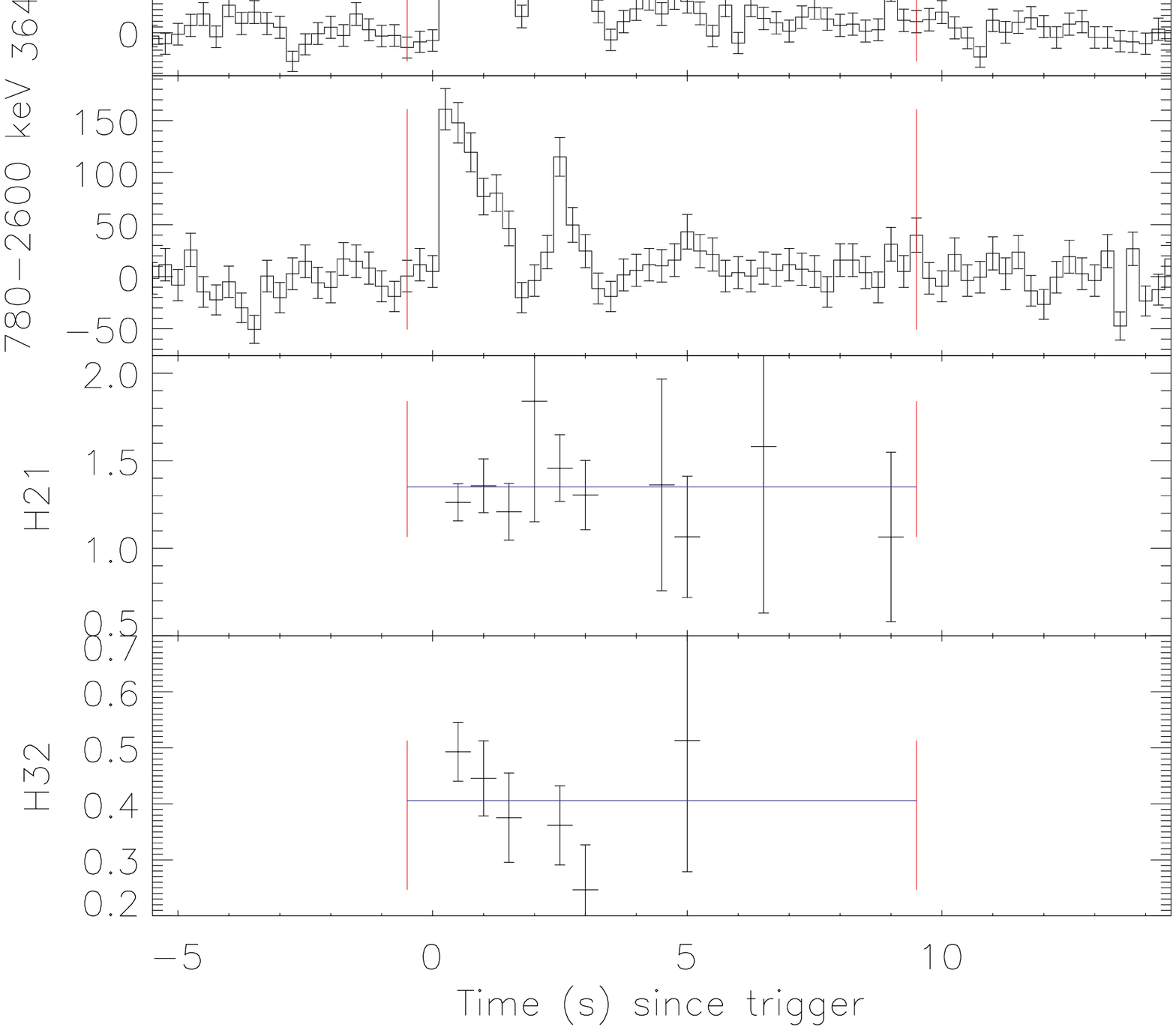}
\caption{GRB~070207 background-subtracted light curve with 0.25~s binning time and hardness ratio with 0.5~s binning time. See caption of Fig.~\ref{lcfirst} for a description of the lines.}
\end{figure}
}

\onlfig{19}{
\begin{figure}
\includegraphics[angle=90,width=0.45\textwidth]{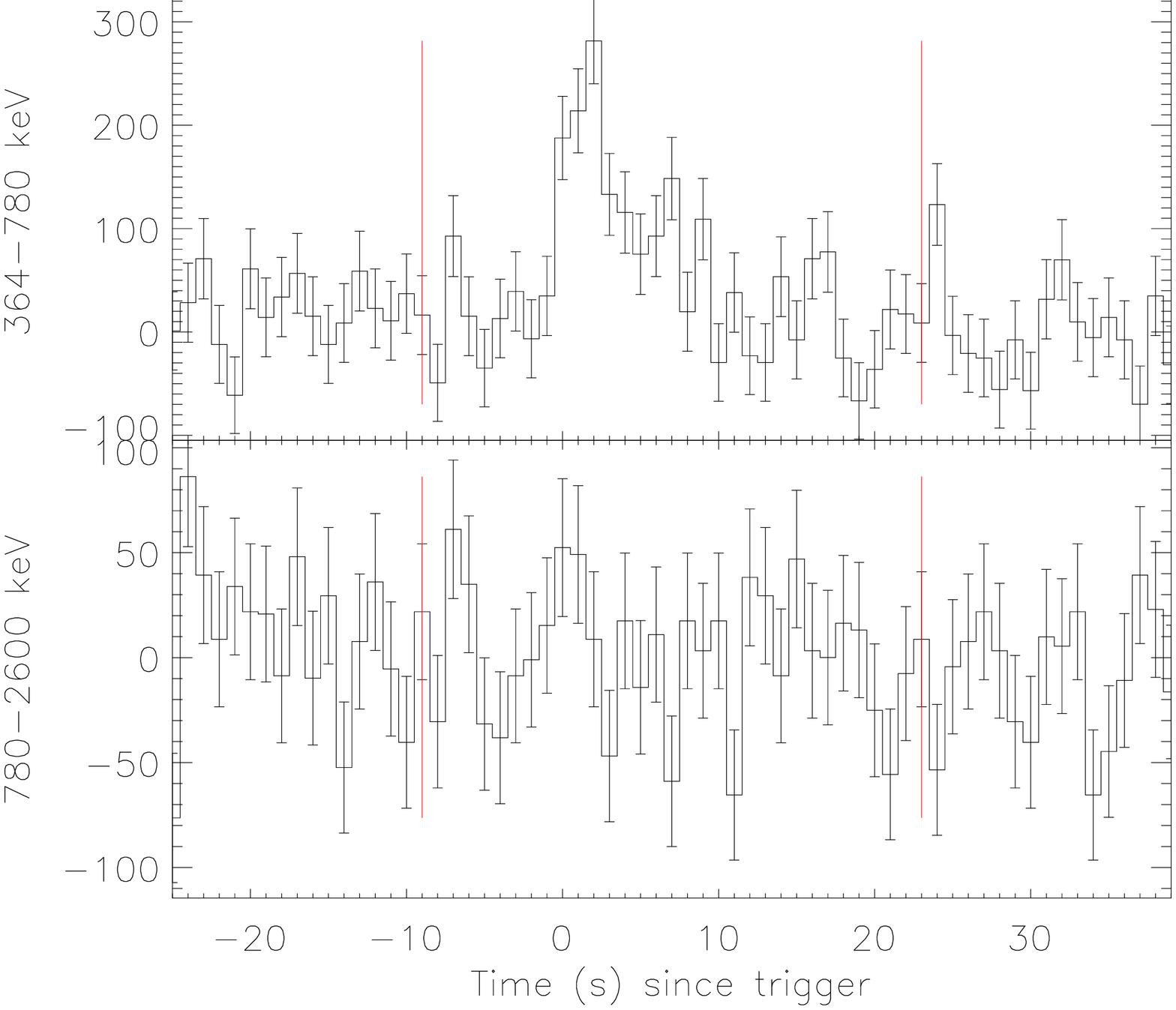}
\caption{Burst~070227B background-subtracted light curve with 1~s binning time.  Vertical (red) lines define the T$_{90}$ interval. }
\end{figure}
}

\onlfig{20}{
\begin{figure}
\includegraphics[angle=90,width=0.45\textwidth]{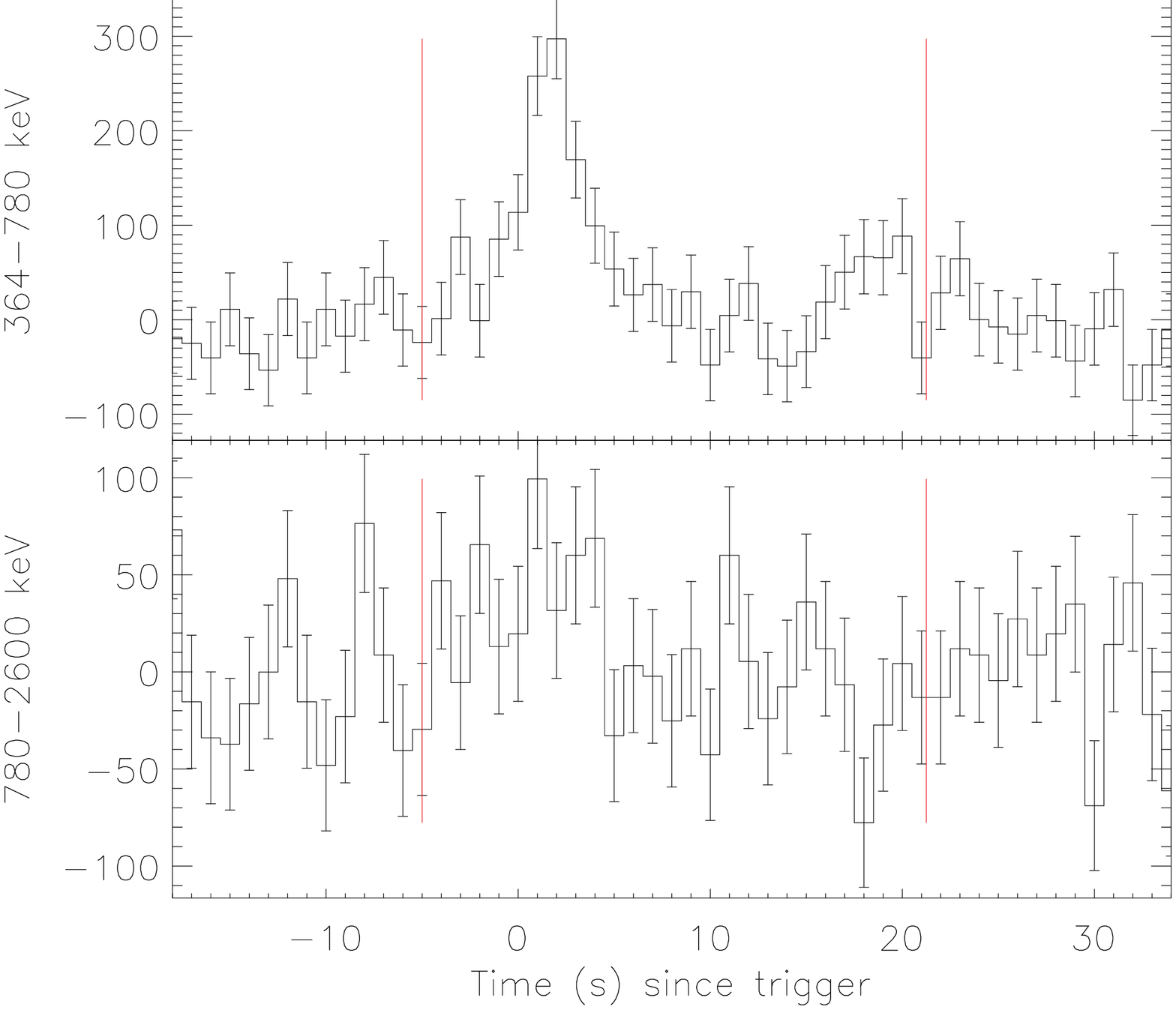}
\caption{Burst~070326 background-subtracted light curve with 1~s binning time.  Vertical (red) lines define the T$_{90}$ interval. }
\end{figure}
}

\onlfig{21}{
\begin{figure}
\includegraphics[angle=90,width=0.45\textwidth]{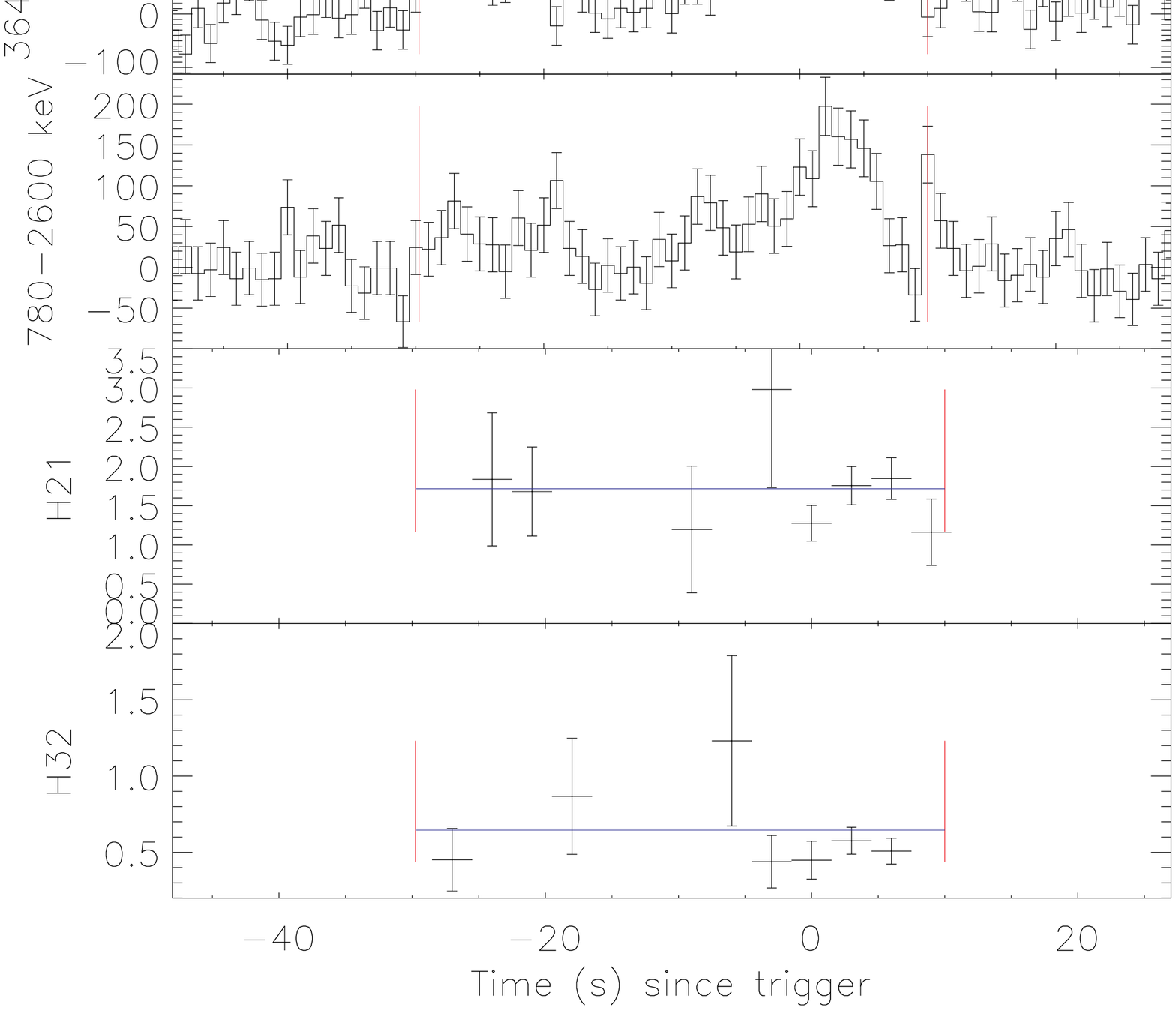}
\caption{Burst~070329 background-subtracted light curve with 1~s binning time and hardness ratio with 3~s binning time. See caption of Fig.~\ref{lcfirst} for a description of the lines. }
\end{figure}
}

\onlfig{22}{
\begin{figure}
\includegraphics[angle=90,width=0.45\textwidth]{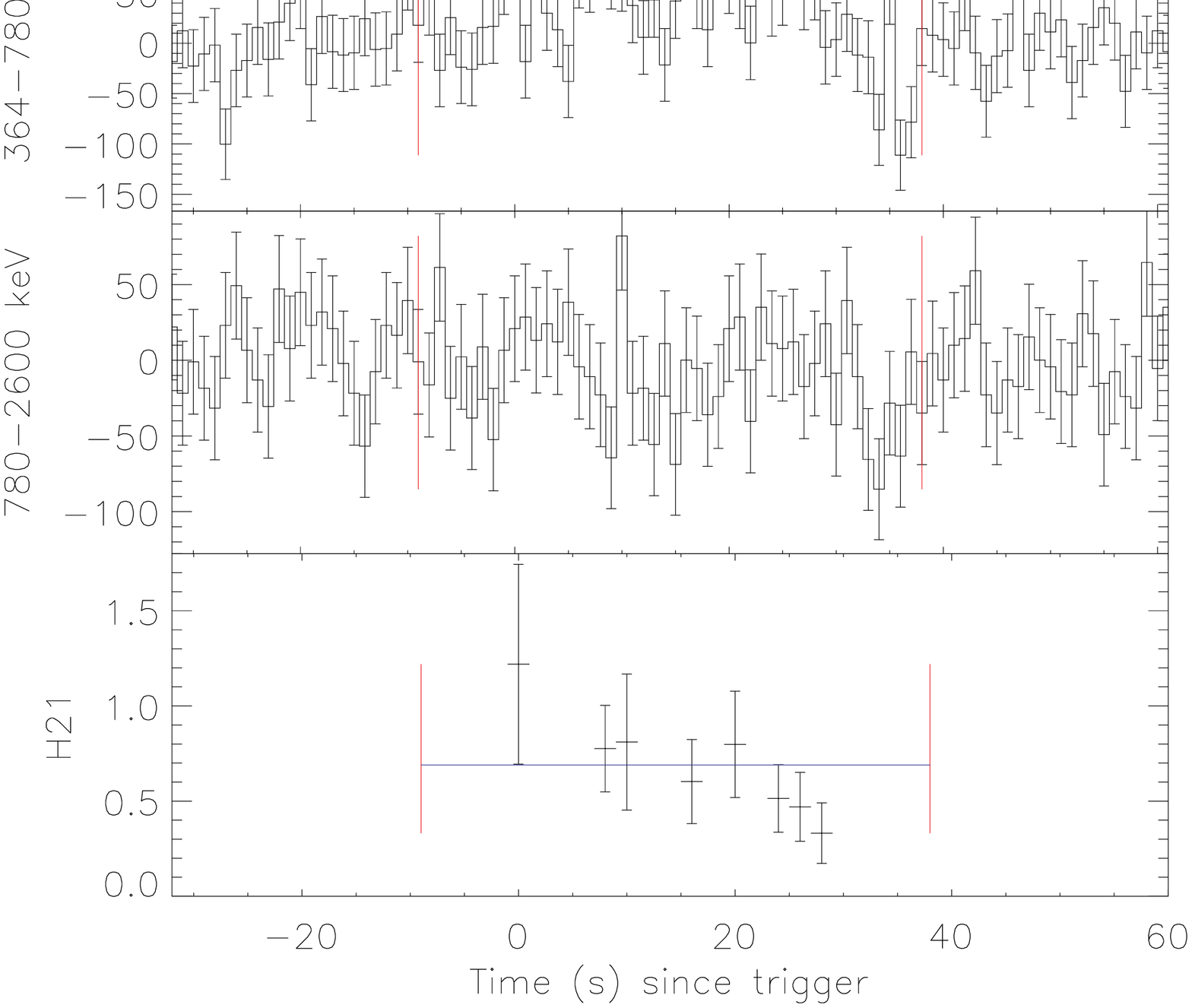}
\caption{Burst~070403 background-subtracted light curve with 1~s binning time and hardness ratio with 2~s binning time. See caption of Fig.~\ref{lcfirst} for a description of the lines.}
\end{figure}
}

\onlfig{23}{
\begin{figure}
\includegraphics[angle=90,width=0.45\textwidth]{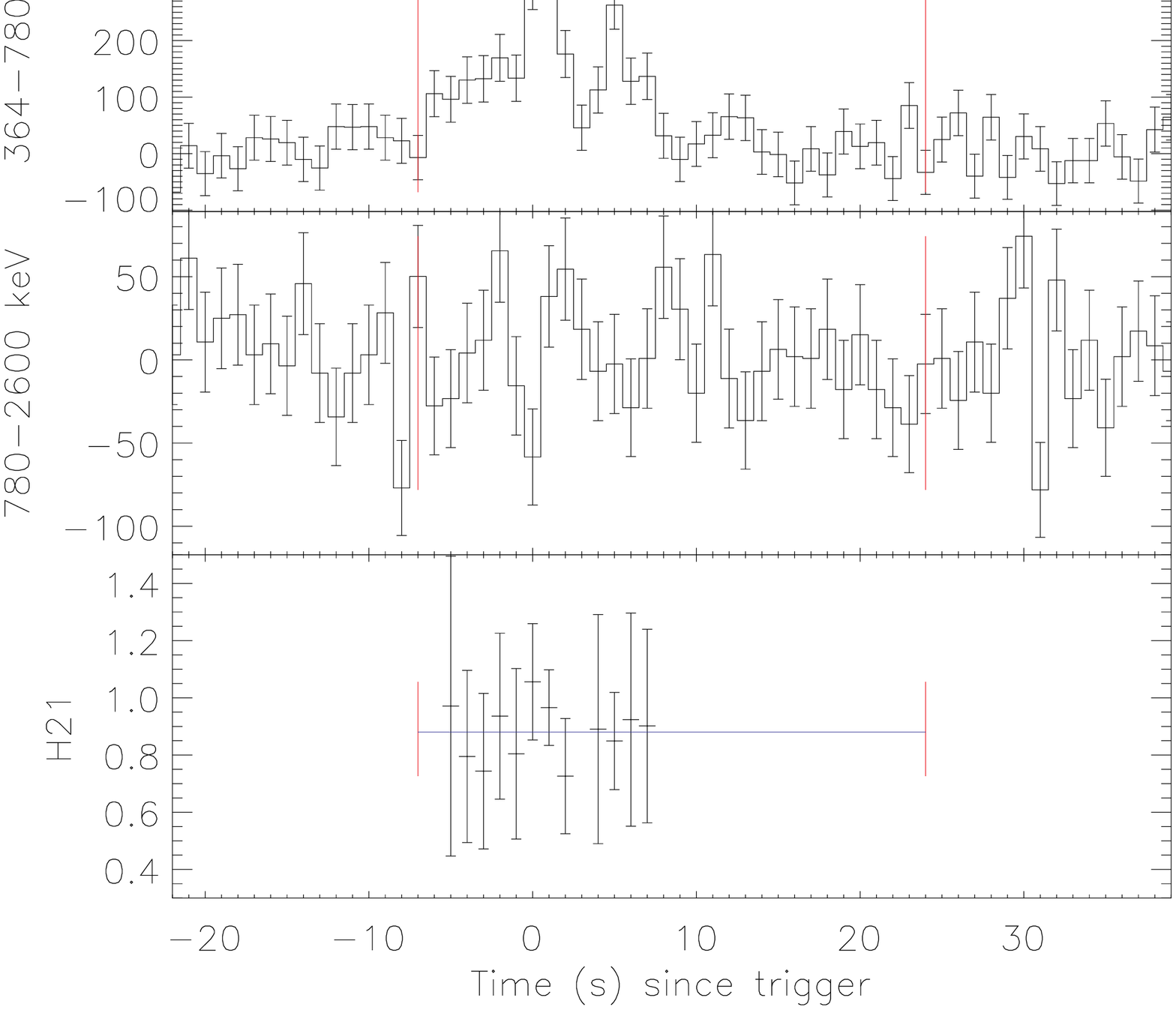}
\caption{Burst~070418 background-subtracted light curve with 1~s binning time and hardness ratio with 1~s binning time. See caption of Fig.~\ref{lcfirst} for a description of the lines.}
\end{figure}
}

\onlfig{24}{
\begin{figure}
\includegraphics[angle=90,width=0.45\textwidth]{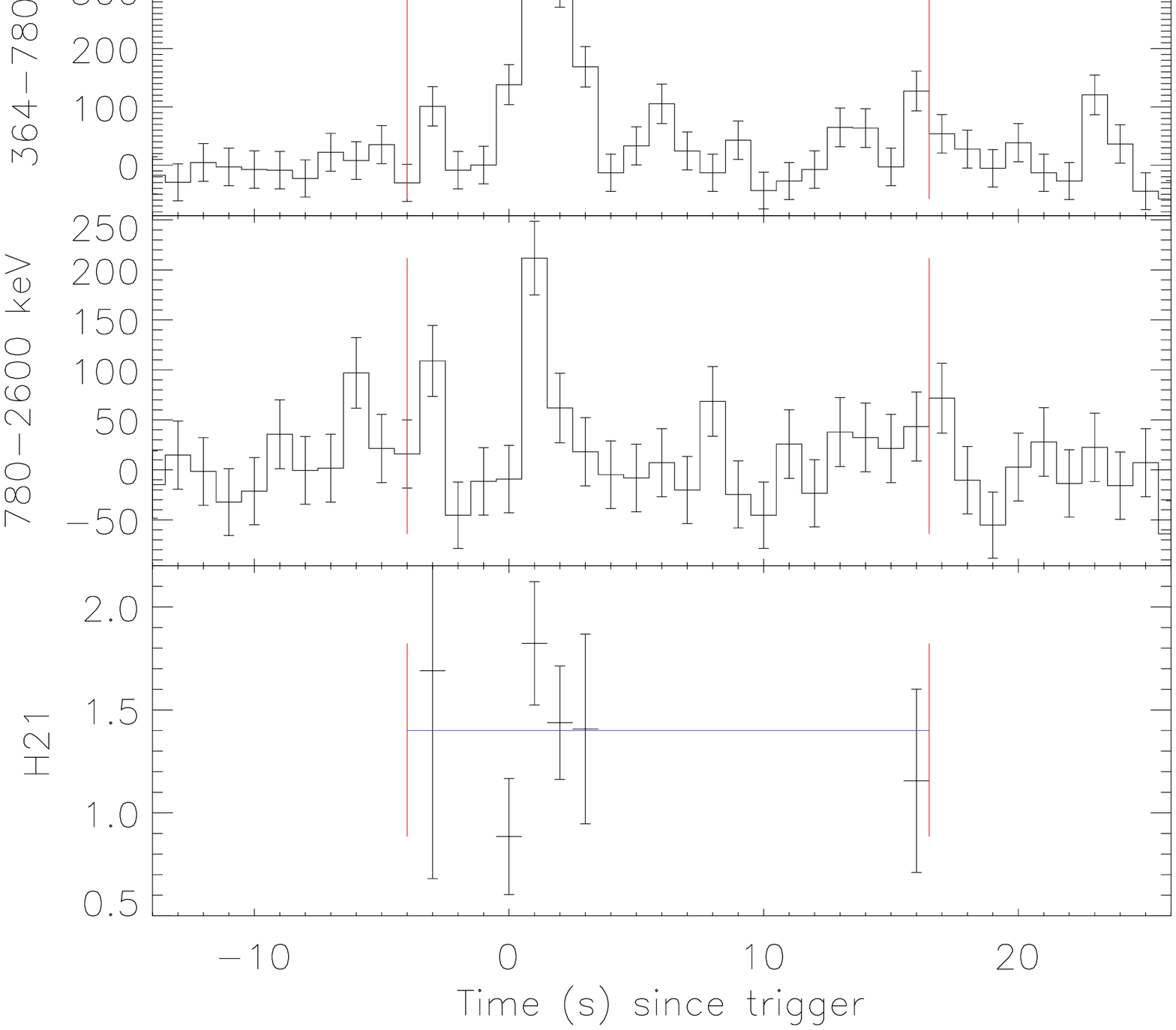}
\caption{Burst~070429 background-subtracted light curve with 1~s binning time and hardness ratio with 1~s binning time. See caption of Fig.~\ref{lcfirst} for a description of the lines.}
\end{figure}
}

\onlfig{25}{
\begin{figure}
\includegraphics[angle=90,width=0.45\textwidth]{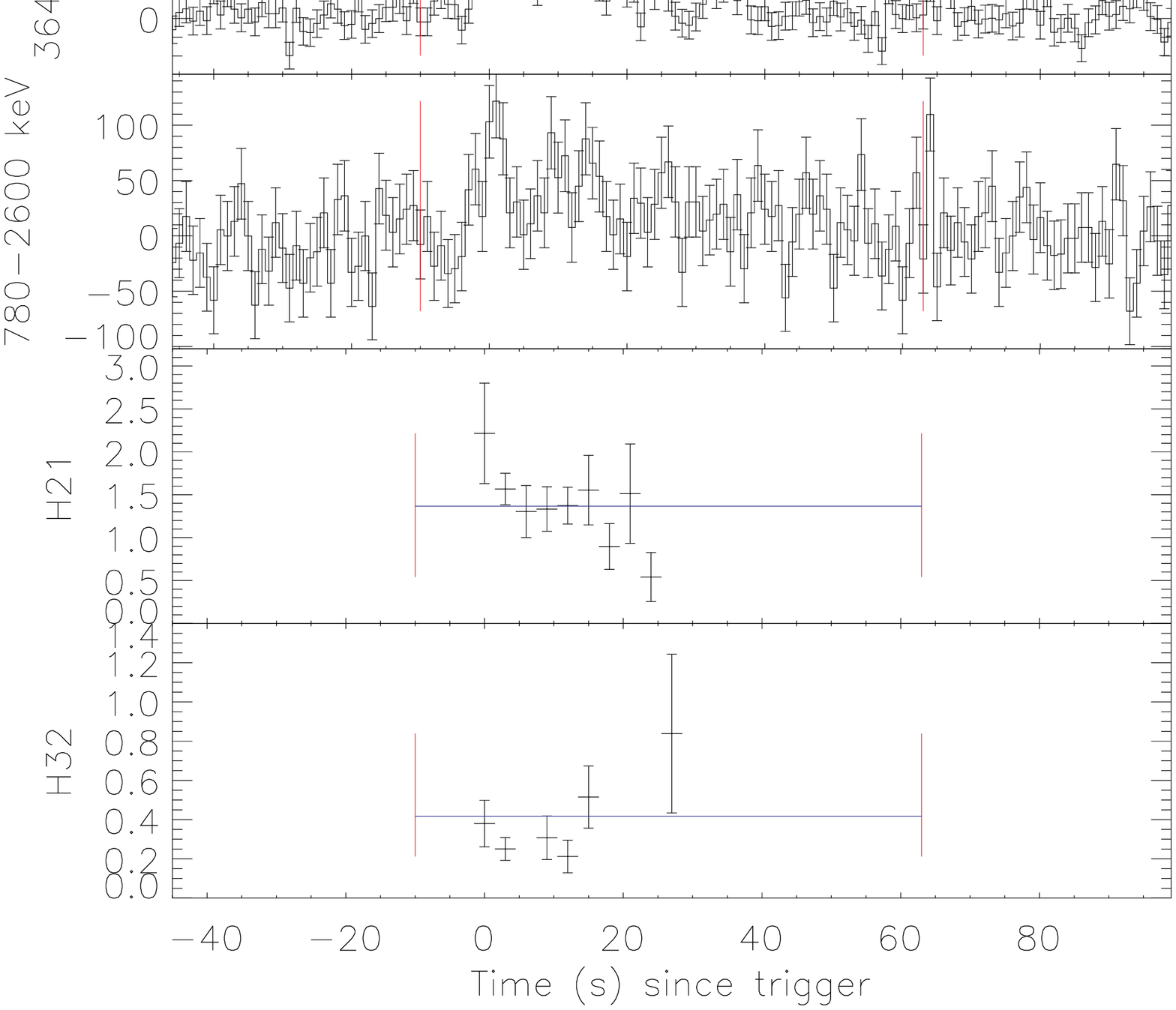}
\caption{Burst~070829 background-subtracted light curve with 1~s binning time and hardness ratio with 3~s binning time.  See caption of Fig.~\ref{lcfirst} for a description of the lines.}
\end{figure}
}

\onlfig{26}{
\begin{figure}
\includegraphics[angle=90,width=0.45\textwidth]{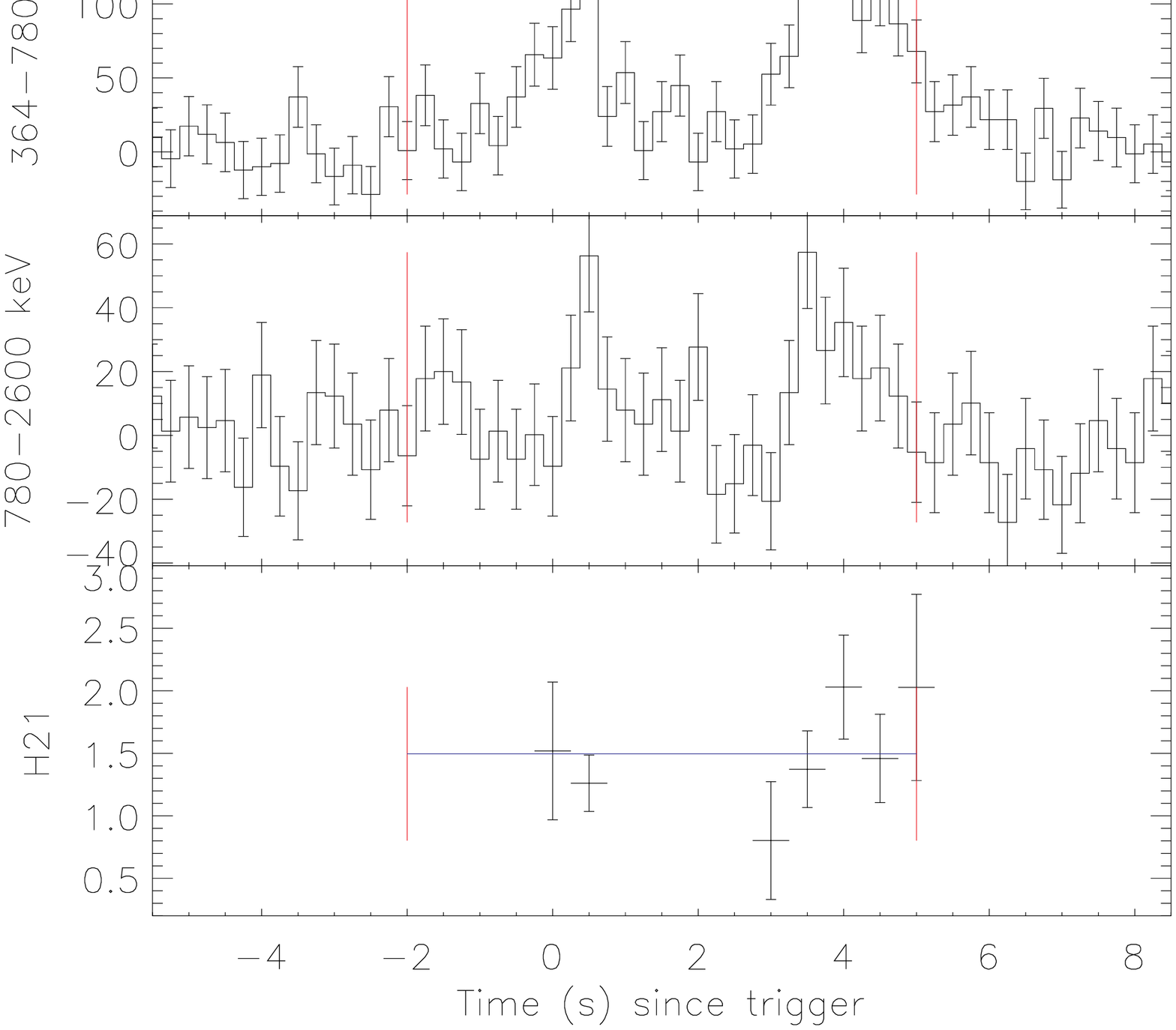}
\caption{Burst~070917 background-subtracted light curve with 0.25~s binning time and hardness ratio with 0.5~s binning time. See caption of Fig.~\ref{lcfirst} for a description of the lines.}
\end{figure}
}

\onlfig{27}{
\begin{figure}
\includegraphics[angle=90,width=0.45\textwidth]{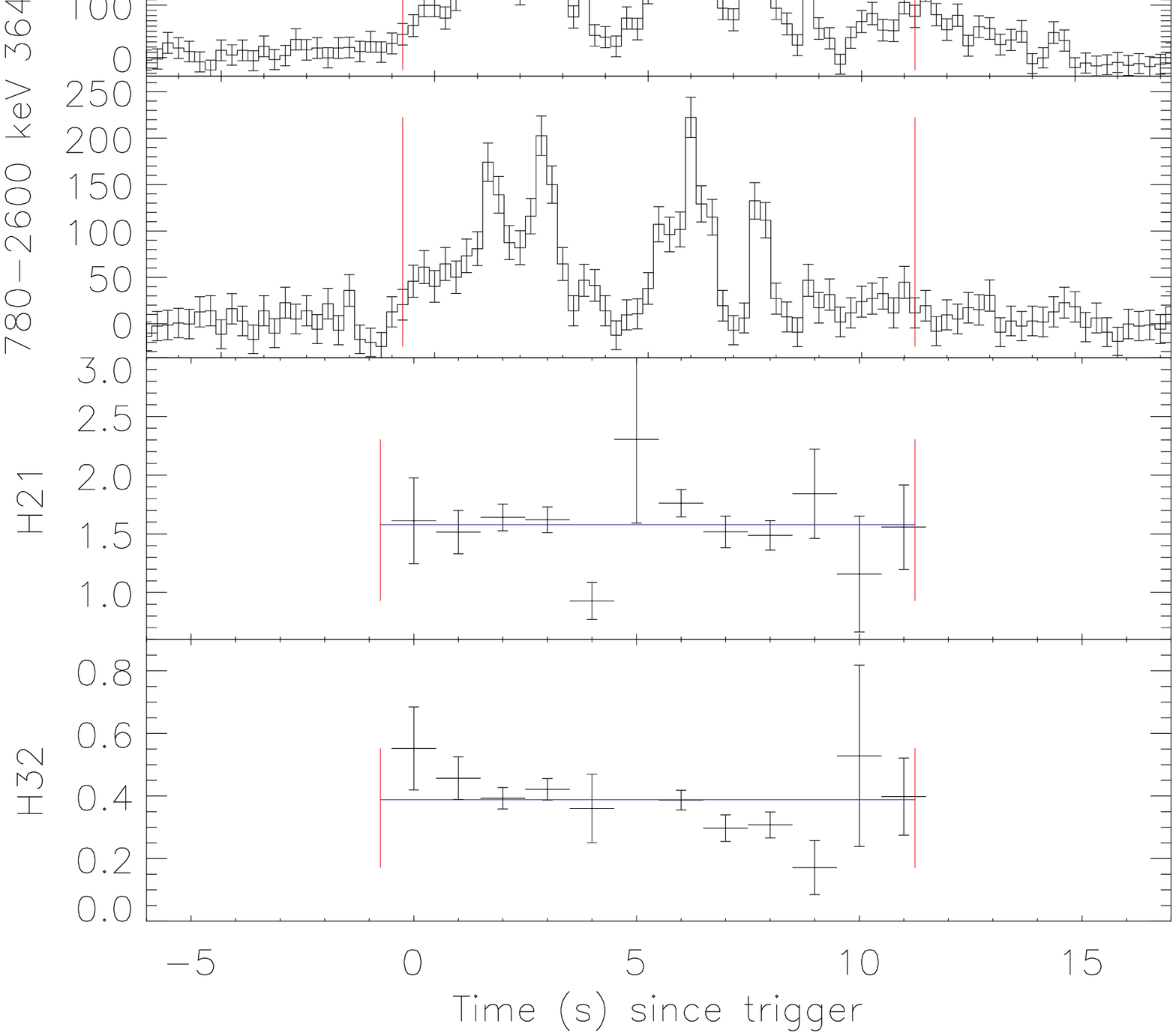}
\caption{GRB~071003 background-subtracted light curve with 0.25~s binning time and hardness ratio with 1~s binning time. See caption of Fig.~\ref{lcfirst} for a description of the lines. }
\end{figure}
}

\onlfig{28}{
\begin{figure}
\includegraphics[angle=90,width=0.45\textwidth]{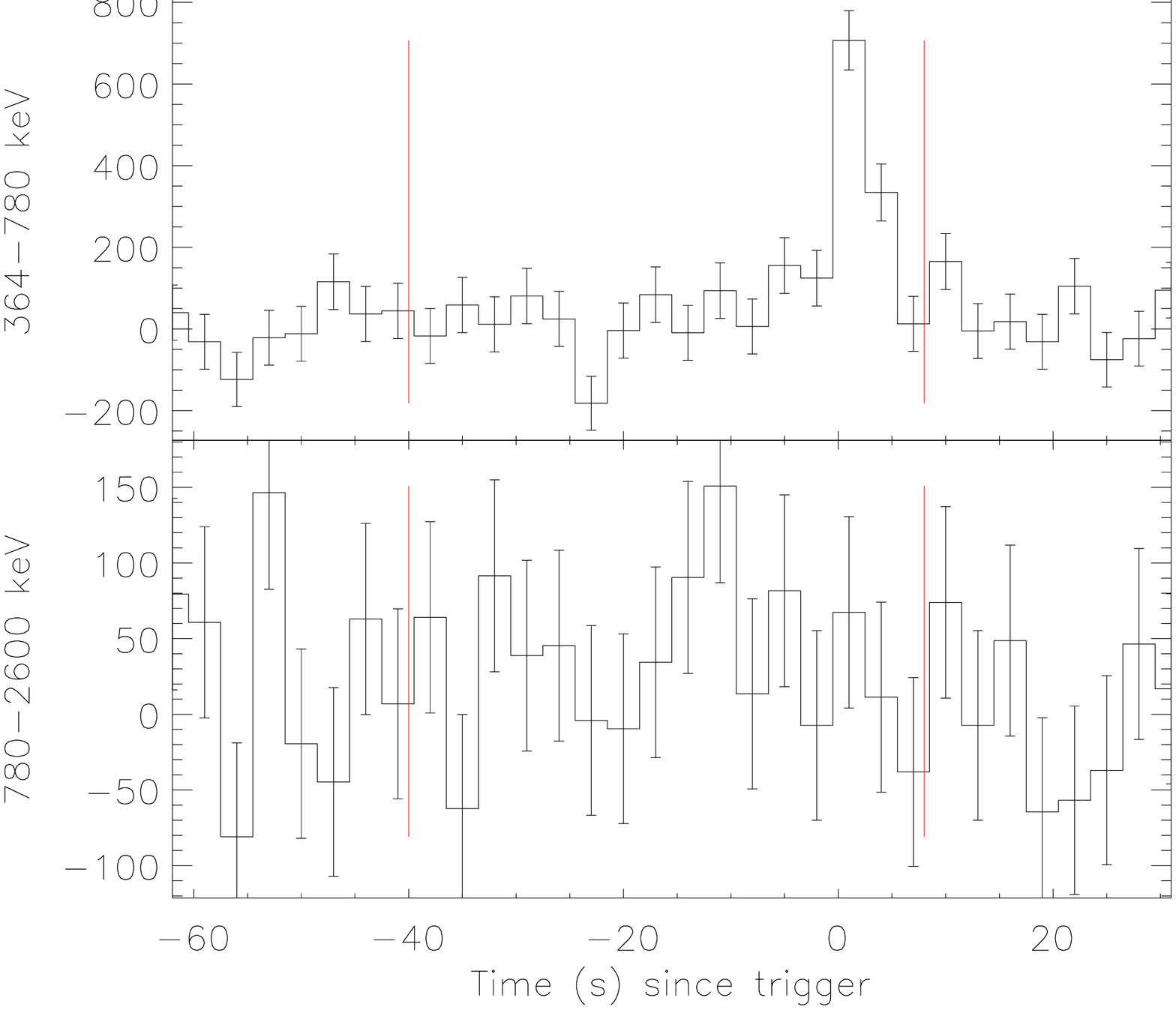}
\caption{GRB~071006 background-subtracted light curve with 3~s binning time.  Vertical (red) lines define the T$_{90}$ interval. }
\end{figure}
}

\onlfig{29}{
\begin{figure}
\includegraphics[angle=90,width=0.45\textwidth]{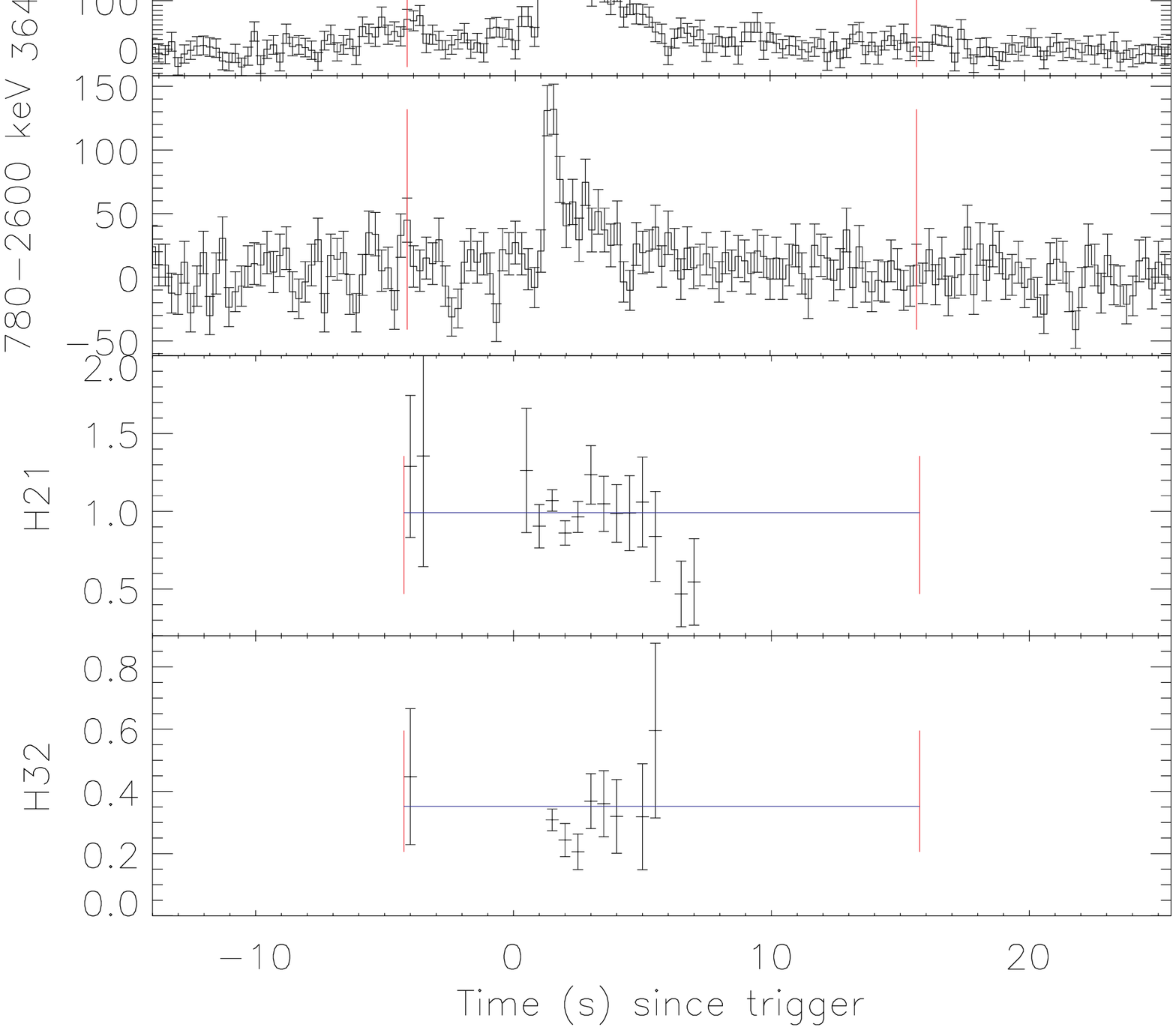}
\caption{Burst~071108 background-subtracted light curve with 0.25~s binning time and hardness ratio with 0.5~s binning time. See caption of Fig.~\ref{lcfirst} for a description of the lines.}
\end{figure}
}

\onlfig{30}{
\begin{figure}
\includegraphics[angle=90,width=0.45\textwidth]{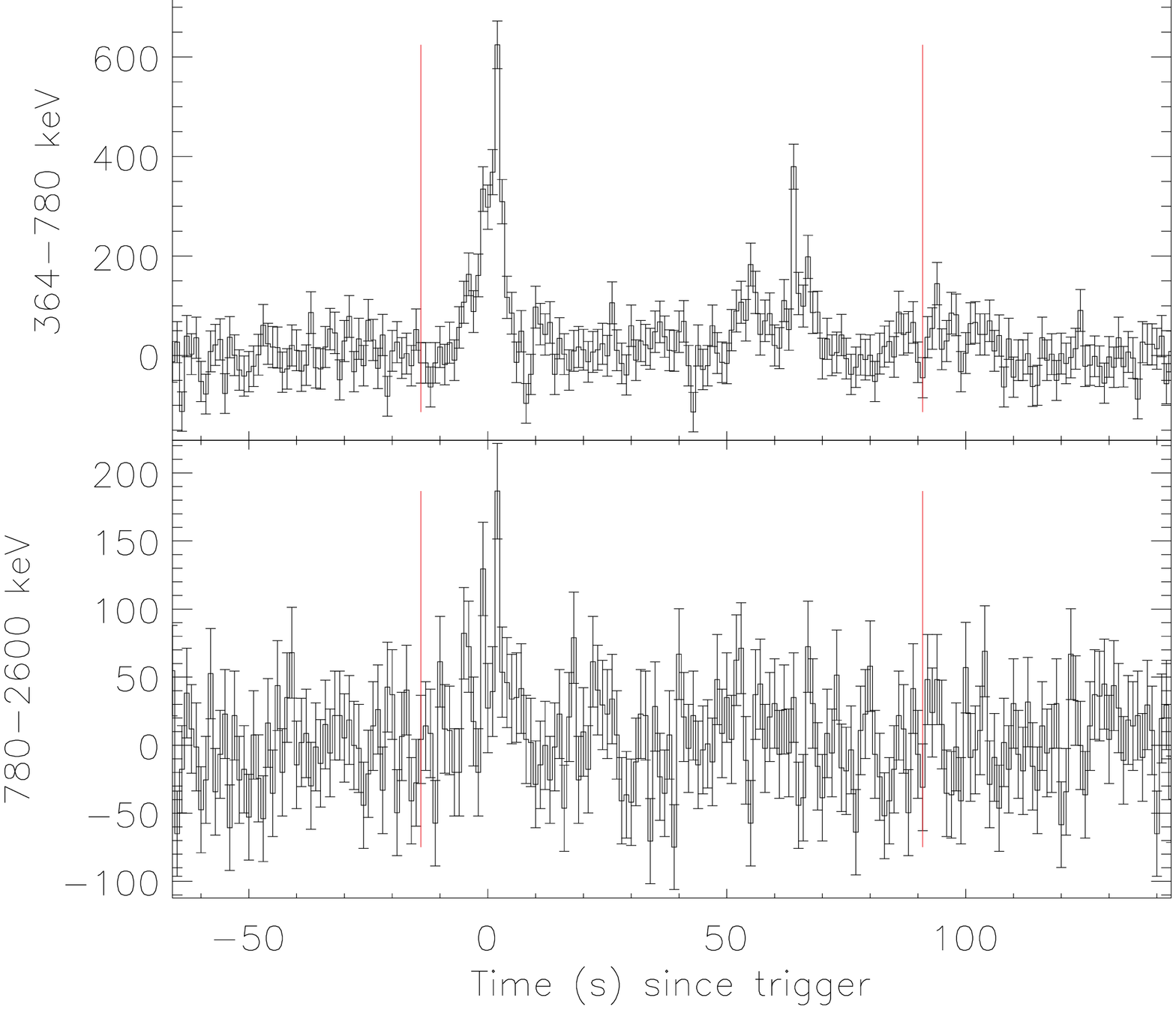}
\caption{GRB~080122 background-subtracted light curve with 1~s binning time.  Vertical (red) lines define the T$_{90}$ interval. }
\end{figure}
}

\onlfig{31}{
\begin{figure}
\includegraphics[angle=90,width=0.45\textwidth]{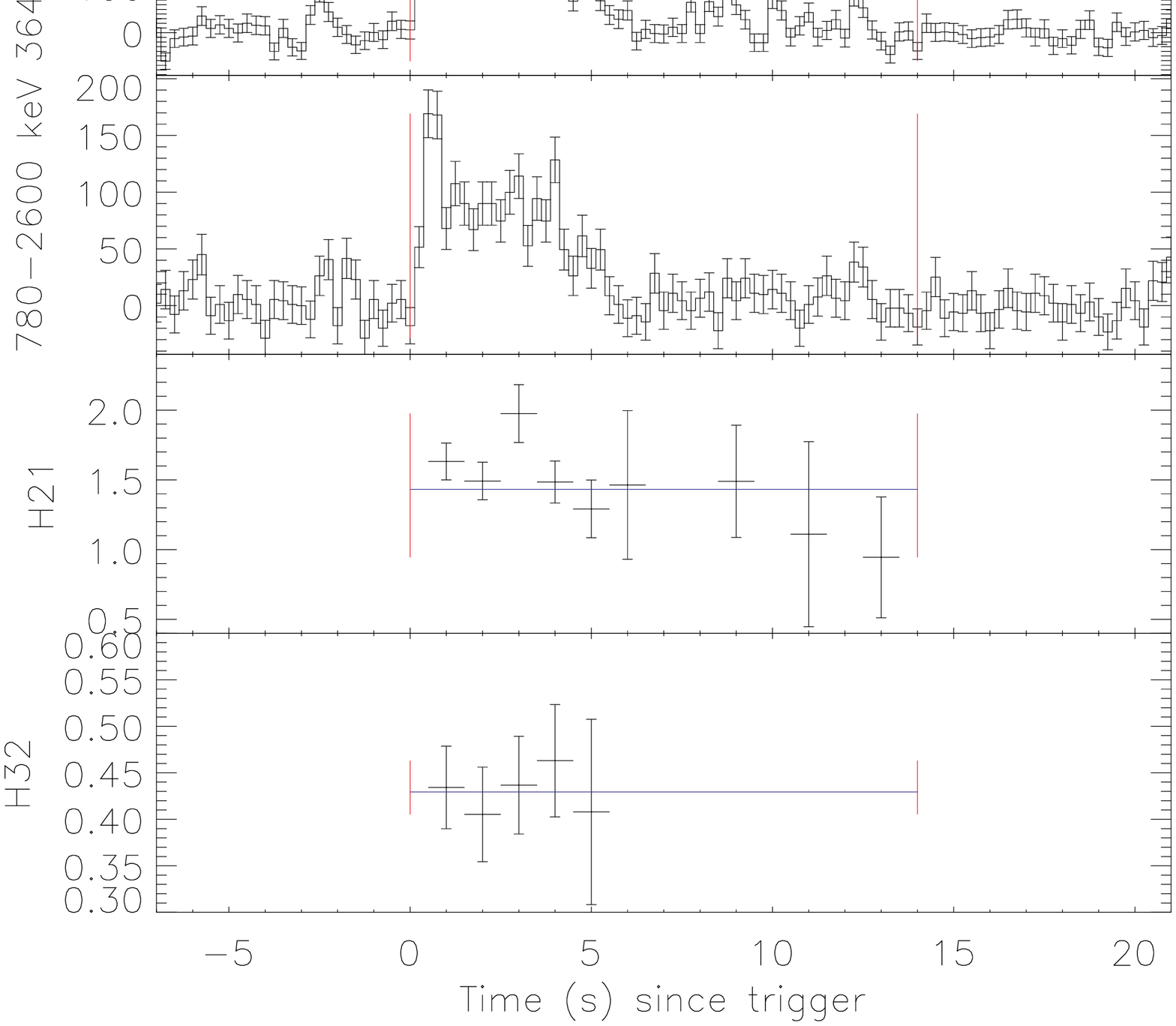}
\caption{GRB~080204 background-subtracted light curve with 0.25~s binning time and hardness ratio with 1~s binning time. See caption of Fig.~\ref{lcfirst} for a description of the lines.}
\end{figure}
}

\onlfig{32}{
\begin{figure}
\includegraphics[angle=90,width=0.45\textwidth]{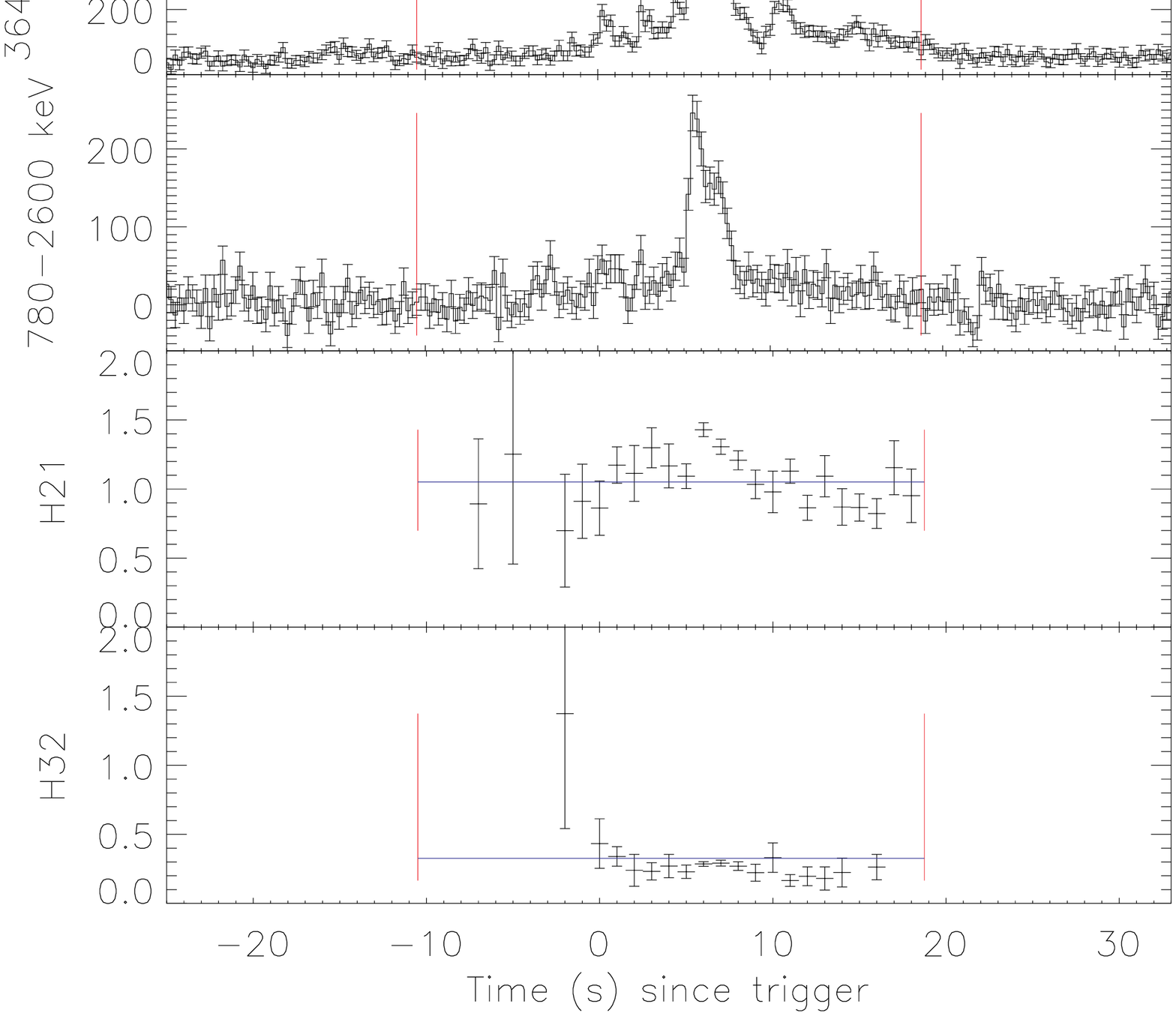}
\caption{Burst~080303 background-subtracted light curve with 0.25~s binning time and hardness ratio with 1~s binning time. See caption of Fig.~\ref{lcfirst} for a description of the lines.}
\end{figure}
}

\onlfig{33}{
\begin{figure}
\includegraphics[angle=90,width=0.45\textwidth]{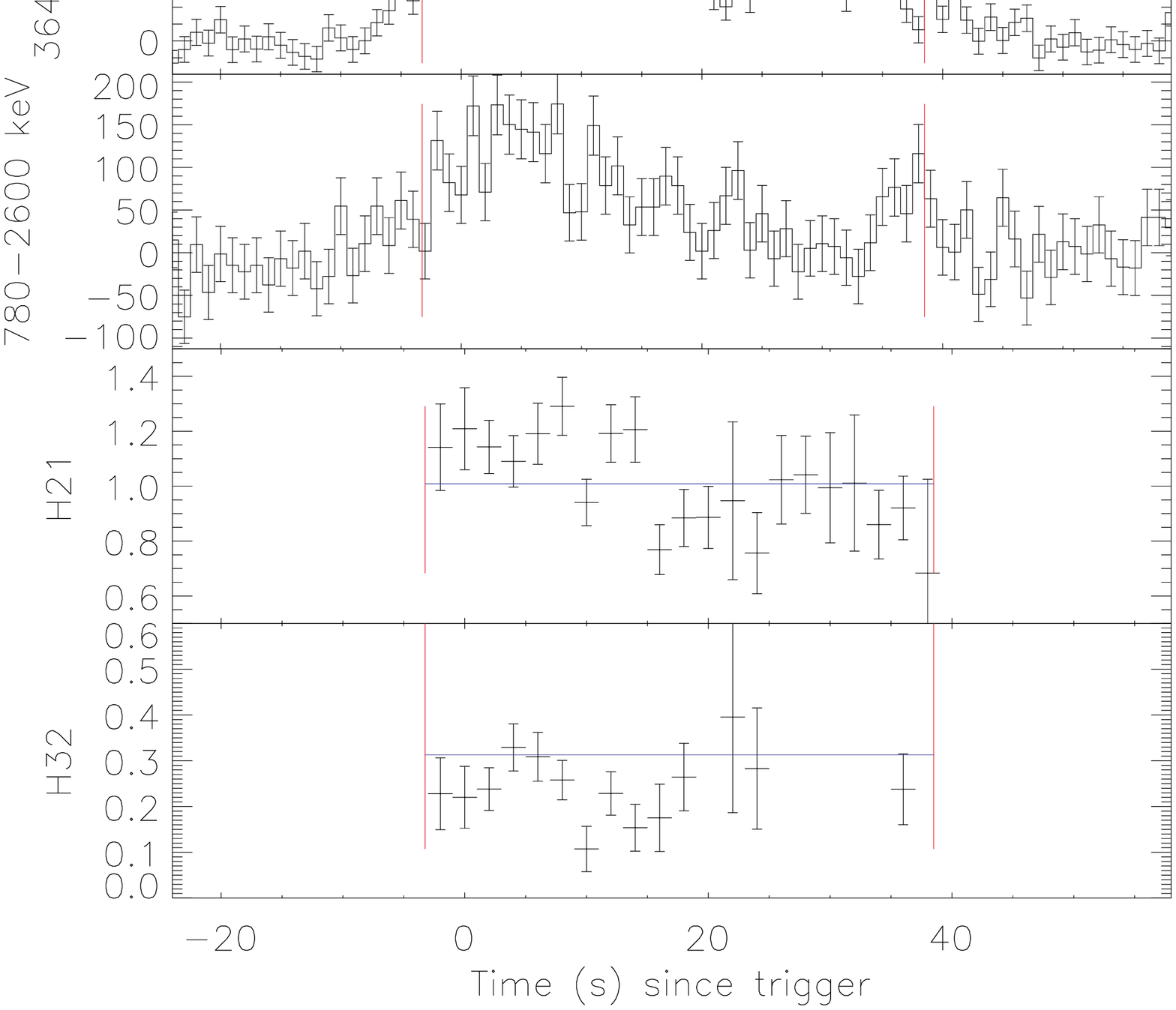}
\caption{GRB~080319B background-subtracted light curve with 1~s binning time and hardness ratio with 2~s binning time. See caption of Fig.~\ref{lcfirst} for a description of the lines.}
\end{figure}
}

\onlfig{34}{
\begin{figure}
\includegraphics[angle=90,width=0.45\textwidth]{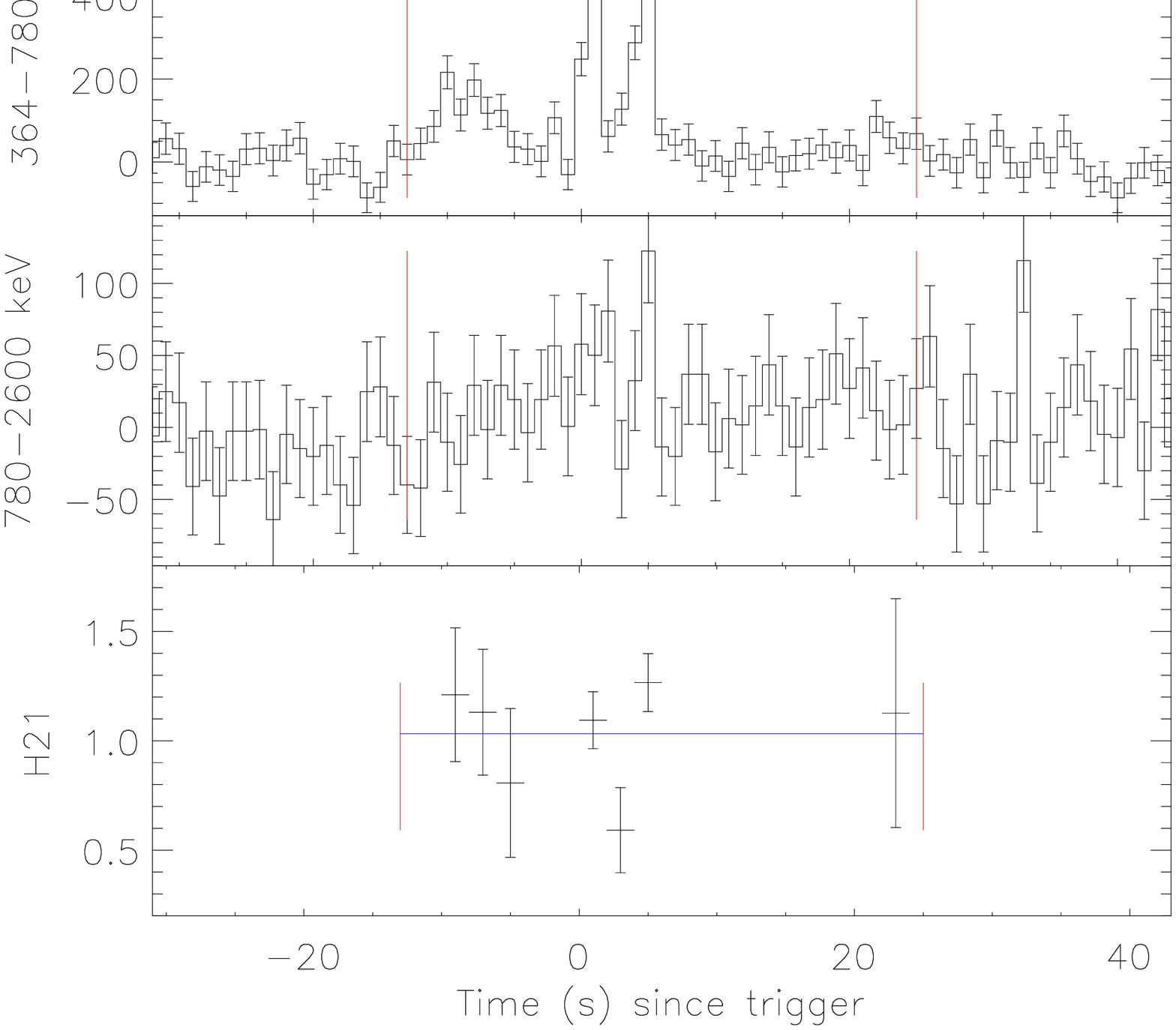}
\caption{GRB~080328 background-subtracted light curve with 1~s binning time and hardness ratio with 2~s binning time. See caption of Fig.~\ref{lcfirst} for a description of the lines.}
\end{figure}
}

\onlfig{35}{
\begin{figure}
\includegraphics[angle=90,width=0.45\textwidth]{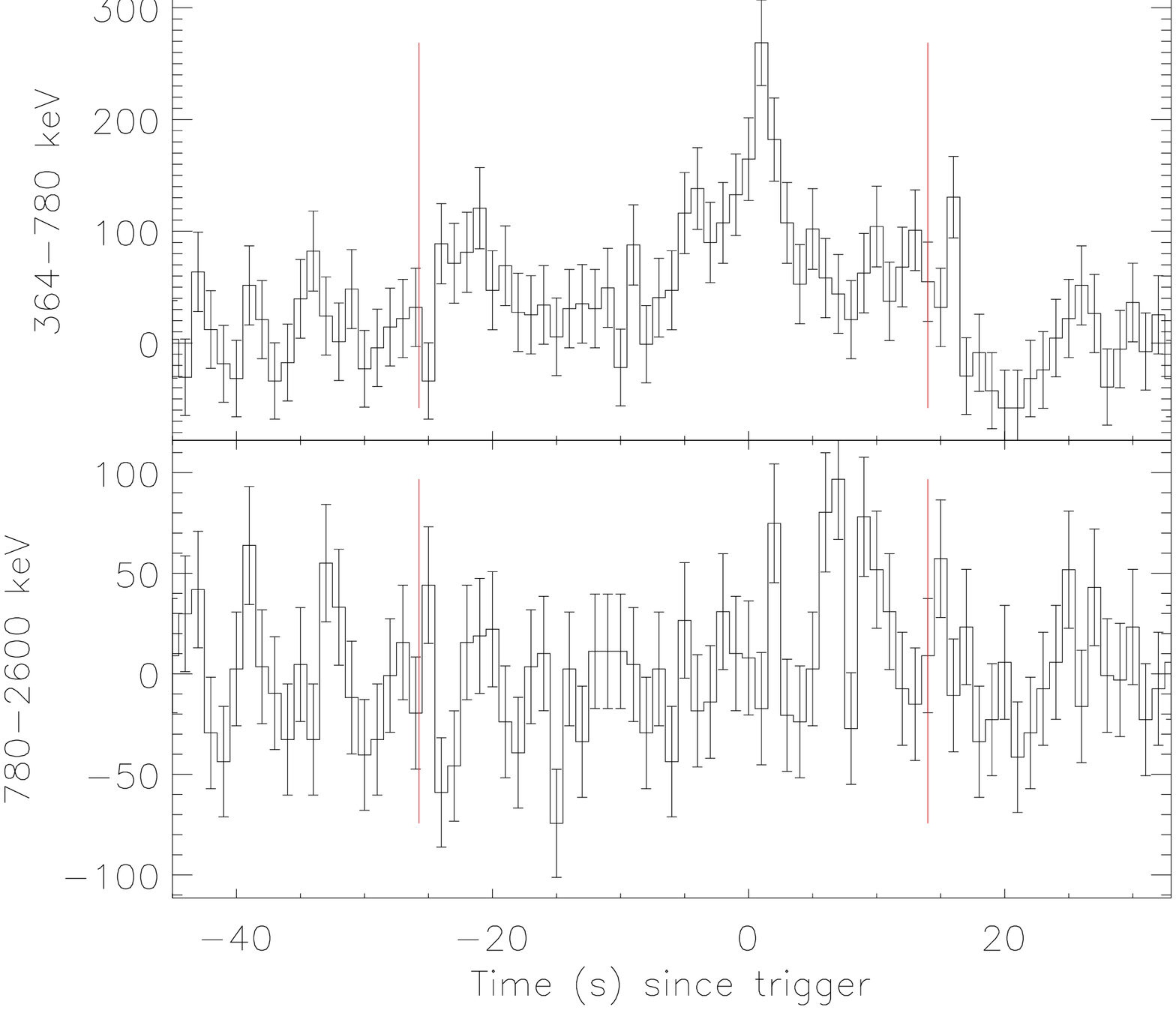}
\caption{Burst~080408 background-subtracted light curve with 1~s binning time.  Vertical (red) lines define the T$_{90}$ interval. }
\end{figure}
}

\onlfig{36}{
\begin{figure}
\includegraphics[angle=90,width=0.45\textwidth]{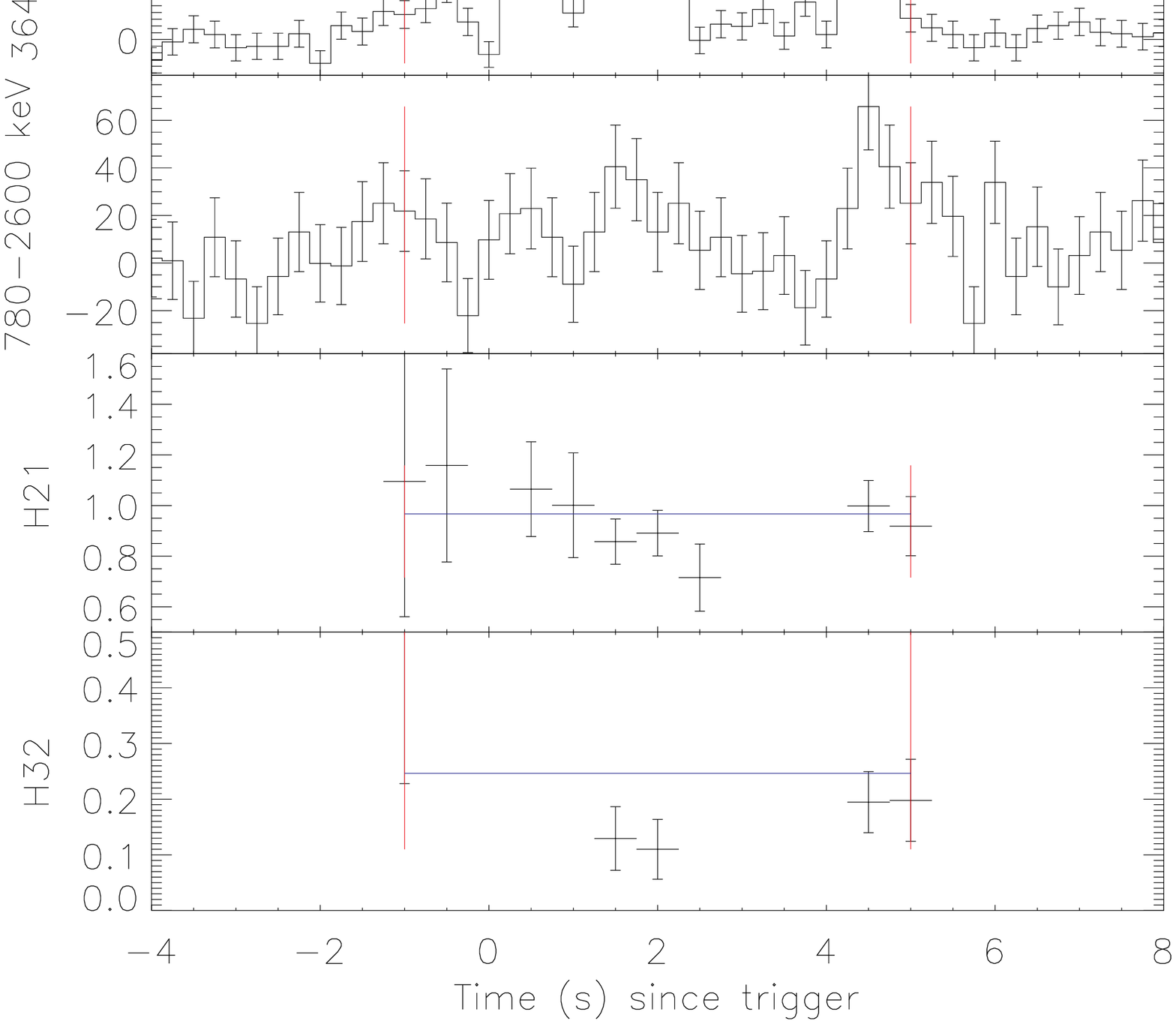}
\caption{GRB~080514B background-subtracted light curve with 0.25~s binning time and hardness ratio with 0.5~s binning time. See caption of Fig.~\ref{lcfirst} for a description of the lines.}
\end{figure}
}

\onlfig{37}{
\begin{figure}
\includegraphics[angle=90,width=0.45\textwidth]{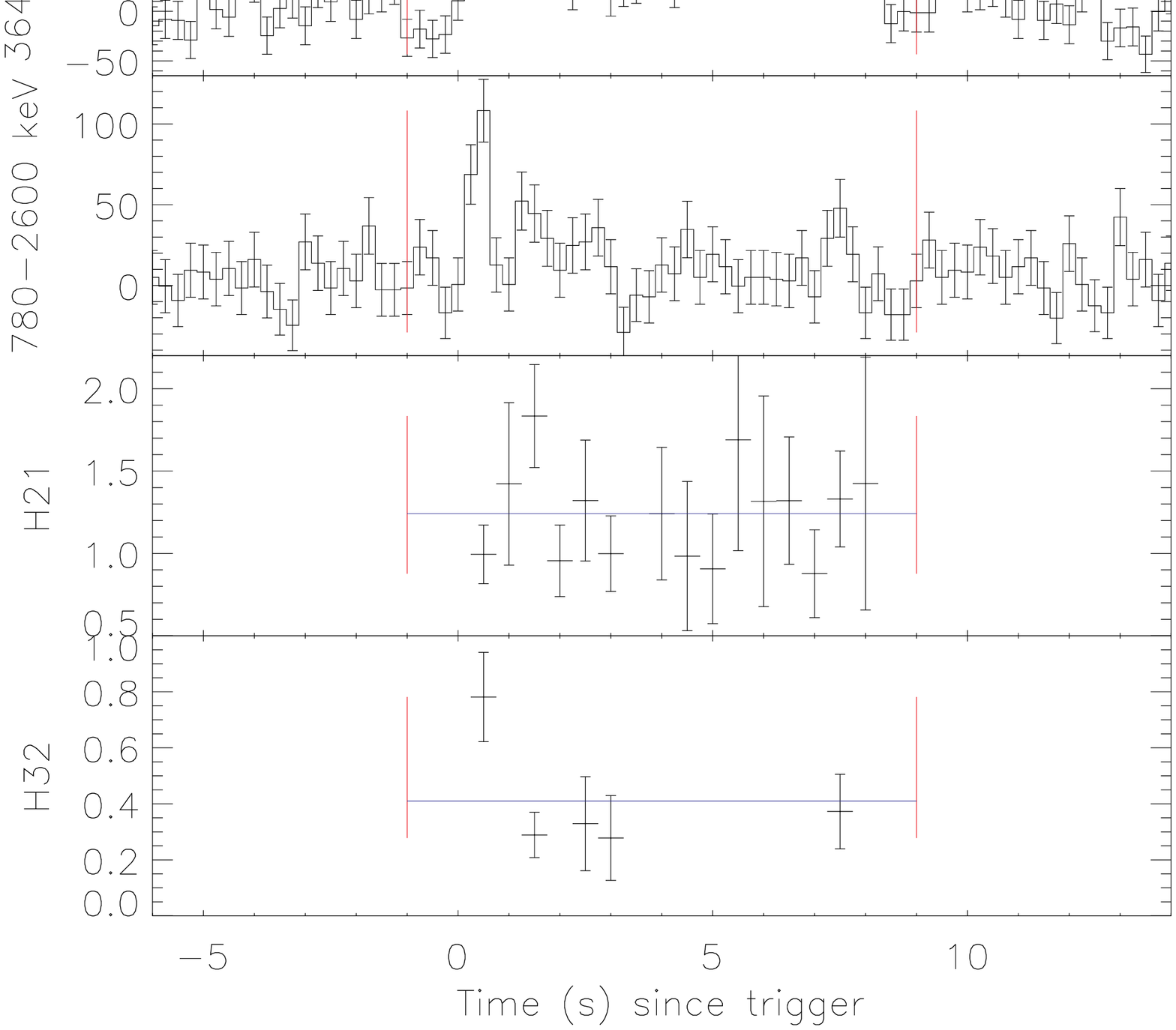}
\caption{GRB~080607 background-subtracted light curve with -.25~s binning time and hardness ratio with 0.5~s binning time. See caption of Fig.~\ref{lcfirst} for a description of the lines.}
\end{figure}
}

\onlfig{38}{
\begin{figure}
\includegraphics[angle=90,width=0.45\textwidth]{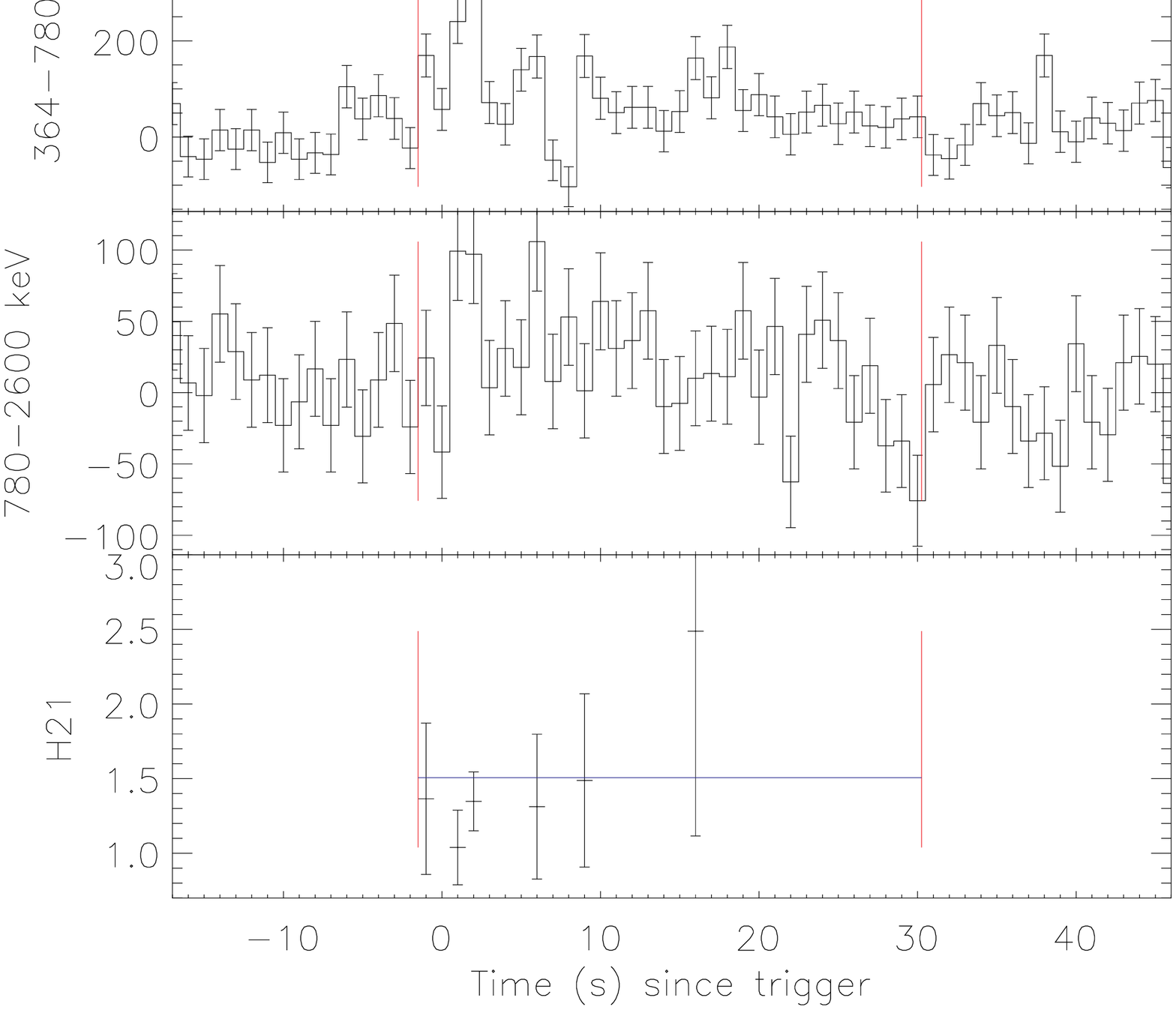}
\caption{GRB~080613B background-subtracted light curve with 1~s binning time and hardness ratio with 1~s binning time.  See caption of Fig.~\ref{lcfirst} for a description of the lines. }
\end{figure}
}

\onlfig{39}{
\begin{figure}
\includegraphics[angle=90,width=0.45\textwidth]{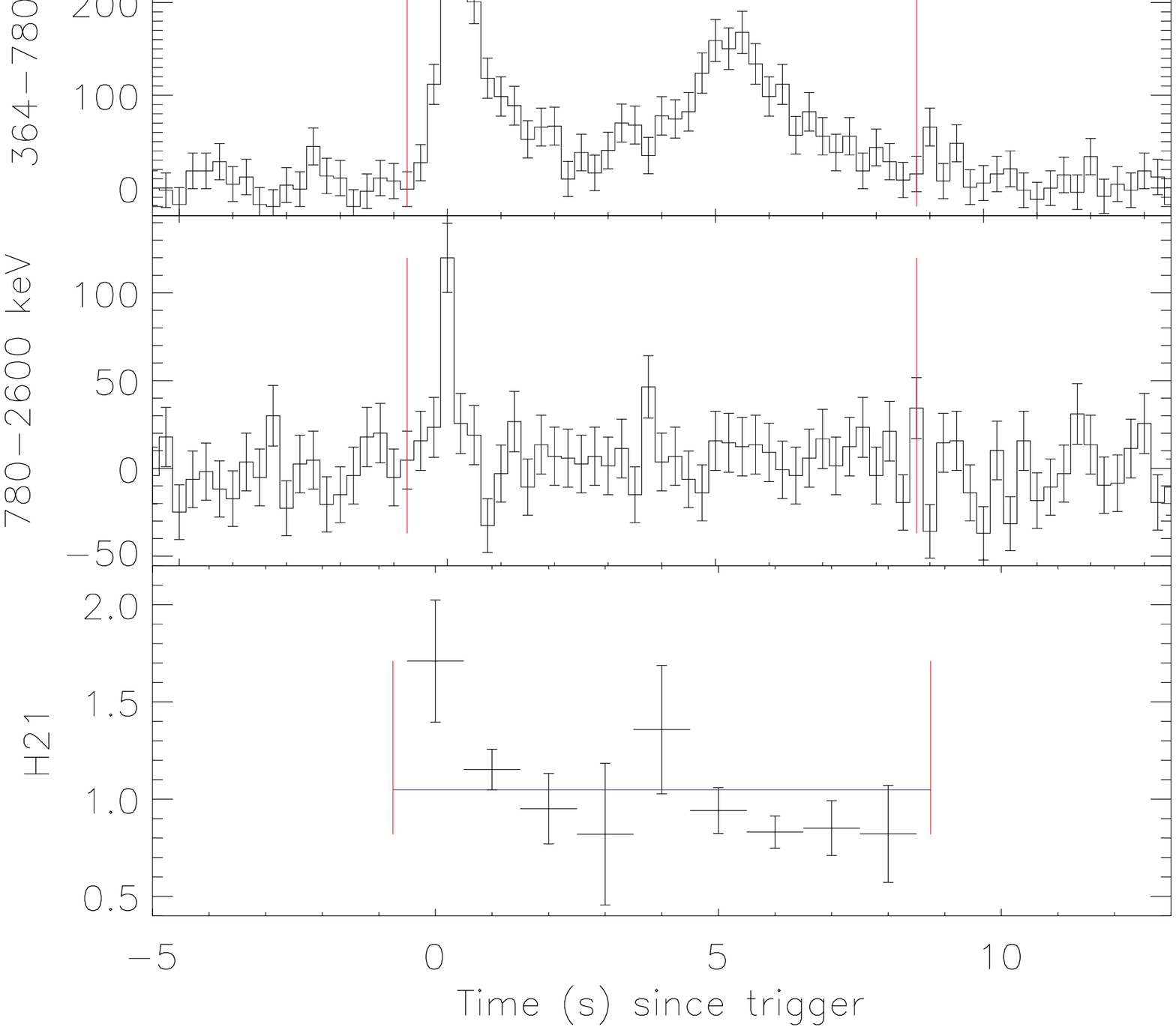}
\caption{Burst~080615 background-subtracted light curve with 0.25~s binning time and hardness ratio with 1~s binning time. See caption of Fig.~\ref{lcfirst} for a description of the lines.}
\end{figure}
}

\onlfig{40}{
\begin{figure}
\includegraphics[angle=90,width=0.45\textwidth]{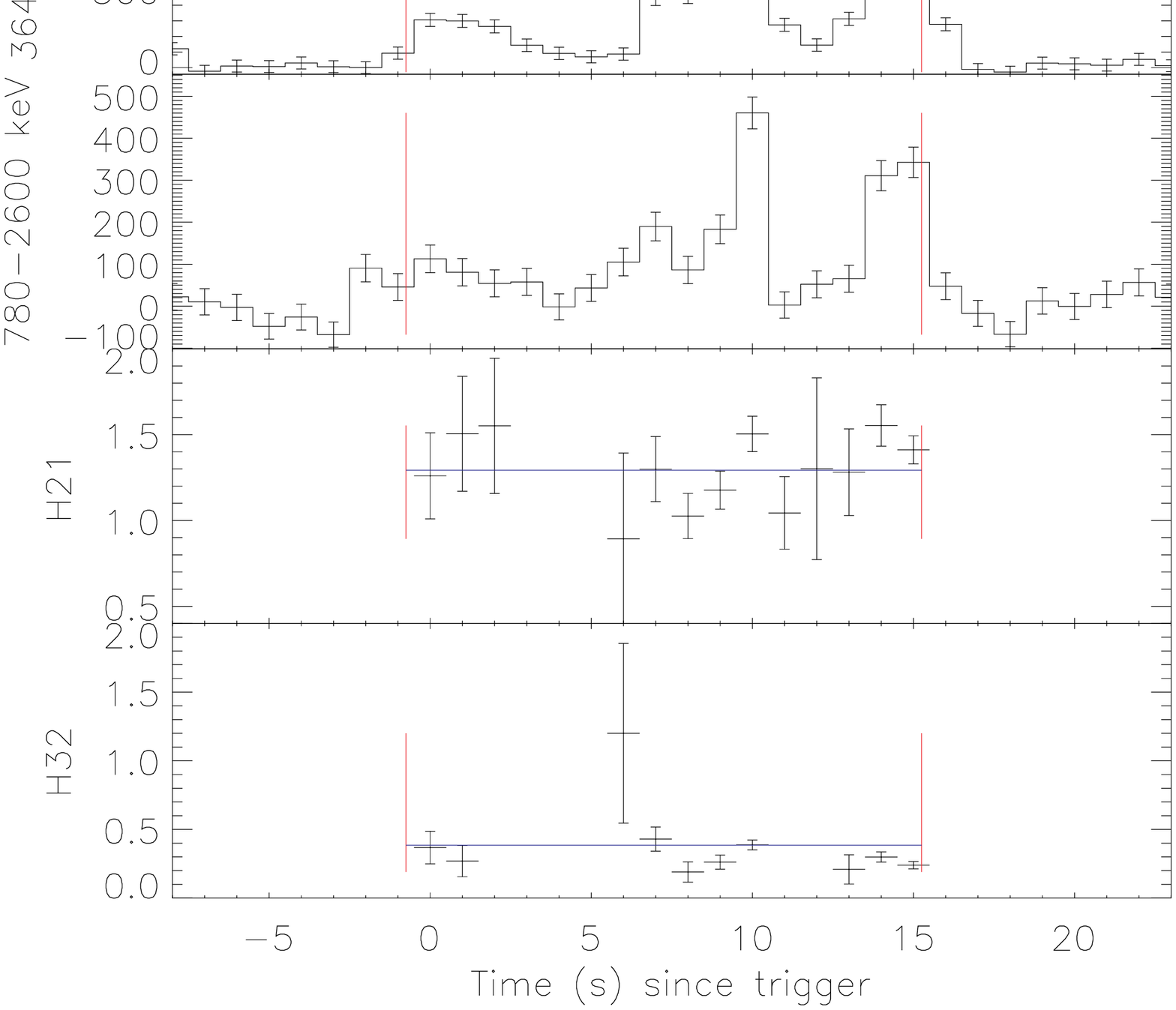}
\caption{GRB~080721 background-subtracted light curve with 1~s binning time and hardness ratio with 1~s binning time. See caption of Fig.~\ref{lcfirst} for a description of the lines.}
\end{figure}
}

\onlfig{41}{
\begin{figure}
\includegraphics[angle=90,width=0.45\textwidth]{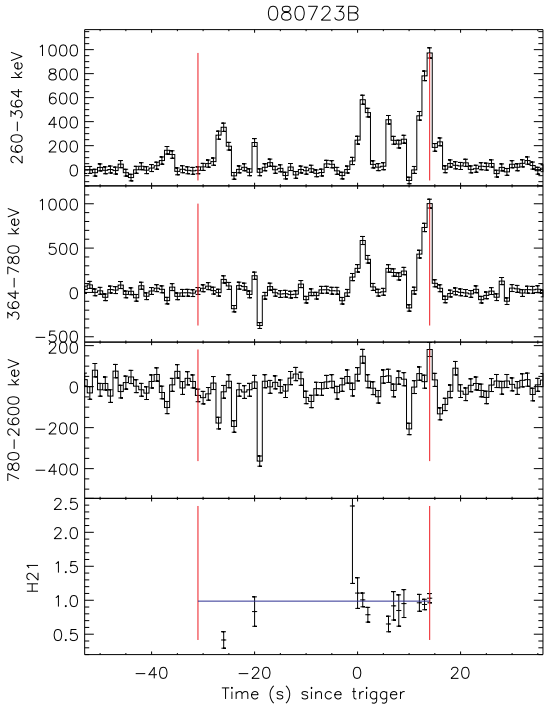}
\caption{GRB~080723B background-subtracted light curve with 1~s binning time and hardness ratio with 1~s binning time. Data gaps are visible in all energy bands. See caption of Fig.~\ref{lcfirst} for a description of the lines.}
\end{figure}
}

\onlfig{42}{
\begin{figure}
\includegraphics[angle=90,width=0.45\textwidth]{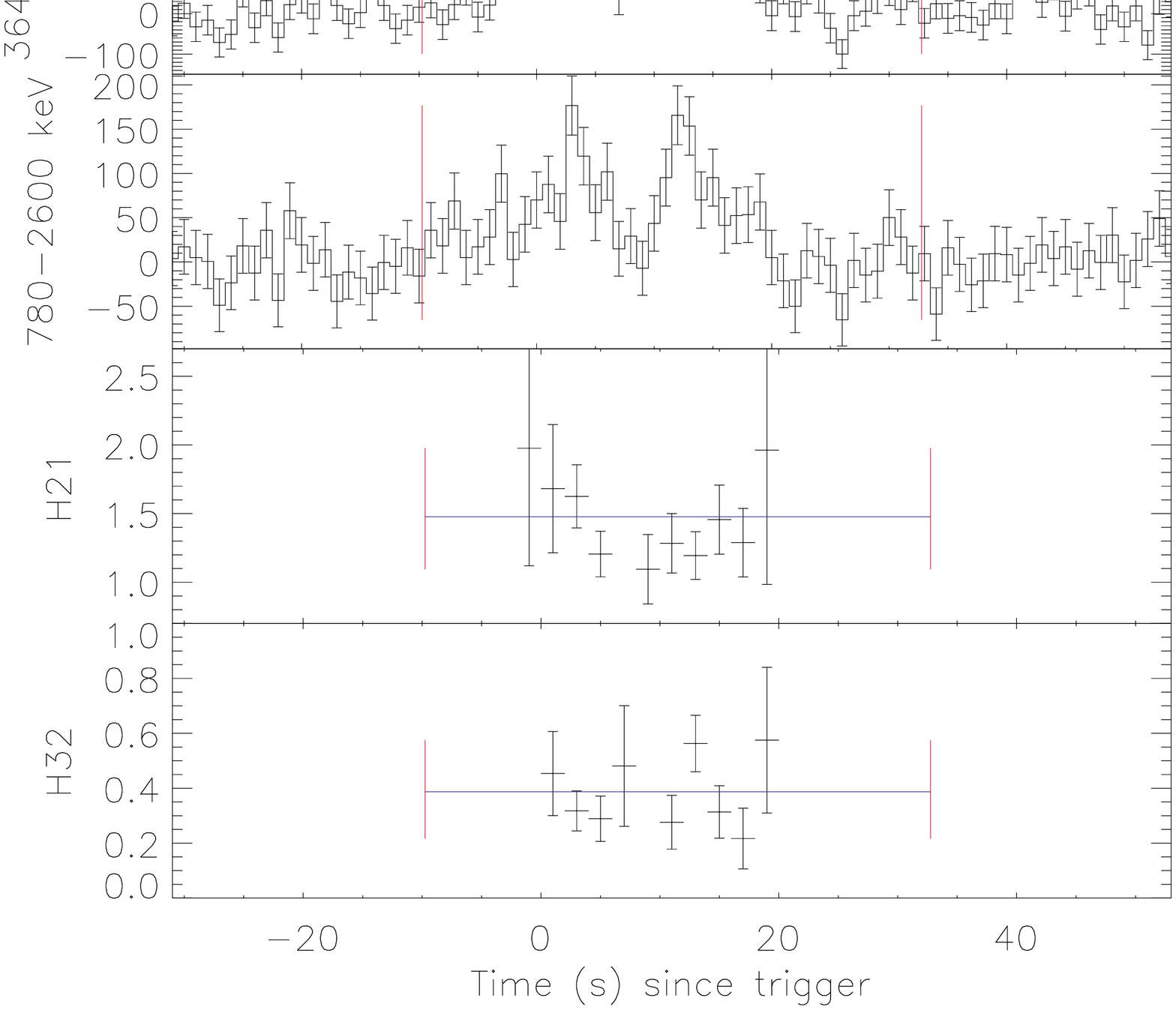}
\caption{GRB~080817 background-subtracted light curve with 1~s binning time and hardness ratio with 2~s binning time. See caption of Fig.~\ref{lcfirst} for a description of the lines.}
\end{figure}
}

\onlfig{43}{
\begin{figure}
\includegraphics[angle=90,width=0.45\textwidth]{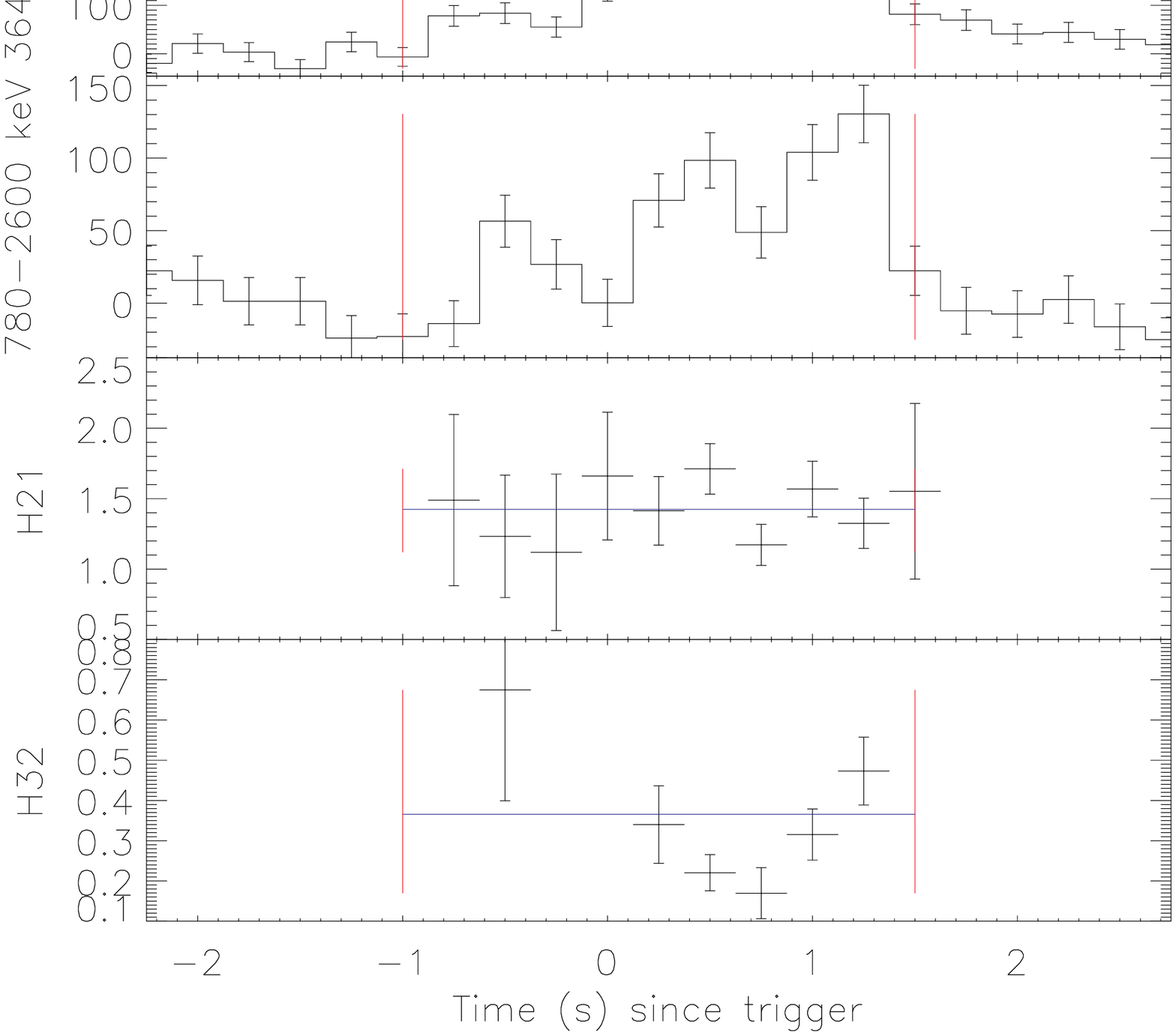}
\caption{Burst~080918 background-subtracted light curve with 0.25~s binning time and hardness ratio with 0.25~s binning time. See caption of Fig.~\ref{lcfirst} for a description of the lines.}
\end{figure}
}

\onlfig{44}{
\begin{figure}
\includegraphics[angle=90,width=0.45\textwidth]{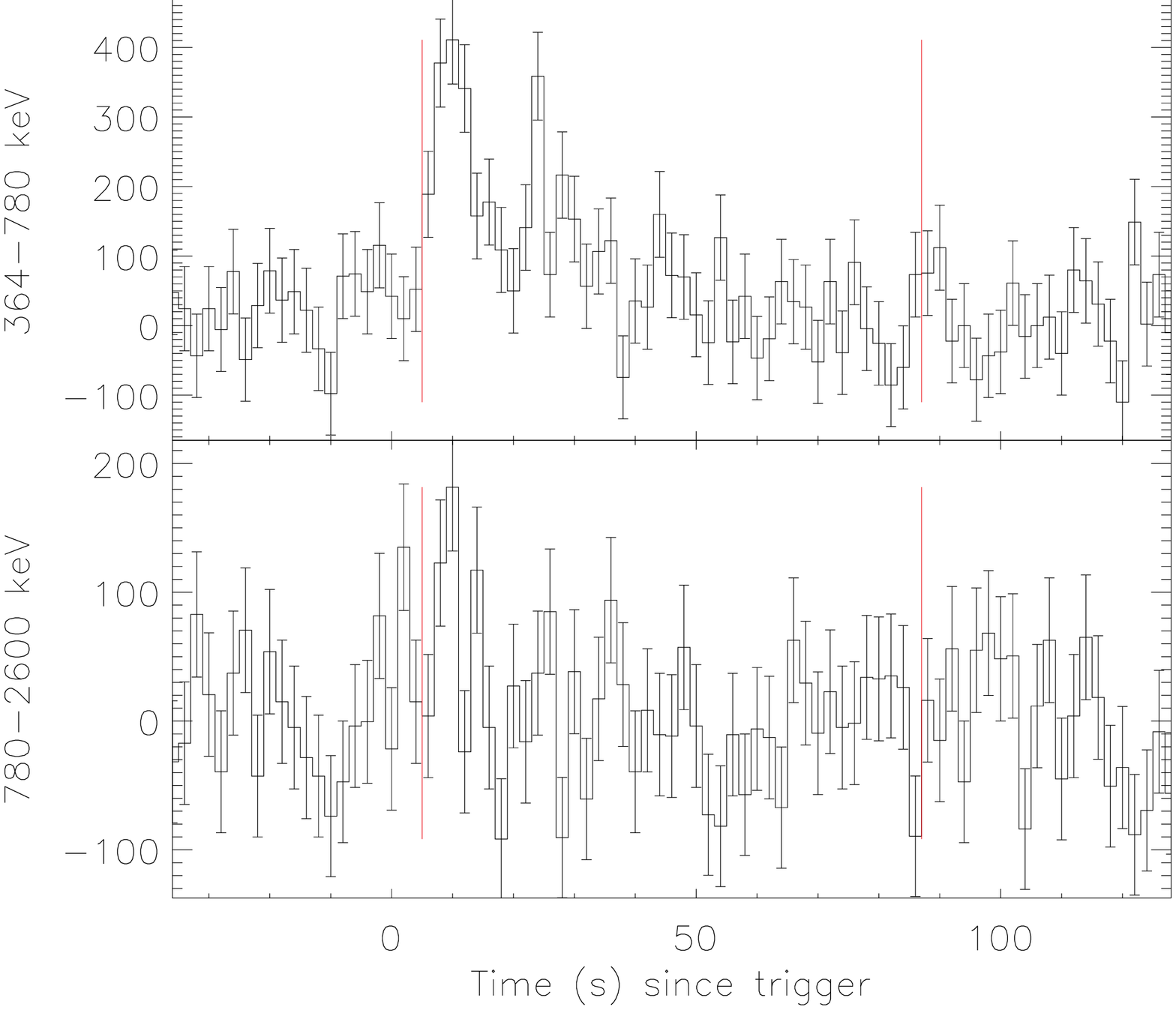}
\caption{GRB~090528B background-subtracted light curve with 2~s binning time.  Vertical (red) lines define the T$_{90}$ interval. }
\end{figure}
}

\onlfig{45}{
\begin{figure}
\includegraphics[angle=90,width=0.45\textwidth]{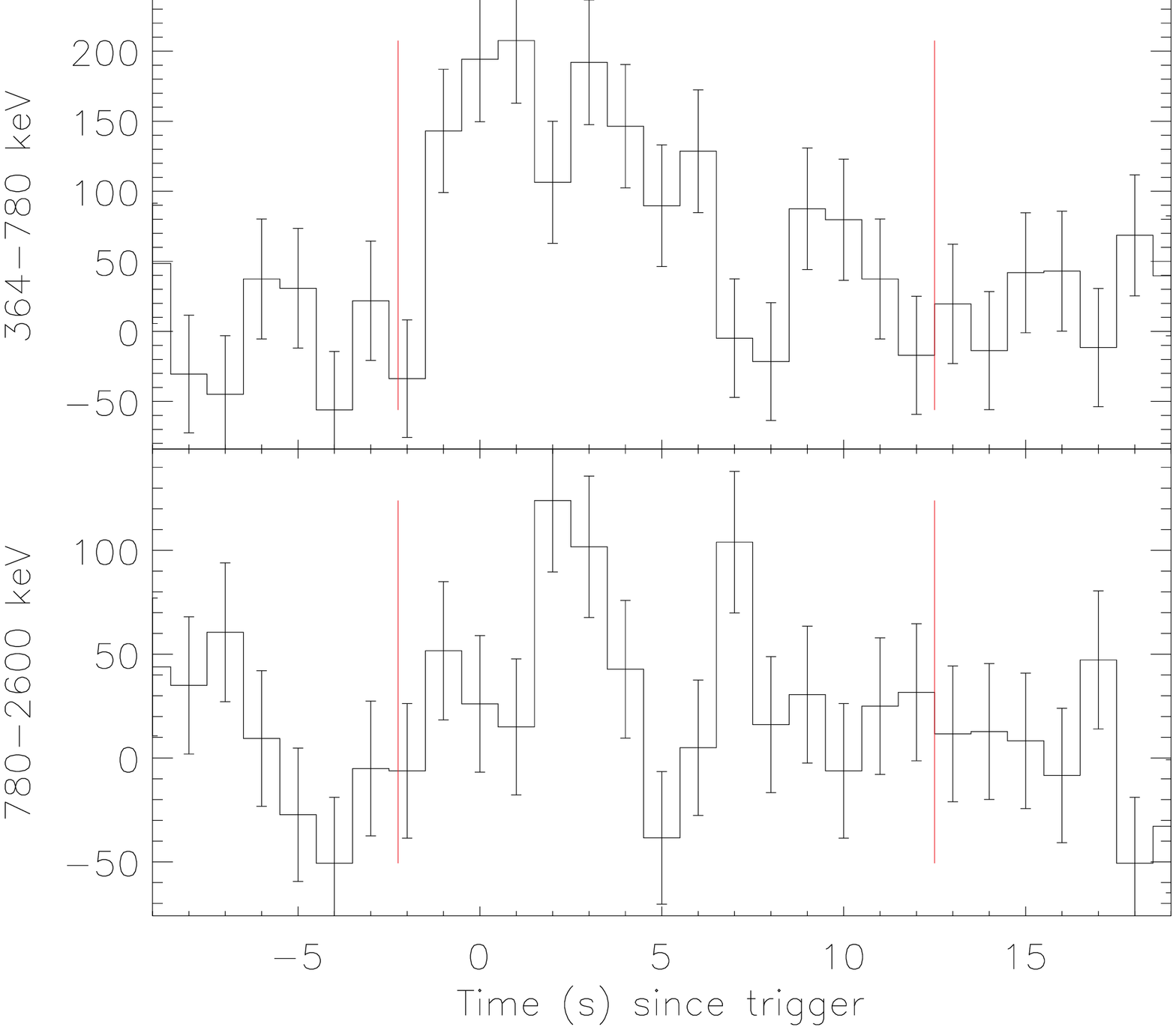}
\caption{GRB~090618A background-subtracted light curve with 1~s binning time.  Vertical (red) lines define the T$_{90}$ interval. }
\end{figure}
}

\onlfig{46}{
\begin{figure}
\includegraphics[angle=90,width=0.45\textwidth]{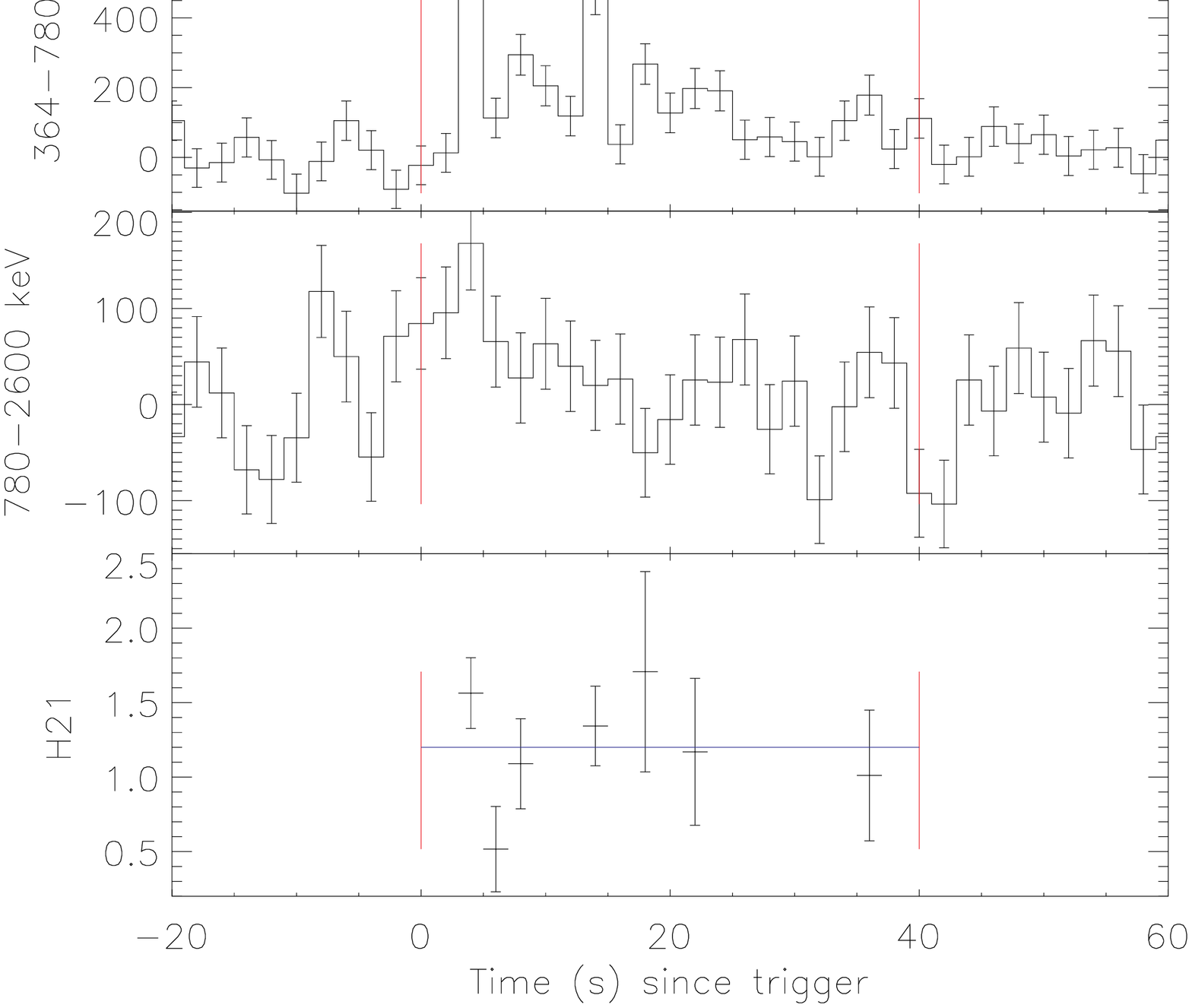}
\caption{GRB~090623 background-subtracted light curve with 2~s binning time and hardness ratio with 2~s binning time. See caption of Fig.~\ref{lcfirst} for a description of the lines.}
\end{figure}
}

\onlfig{47}{
\begin{figure}
\includegraphics[angle=90,width=0.45\textwidth]{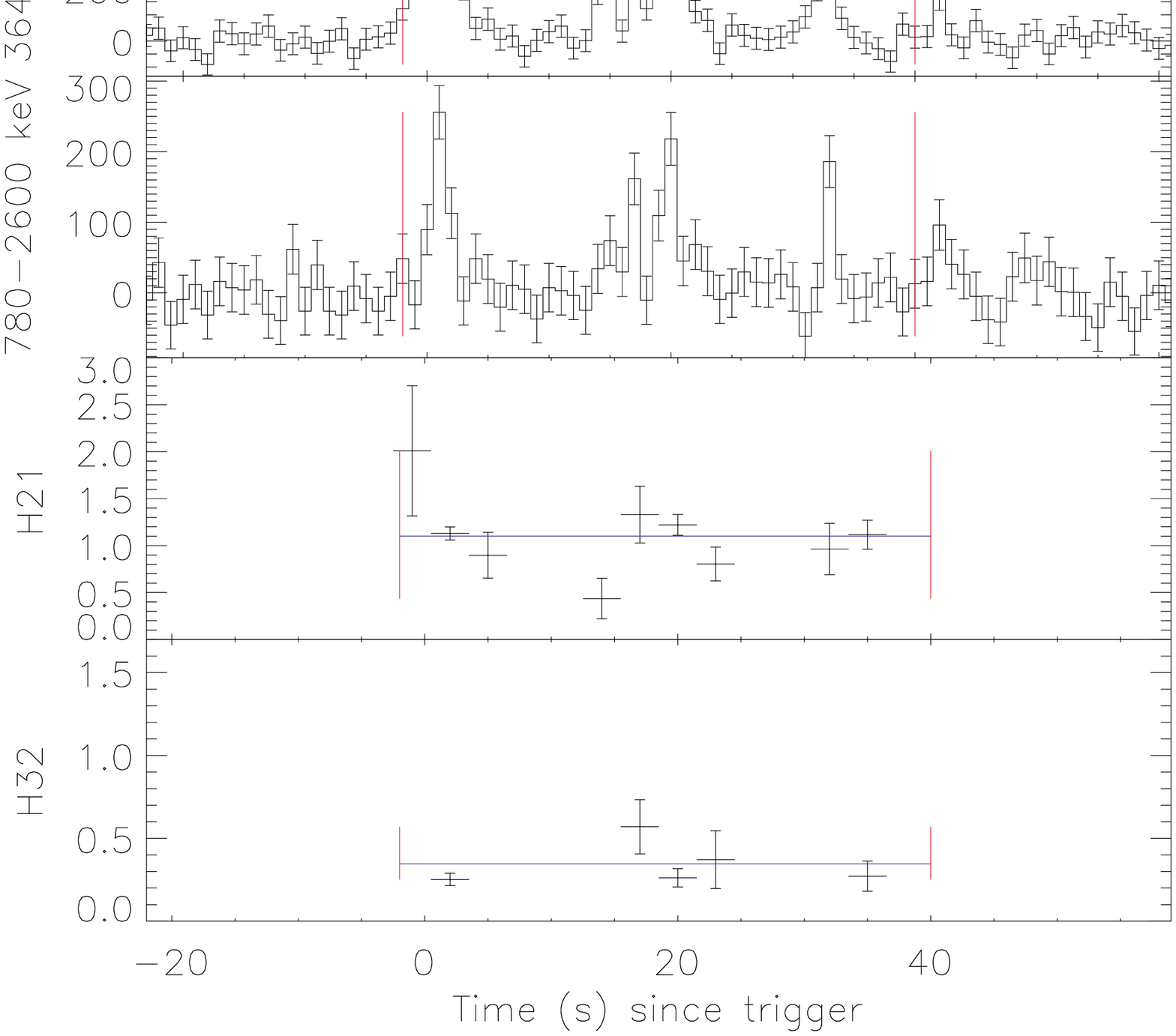}
\caption{GRB~090626A background-subtracted light curve with 1~s binning time and hardness ratio with 3~s binning time. See caption of Fig.~\ref{lcfirst} for a description of the lines.}
\end{figure}
}

\onlfig{48}{
\begin{figure}
\includegraphics[angle=90,width=0.45\textwidth]{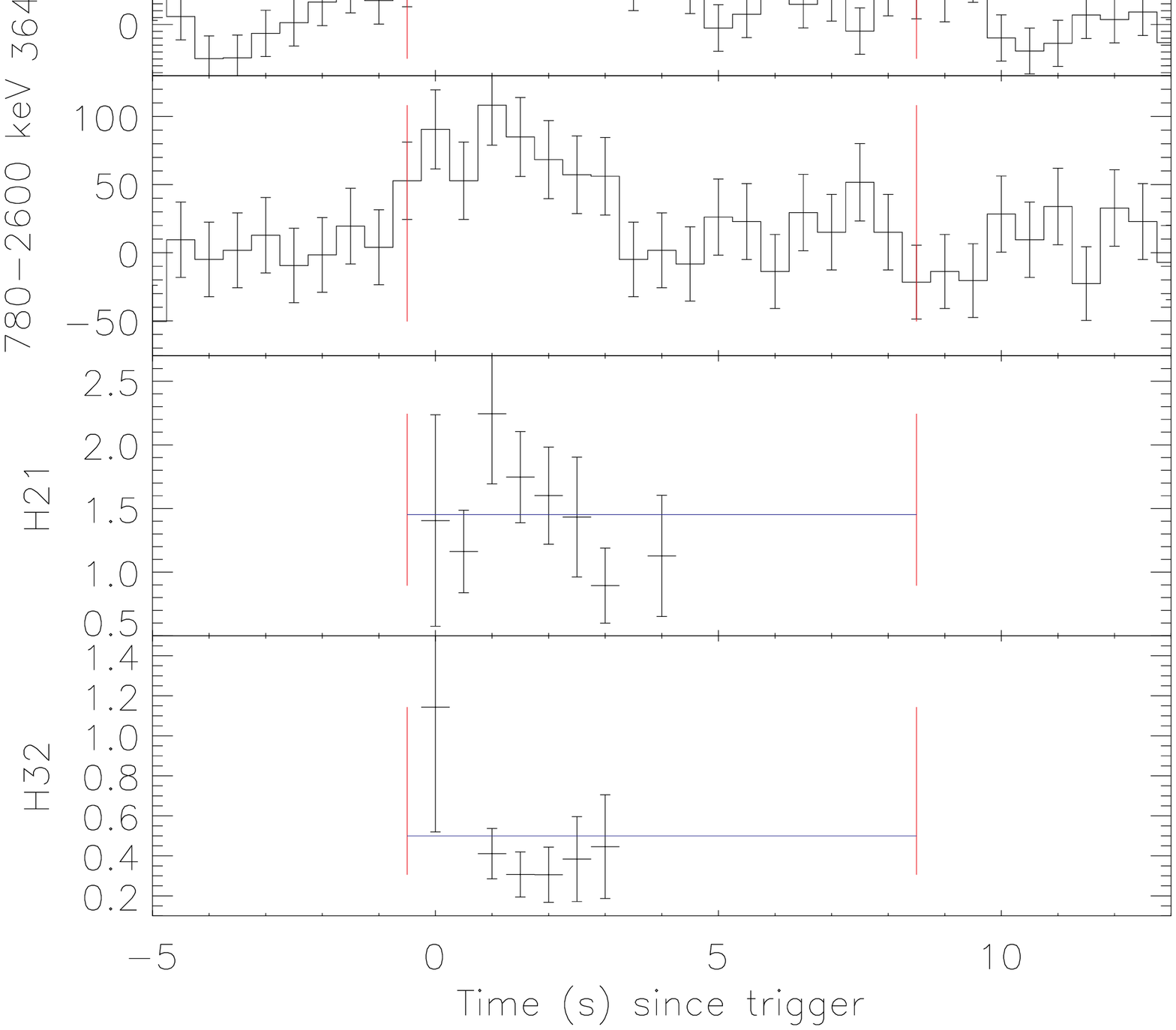}
\caption{\label{lclast} GRB~090809B background-subtracted light curve with 0.5~s binning time and hardness ratio with 0.5~s binning time. See caption of Fig.~\ref{lcfirst} for a description of the lines.}
\end{figure}
}


\begin{thebibliography}{64}
\expandafter\ifx\csname natexlab\endcsname\relax\def\natexlab#1{#1}\fi

\bibitem[{{Bellm} {et~al.}(2006){Bellm}, {Bandstra}, {Boggs}, {Smith}, {Lin},
  {McTiernan}, {Schwartz}, {Wigger}, {Hajdas}, {Zehnder}, \&
  {Hurley}}]{bellm06}
{Bellm}, E., {Bandstra}, M., {Boggs}, S., {et~al.} 2006, GRB Coordinates
  Network, 5418, 1

\bibitem[{{Bissaldi} {et~al.}(2008){Bissaldi}, {Briggs}, {von Kienlin}, \&
  {McBreen}}]{bissaldi08}
{Bissaldi}, E., {Briggs}, M., {von Kienlin}, A., \& {McBreen}, S. 2008, GRB
  Coordinates Network, 8108, 1

\bibitem[{{Cummings}(2007)}]{cummings07}
{Cummings}, J.~R. 2007, GRB Coordinates Network, 6858, 1

\bibitem[{{Di Cocco} {et~al.}(2007){Di Cocco}, {Bianchin}, {Foschini},
  {Gianotti}, {Laurent}, {Malaguti}, {Natalucci}, \& {Schiavone}}]{dicocco07}
{Di Cocco}, G., {Bianchin}, V., {Foschini}, L., {et~al.} 2007, in ESA Special
  Publication, Vol. 622, ESA Special Publication, 619--622

\bibitem[{{Di Cocco} {et~al.}(2003){Di Cocco}, {Caroli}, {Celesti}, {Foschini},
  {Gianotti}, {Labanti}, {Malaguti}, {Mauri}, {Rossi}, {Schiavone},
  {Spizzichino}, {Stephen}, {Traci}, \& {Trifoglio}}]{dicocco03}
{Di Cocco}, G., {Caroli}, E., {Celesti}, E., {et~al.} 2003, \aap, 411, L189

\bibitem[{{Fenimore} \& {Bloom}(1995)}]{fenimore95}
{Fenimore}, E.~E. \& {Bloom}, J.~S. 1995, \apj, 453, 25

\bibitem[{{Foley} {et~al.}(2008){Foley}, {McGlynn}, {Hanlon}, {McBreen}, \&
  {McBreen}}]{foley08}
{Foley}, S., {McGlynn}, S., {Hanlon}, L., {McBreen}, S., \& {McBreen}, B. 2008,
  \aap, 484, 143

\bibitem[{{Goldwurm} {et~al.}(2003){Goldwurm}, {David}, {Foschini}, {Gros},
  {Laurent}, {Sauvageon}, {Bird}, {Lerusse}, \& {Produit}}]{goldwurm03}
{Goldwurm}, A., {David}, P., {Foschini}, L., {et~al.} 2003, \aap, 411, L223

\bibitem[{{Golenetskii} {et~al.}(2006{\natexlab{a}}){Golenetskii}, {Aptekar},
  {Mazets}, {Pal'Shin}, {Frederiks}, \& {Cline}}]{golenetskii06a}
{Golenetskii}, S., {Aptekar}, R., {Mazets}, E., {et~al.} 2006{\natexlab{a}},
  GRB Coordinates Network, 5498, 1

\bibitem[{{Golenetskii} {et~al.}(2006{\natexlab{b}}){Golenetskii}, {Aptekar},
  {Mazets}, {Pal'Shin}, {Frederiks}, \& {Cline}}]{golenetskii06b}
{Golenetskii}, S., {Aptekar}, R., {Mazets}, E., {et~al.} 2006{\natexlab{b}},
  GRB Coordinates Network, 5689, 1

\bibitem[{{Golenetskii} {et~al.}(2006{\natexlab{c}}){Golenetskii}, {Aptekar},
  {Mazets}, {Pal'Shin}, {Frederiks}, \& {Cline}}]{golenetskii06c}
{Golenetskii}, S., {Aptekar}, R., {Mazets}, E., {et~al.} 2006{\natexlab{c}},
  GRB Coordinates Network, 5841, 1

\bibitem[{{Golenetskii} {et~al.}(2006{\natexlab{d}}){Golenetskii}, {Aptekar},
  {Mazets}, {Pal'shin}, {Frederiks}, \& {Cline}}]{golenetskii06d}
{Golenetskii}, S., {Aptekar}, R., {Mazets}, E., {et~al.} 2006{\natexlab{d}},
  GRB Coordinates Network, 5984, 1

\bibitem[{{Golenetskii} {et~al.}(2007{\natexlab{a}}){Golenetskii}, {Aptekar},
  {Mazets}, {Pal'Shin}, {Frederiks}, \& {Cline}}]{golenetskii07b}
{Golenetskii}, S., {Aptekar}, R., {Mazets}, E., {et~al.} 2007{\natexlab{a}},
  GRB Coordinates Network, 6849, 1

\bibitem[{{Golenetskii} {et~al.}(2007{\natexlab{b}}){Golenetskii}, {Aptekar},
  {Mazets}, {Pal'Shin}, {Frederiks}, \& {Cline}}]{golenetskii07c}
{Golenetskii}, S., {Aptekar}, R., {Mazets}, E., {et~al.} 2007{\natexlab{b}},
  GRB Coordinates Network, 6867, 1

\bibitem[{{Golenetskii} {et~al.}(2008{\natexlab{a}}){Golenetskii}, {Aptekar},
  {Mazets}, {Pal'Shin}, {Frederiks}, \& {Cline}}]{golenetskii08b}
{Golenetskii}, S., {Aptekar}, R., {Mazets}, E., {et~al.} 2008{\natexlab{a}},
  GRB Coordinates Network, 7219, 1

\bibitem[{{Golenetskii} {et~al.}(2008{\natexlab{b}}){Golenetskii}, {Aptekar},
  {Mazets}, {Pal'Shin}, {Frederiks}, \& {Cline}}]{golenetskii08c}
{Golenetskii}, S., {Aptekar}, R., {Mazets}, E., {et~al.} 2008{\natexlab{b}},
  GRB Coordinates Network, 7263, 1

\bibitem[{{Golenetskii} {et~al.}(2008{\natexlab{c}}){Golenetskii}, {Aptekar},
  {Mazets}, {Pal'Shin}, {Frederiks}, \& {Cline}}]{golenetskii08d}
{Golenetskii}, S., {Aptekar}, R., {Mazets}, E., {et~al.} 2008{\natexlab{c}},
  GRB Coordinates Network, 7482, 1

\bibitem[{{Golenetskii} {et~al.}(2008{\natexlab{d}}){Golenetskii}, {Aptekar},
  {Mazets}, {Pal'Shin}, {Frederiks}, \& {Cline}}]{golenetskii08e}
{Golenetskii}, S., {Aptekar}, R., {Mazets}, E., {et~al.} 2008{\natexlab{d}},
  GRB Coordinates Network, 7548, 1

\bibitem[{{Golenetskii} {et~al.}(2008{\natexlab{e}}){Golenetskii}, {Aptekar},
  {Mazets}, {Pal'Shin}, {Frederiks}, \& {Cline}}]{golenetskii08f}
{Golenetskii}, S., {Aptekar}, R., {Mazets}, E., {et~al.} 2008{\natexlab{e}},
  GRB Coordinates Network, 7751, 1

\bibitem[{{Golenetskii} {et~al.}(2007{\natexlab{c}}){Golenetskii}, {Aptekar},
  {Mazets}, {Pal'Shin}, {Frederiks}, {Cline}, {Cummings}, {Barthelmy},
  {Gehrels}, {Krimm}, {von Kienlin}, {Lichti}, {Rau}, {G\"{o}tz}, {Mereghetti},
  {Yamaoka}, {Ohno}, {Fukazawa}, {Takahashi}, {Tashiro}, {Terada}, {Murakami},
  {Makishima}, {Hurley}, {Smith}, {Lin}, {McTiernan}, {Schwartz}, {Wigger},
  {Hajdas}, \& {Zehnder}}]{golenetskii07a}
{Golenetskii}, S., {Aptekar}, R., {Mazets}, E., {et~al.} 2007{\natexlab{c}},
  GRB Coordinates Network, 6089, 1

\bibitem[{{Golenetskii} {et~al.}(2008{\natexlab{f}}){Golenetskii}, {Aptekar},
  {Mazets}, {Pal'Shin}, {Frederiks}, {Hurley}, {Cline}, {Boynton}, {Fellows},
  {Harshman}, {Shinohara}, {Starr}, {Mitrofanov}, {Golovin}, {Litvak}, {Sanin},
  {Yamaoka}, {Ohno}, {Fukazawa}, {Takahashi}, {Tashiro}, {Terada}, {Murakami},
  {Makishima}, {von Kienlin}, {Lichti}, {Rau}, {Marisaldi}, {Fuschino},
  {Galli}, {Labanti}, {Del}, {Lazzarotto}, {Pacciani}, {Soffitta}, {Cummings},
  {Barthelmy}, {Gehrels}, \& {Krimm}}]{golenetskii08a}
{Golenetskii}, S., {Aptekar}, R., {Mazets}, E., {et~al.} 2008{\natexlab{f}},
  GRB Coordinates Network, 7256, 1

\bibitem[{{Golenetskii} {et~al.}(2008{\natexlab{g}}){Golenetskii}, {Aptekar},
  {Mazets}, {Pal'Shin}, {Frederiks}, {Oleynik}, {Ulanov}, {Svinkin}, \&
  {Cline}}]{golenetskii08g}
{Golenetskii}, S., {Aptekar}, R., {Mazets}, E., {et~al.} 2008{\natexlab{g}},
  GRB Coordinates Network, 7862, 1

\bibitem[{{Golenetskii} {et~al.}(2008{\natexlab{h}}){Golenetskii}, {Aptekar},
  {Mazets}, {Pal'Shin}, {Frederiks}, {Oleynik}, {Ulanov}, {Svinkin}, \&
  {Cline}}]{golenetskii08h}
{Golenetskii}, S., {Aptekar}, R., {Mazets}, E., {et~al.} 2008{\natexlab{h}},
  GRB Coordinates Network, 7884, 1

\bibitem[{{Golenetskii} {et~al.}(2008{\natexlab{i}}){Golenetskii}, {Aptekar},
  {Mazets}, {Pal'Shin}, {Frederiks}, {Oleynik}, {Ulanov}, {Svinkin}, \&
  {Cline}}]{golenetskii08i}
{Golenetskii}, S., {Aptekar}, R., {Mazets}, E., {et~al.} 2008{\natexlab{i}},
  GRB Coordinates Network, 7995, 1

\bibitem[{{Golenetskii} {et~al.}(2008{\natexlab{j}}){Golenetskii}, {Aptekar},
  {Mazets}, {Pal'Shin}, {Frederiks}, {Oleynik}, {Ulanov}, {Svinkin}, \&
  {Cline}}]{golenetskii08j}
{Golenetskii}, S., {Aptekar}, R., {Mazets}, E., {et~al.} 2008{\natexlab{j}},
  GRB Coordinates Network, 8015, 1

\bibitem[{{Golenetskii} {et~al.}(2009{\natexlab{a}}){Golenetskii}, {Aptekar},
  {Mazets}, {Pal'Shin}, {Frederiks}, {Oleynik}, {Ulanov}, {Svinkin}, \&
  {Cline}}]{golenetskii09a}
{Golenetskii}, S., {Aptekar}, R., {Mazets}, E., {et~al.} 2009{\natexlab{a}},
  GRB Coordinates Network, 9553, 1

\bibitem[{{Golenetskii} {et~al.}(2009{\natexlab{b}}){Golenetskii}, {Aptekar},
  {Mazets}, {Pal'Shin}, {Frederiks}, {Oleynik}, {Ulanov}, {Svinkin}, \&
  {Cline}}]{golenetskii09b}
{Golenetskii}, S., {Aptekar}, R., {Mazets}, E., {et~al.} 2009{\natexlab{b}},
  GRB Coordinates Network, 9595, 1

\bibitem[{{G\"{o}tz} {et~al.}(2008){G\"{o}tz}, {Paizis}, {Mereghetti},
  {Beckmann}, {Beck}, {Manousakis}, \& {Borkowski}}]{gotz08}
{G\"{o}tz}, D., {Paizis}, A., {Mereghetti}, S., {et~al.} 2008, GRB Coordinates
  Network, 8002, 1

\bibitem[{{Grupe} {et~al.}(2006){Grupe}, {Barbier}, {Brown}, {Chester},
  {Cummings}, {Gehrels}, {Holland}, {Kennea}, {Marshall}, {McLean}, {Palmer},
  {Racusin}, \& {Stamatikos}}]{grupe06}
{Grupe}, D., {Barbier}, L.~M., {Brown}, P.~J., {et~al.} 2006, GRB Coordinates
  Network, 5954, 1

\bibitem[{{Guidorzi} {et~al.}(2011){Guidorzi}, {Lacapra}, {Frontera},
  {Montanari}, {Amati}, {Calura}, {Nicastro}, \& {Orlandini}}]{guidorzi11}
{Guidorzi}, C., {Lacapra}, M., {Frontera}, F., {et~al.} 2011, \aap, 526, A49

\bibitem[{{Hajdas} {et~al.}(2003){Hajdas}, {B{\"u}hler}, {Eggel}, {Favre},
  {Mchedlishvili}, \& {Zehnder}}]{hajdas03}
{Hajdas}, W., {B{\"u}hler}, P., {Eggel}, C., {et~al.} 2003, \aap, 411, L43

\bibitem[{{Hurley} {et~al.}(2008){Hurley}, {Cline}, {Boynton}, {Fellows},
  {Harshman}, {Shinohara}, {Starr}, {Mitrofanov}, {Golovin}, {Litvak}, {Sanin},
  {Golenetskii}, {Aptekar}, {Mazets}, {Pal'Shin}, {Frederiks}, {Goldsten},
  {Yamaoka}, {Ohno}, {Fukazawa}, {Takahashi}, {Tashiro}, {Terada}, {Murakami},
  {Makishima}, {Cummings}, {Barthelmy}, {Gehrels}, {Krimm}, {Del},
  {Evangelista}, {Lapshov}, {Rapisarda}, {Giuliani}, {Fuschino}, {Marisaldi},
  {von Kienlin}, {Lichti}, \& {Rau}}]{hurley08}
{Hurley}, K., {Cline}, T., {Boynton}, W., {et~al.} 2008, GRB Coordinates
  Network, 7211, 1

\bibitem[{{Hurley} {et~al.}(2006{\natexlab{a}}){Hurley}, {Cline},
  {Golenetskii}, {Mazets}, {Pal'Shin}, {Frederiks}, {Smith}, {Lin},
  {McTiernan}, {Bellm}, {Schwartz}, {Wigger}, {Hajdas}, {Zehnder}, {von
  Kienlin}, {Lichti}, {Rau}, {Yamaoka}, {Ohno}, {Fukazawa}, {Takahashi},
  {Tashiro}, {Terada}, {Murakami}, {Makishima}, {Barthelmy}, {Cummings},
  {Palmer}, {Gehrels}, \& {Krimm}}]{hurley06b}
{Hurley}, K., {Cline}, T., {Golenetskii}, S., {et~al.} 2006{\natexlab{a}}, GRB
  Coordinates Network, 5684, 1

\bibitem[{{Hurley} {et~al.}(2006{\natexlab{b}}){Hurley}, {Cline}, {Mitrofanov},
  {Kozyrev}, {Litvak}, {Sanin}, {Tret'yakov}, {Parshukov}, {Boynton},
  {Fellows}, {Harshman}, {Shinohara}, {Starr}, {Golenetskii}, {Mazets},
  {Pal'Shin}, {Frederiks}, {Smith}, {Lin}, {McTiernan}, {Schwartz}, {Wigger},
  {Hajdas}, {Zehnder}, {von Kienlin}, {Lichti}, \& {Rau}}]{hurley06a}
{Hurley}, K., {Cline}, T., {Mitrofanov}, I., {et~al.} 2006{\natexlab{b}}, GRB
  Coordinates Network, 5407, 1

\bibitem[{{Hurley} {et~al.}(2004){Hurley}, {Rau}, {von Kienlin}, \&
  {Lichti}}]{hurley04}
{Hurley}, K., {Rau}, A., {von Kienlin}, A., \& {Lichti}, G. 2004, in ESA
  Special Publication, Vol. 552, 5th INTEGRAL Workshop on the INTEGRAL
  Universe, ed. {V.~Schoenfelder, G.~Lichti, \& C.~Winkler}, 645

\bibitem[{{Kaneko} {et~al.}(2006){Kaneko}, {Preece}, {Briggs}, {Paciesas},
  {Meegan}, \& {Band}}]{kaneko06}
{Kaneko}, Y., {Preece}, R.~D., {Briggs}, M.~S., {et~al.} 2006, \apjs, 166, 298

\bibitem[{{Kommers} {et~al.}(2000){Kommers}, {Lewin}, {Kouveliotou}, {van
  Paradijs}, {Pendleton}, {Meegan}, \& {Fishman}}]{kommers00}
{Kommers}, J.~M., {Lewin}, W.~H.~G., {Kouveliotou}, C., {et~al.} 2000, \apj,
  533, 696

\bibitem[{{Koshut} {et~al.}(1996){Koshut}, {Paciesas}, {Kouveliotou}, {van
  Paradijs}, {Pendleton}, {Fishman}, \& {Meegan}}]{koshut}
{Koshut}, T.~M., {Paciesas}, W.~S., {Kouveliotou}, C., {et~al.} 1996, \apj,
  463, 570

\bibitem[{{Lin} {et~al.}(2002){Lin}, {Dennis}, {Hurford}, {Smith}, {Zehnder},
  {Harvey}, {Curtis}, {Pankow}, {Turin}, {Bester}, {Csillaghy}, {Lewis},
  {Madden}, {van Beek}, {Appleby}, {Raudorf}, {McTiernan}, {Ramaty}, {Schmahl},
  {Schwartz}, {Krucker}, {Abiad}, {Quinn}, {Berg}, {Hashii}, {Sterling},
  {Jackson}, {Pratt}, {Campbell}, {Malone}, {Landis}, {Barrington-Leigh},
  {Slassi-Sennou}, {Cork}, {Clark}, {Amato}, {Orwig}, {Boyle}, {Banks},
  {Shirey}, {Tolbert}, {Zarro}, {Snow}, {Thomsen}, {Henneck}, {McHedlishvili},
  {Ming}, {Fivian}, {Jordan}, {Wanner}, {Crubb}, {Preble}, {Matranga}, {Benz},
  {Hudson}, {Canfield}, {Holman}, {Crannell}, {Kosugi}, {Emslie}, {Vilmer},
  {Brown}, {Johns-Krull}, {Aschwanden}, {Metcalf}, \& {Conway}}]{lin02}
{Lin}, R.~P., {Dennis}, B.~R., {Hurford}, G.~J., {et~al.} 2002, \solphys, 210,
  3

\bibitem[{{Malaguti} {et~al.}(2003){Malaguti}, {Bazzano}, {Beckmann}, {Bird},
  {Del Santo}, {Di Cocco}, {Foschini}, {Goldoni}, {G{\"o}tz}, {Mereghetti},
  {Paizis}, {Segreto}, {Skinner}, {Ubertini}, \& {von Kienlin}}]{malaguti03}
{Malaguti}, G., {Bazzano}, A., {Beckmann}, V., {et~al.} 2003, \aap, 411, L307

\bibitem[{{Mangano} {et~al.}(2008){Mangano}, {Cummings}, {Cusumano}, {Gehrels},
  {La Parola}, {Markwardt}, {Osborne}, {Sbarufatti}, \& {vanden
  Berk}}]{mangano08}
{Mangano}, V., {Cummings}, J.~R., {Cusumano}, G., {et~al.} 2008, GRB
  Coordinates Network, 7847, 1

\bibitem[{{Markwardt} {et~al.}(2008){Markwardt}, {Barthelmy}, {Baumgartner},
  {Cusumano}, {Gronwall}, {Guidorzi}, {Holland}, {Mao}, {Marshall}, {O'Brien},
  {Osborne}, {Page}, {Stamatikos}, {Starling}, {Tagliaferri}, \&
  {Ziaeepour}}]{markwardt08}
{Markwardt}, C.~B., {Barthelmy}, S.~D., {Baumgartner}, W.~H., {et~al.} 2008,
  GRB Coordinates Network, 7873, 1

\bibitem[{{Marshall} {et~al.}(2008){Marshall}, {Beardmore}, {Burrows}, {Evans},
  {Gehrels}, {Gronwall}, {Guidorzi}, {Holland}, {Mangano}, {O'Brien},
  {Osborne}, {Page}, {Palmer}, {Perez}, {Perri}, {Preger}, {Romano},
  {Stamatikos}, {Starling}, {Ukwatta}, {Ward}, \& {Ziaeepour}}]{marshall08}
{Marshall}, F.~E., {Beardmore}, A.~P., {Burrows}, D.~N., {et~al.} 2008, GRB
  Coordinates Network, 7988, 1

\bibitem[{{McGlynn} {et~al.}(2009){McGlynn}, {Foley}, {McBreen}, {Hanlon},
  {McBreen}, {Clark}, {Dean}, {Martin-Carrillo}, \& {O'Connor}}]{mcglynn09}
{McGlynn}, S., {Foley}, S., {McBreen}, B., {et~al.} 2009, \aap, 499, 465

\bibitem[{{Mereghetti} {et~al.}(2003){Mereghetti}, {G{\"o}tz}, {Borkowski},
  {Walter}, \& {Pedersen}}]{mereghetti03}
{Mereghetti}, S., {G{\"o}tz}, D., {Borkowski}, J., {Walter}, R., \& {Pedersen},
  H. 2003, \aap, 411, L291

\bibitem[{{Mereghetti} {et~al.}(2006{\natexlab{a}}){Mereghetti}, {Paizis},
  {G\"{o}tz}, {Petry}, {Mowlavi}, {Beck}, \& {Borkowski}}]{mereghetti06b}
{Mereghetti}, S., {Paizis}, A., {G\"{o}tz}, D., {et~al.} 2006{\natexlab{a}},
  GRB Coordinates Network, 5834, 1

\bibitem[{{Mereghetti} {et~al.}(2006{\natexlab{b}}){Mereghetti}, {Paizis},
  {G\"{o}tz}, {Shaw}, {Beck}, \& {Borkowski}}]{mereghetti06a}
{Mereghetti}, S., {Paizis}, A., {G\"{o}tz}, D., {et~al.} 2006{\natexlab{b}},
  GRB Coordinates Network, 5491, 1

\bibitem[{{Nakagawa} {et~al.}(2009){Nakagawa}, {Tamagawa}, {Ohmori},
  {Daikyuji}, {Sonoda}, {Kono}, {Hayashi}, {Noda}, {Nishioka}, {Yamauchi},
  {Ohno}, {Suzuki}, {Kokubun}, {Takahashi}, {Yamaoka}, {Sugita}, {Hong},
  {Vasquez}, {Hanabata}, {Uehara}, {Fukazawa}, {Iwakiri}, {Tashiro}, {Terada},
  {Endo}, {Onda}, {Sugasahara}, {Urata}, {Enoto}, {Nakazawa}, \&
  {Makishima}}]{nakagawa09}
{Nakagawa}, Y.~E., {Tamagawa}, T., {Ohmori}, N., {et~al.} 2009, GRB Coordinates
  Network, 9577, 1

\bibitem[{{Perri} {et~al.}(2008){Perri}, {Burrows}, {Campana}, {Evans},
  {Gehrels}, {Guidorzi}, {Markwardt}, {Page}, {Palmer}, {Perez}, {Romano},
  {Starling}, {Stratta}, {vanden Berk}, {Ward}, \& {Ziaeepour}}]{perri08}
{Perri}, M., {Burrows}, D.~N., {Campana}, S., {et~al.} 2008, GRB Coordinates
  Network, 7525, 1

\bibitem[{{Racusin} {et~al.}(2008){Racusin}, {Gehrels}, {Holland}, {Kennea},
  {Markwardt}, {Pagani}, {Palmer}, \& {Stamatikos}}]{racusin08}
{Racusin}, J.~L., {Gehrels}, N., {Holland}, S.~T., {et~al.} 2008, GRB
  Coordinates Network, 7427, 1

\bibitem[{{Rapisarda} {et~al.}(2008){Rapisarda}, {Costa}, {Del Monte},
  {Donnarumma}, {Evangelista}, {Feroci}, {Lapshov}, {Lazzarotto}, {Pacciani},
  {Soffitta}, {Giuliani}, {Vercellone}, {Chen}, {Mereghetti}, {Pellizzoni},
  {Perotti}, {Fornari}, {Fiorini}, {Caraveo}, {Bulgarelli}, {Gianotti},
  {Trifoglio}, {Di Cocco}, {Labanti}, {Fuschino}, {Marisaldi}, {Galli},
  {Tavani}, {Pucella}, {D'Ammando}, {Vittorini}, {Argan}, {Trois},
  {Barbiellini}, {Longo}, {Cattaneo}, {Picozza}, {Morselli}, {Prest},
  {Vallazza}, {Lipari}, {Zanello}, {Giommi}, {Pittori}, {Preger}, {Verrecchia},
  {Salotti}, {Hurley}, {Mitrofanov}, {Golovin}, {Litvak}, {Sanin}, {Boynton},
  {Fellows}, {Harshman}, {Shinohara}, \& {Starr}}]{rapisarda08}
{Rapisarda}, M., {Costa}, E., {Del Monte}, E., {et~al.} 2008, GRB Coordinates
  Network, 7715, 1

\bibitem[{{Rau}(2009)}]{rau09}
{Rau}, A. 2009, GRB Coordinates Network, 9566, 1

\bibitem[{{Rau} {et~al.}(2005){Rau}, {Kienlin}, {Hurley}, \& {Lichti}}]{rau}
{Rau}, A., {Kienlin}, A.~V., {Hurley}, K., \& {Lichti}, G.~G. 2005, \aap, 438,
  1175

\bibitem[{{Schady} {et~al.}(2009){Schady}, {Baumgartner}, {Beardmore},
  {Campana}, {Curran}, {Guidorzi}, {Kennea}, {Mao}, {Margutti}, {Osborne},
  {Page}, {Romano}, {Siegel}, {Stratta}, \& {Ukwatta}}]{schady09}
{Schady}, P., {Baumgartner}, W.~H., {Beardmore}, A.~P., {et~al.} 2009, GRB
  Coordinates Network, 9512, 1

\bibitem[{{Schady} {et~al.}(2007){Schady}, {Evans}, {Gehrels}, {Guidorzi},
  {Holland}, {Kennea}, {Markwardt}, {O'Brien}, {Page}, {Palmer}, {Sato}, \&
  {Starling}}]{schady07}
{Schady}, P., {Evans}, P.~A., {Gehrels}, N., {et~al.} 2007, GRB Coordinates
  Network, 6837, 1

\bibitem[{{Segreto} {et~al.}(2003){Segreto}, {Labanti}, {Bazzano}, {Bird},
  {Celesti}, \& {Marisaldi}}]{segreto}
{Segreto}, A., {Labanti}, C., {Bazzano}, A., {et~al.} 2003, \aap, 411, L215

\bibitem[{{Ubertini} {et~al.}(2003){Ubertini}, {Lebrun}, {Di Cocco}, {Bazzano},
  {Bird}, {Broenstad}, {Goldwurm}, {La Rosa}, {Labanti}, {Laurent}, {Mirabel},
  {Quadrini}, {Ramsey}, {Reglero}, {Sabau}, {Sacco}, {Staubert}, {Vigroux},
  {Weisskopf}, \& {Zdziarski}}]{ubertini}
{Ubertini}, P., {Lebrun}, F., {Di Cocco}, G., {et~al.} 2003, \aap, 411, L131

\bibitem[{{van der Horst}(2009)}]{horst09}
{van der Horst}, A.~J. 2009, GRB Coordinates Network, 9760, 1

\bibitem[{{Vedrenne} {et~al.}(2003){Vedrenne}, {Roques}, {Sch{\"o}nfelder},
  {Mandrou}, {Lichti}, {von Kienlin}, {Cordier}, {Schanne}, {Kn{\"o}dlseder},
  {Skinner}, {Jean}, {Sanchez}, {Caraveo}, {Teegarden}, {von Ballmoos},
  {Bouchet}, {Paul}, {Matteson}, {Boggs}, {Wunderer}, {Leleux},
  {Weidenspointner}, {Durouchoux}, {Diehl}, {Strong}, {Cass{\'e}}, {Clair}, \&
  {Andr{\'e}}}]{vedrenne03}
{Vedrenne}, G., {Roques}, J., {Sch{\"o}nfelder}, V., {et~al.} 2003, \aap, 411,
  L63

\bibitem[{{Vianello} {et~al.}(2009){Vianello}, {G{\"o}tz}, \&
  {Mereghetti}}]{vianello09}
{Vianello}, G., {G{\"o}tz}, D., \& {Mereghetti}, S. 2009, \aap, 495, 1005

\bibitem[{{Vigan{\`o}} \& {Mereghetti}(2009)}]{vigano09}
{Vigan{\`o}}, D. \& {Mereghetti}, S. 2009, ArXiv e-prints 0912.5329

\bibitem[{{von Kienlin}(2009{\natexlab{a}})}]{vkienlin09a}
{von Kienlin}, A. 2009{\natexlab{a}}, GRB Coordinates Network, 9447, 1

\bibitem[{{von Kienlin}(2009{\natexlab{b}})}]{vkienlin09b}
{von Kienlin}, A. 2009{\natexlab{b}}, GRB Coordinates Network, 9579, 1

\bibitem[{{Winkler} {et~al.}(2003){Winkler}, {Courvoisier}, {Di Cocco},
  {Gehrels}, {Gim{\'e}nez}, {Grebenev}, {Hermsen}, {Mas-Hesse}, {Lebrun},
  {Lund}, {Palumbo}, {Paul}, {Roques}, {Schnopper}, {Sch{\"o}nfelder},
  {Sunyaev}, {Teegarden}, {Ubertini}, {Vedrenne}, \& {Dean}}]{winkler03}
{Winkler}, C., {Courvoisier}, T., {Di Cocco}, G., {et~al.} 2003, \aap, 411, L1

\end{thebibliography}
\end{document}